\documentclass[reprint, nofootinbib]{revtex4-2}
\PassOptionsToPackage{table}{xcolor}
\usepackage{amsmath}
\usepackage{amssymb}
\usepackage{multirow}
\usepackage{booktabs}

\usepackage{siunitx}

\usepackage[table]{xcolor} 
\usepackage{tabularx}

\usepackage{soul}

\usepackage{graphicx}% Include figure files
\usepackage{dcolumn}% Align table columns on decimal point
\usepackage{bm}% bold math
\usepackage{float}
\usepackage{hyperref} % adds hyper links inside the generated pdf file
\hypersetup{
	colorlinks=true,       % false: boxed links; true: colored links
	linkcolor=black,        % color of internal links
	citecolor=black,        % color of links to bibliography
	filecolor=magenta,     % color of file links
	urlcolor=black         
}
\begin{document}
\title{Using Large Language Models in Physics Education}

\author{Jonah L. Donaldson}

\author{Aliya Nawaz}

\author{Konstantinos Doran}

\author{Alysta Lim}

\author{Mario Campanelli}
 \altaffiliation{Corresponding author, Professor of Physics University College London}
 \email{m.campanelli@ucl.ac.uk}
 
\affiliation{%
 Department of Physics and Astronomy\\
 University College London
}%

%\begin{center}
%    \includegraphics[width=0.8\textwidth]%{Figures/Maxwell.jpg}
%\end{center}

\begin{abstract}
The rapid advancement of Large Language Models (LLMs) has introduced new possibilities and challenges in physics education, necessitating rigorous evaluation of their capabilities as both problem solvers and automated assessors. This paper presents the results of three complementary studies that evaluated frontier models released between mid-2024 and late-2025. Models were assessed on their ability to generate accurate, step-by-step solutions to university-level physics problems in Classical Mechanics, Electromagnetism, and Quantum Mechanics, and subsequently on their reliability in grading student solutions against a formal mark scheme. The results indicate a clear trajectory toward benchmark saturation in text-based reasoning, with recent architectures (such as ChatGPT-5.1 and Gemini 3.0 Pro) achieving near-perfect scores. Furthermore, recent advances in native multimodal integration have resolved previous limitations in spatial geometry and topological interpretation, enabling models to accurately process accompanying diagrams. As automated assessors, newer models demonstrated significant improvements in alignment with human grading, heavily mitigating the systemic over-marking observed in earlier iterations. However, while models reliably evaluate fully correct handwritten work, assigning partial credit to flawed or incomplete reasoning remains a persistent challenge. These findings suggest that as of late 2025, LLMs offer viable support for both independent student learning and instructional automation, provided their limitations in evaluating ambiguous reasoning are actively managed.
\end{abstract}

\maketitle

% Table of Contents
%\tableofcontents
\newpage

\section{Introduction}
\subsection{Background}

Prior to the recent surge in artificial intelligence development, research had already established LLM proficiency across various natural language processing (NLP) tasks, including summarisation, translation, and common-sense reasoning \cite{brown2020language, rae2021scaling, ouyang2022training}. Concurrently, investigators began applying these models to educational frameworks, exploring their utility in automated tutoring, lesson plan generation, and problem-solving strategies \cite{zawacki2019systematic, steenbergen2014meta}. While initial findings indicated that LLMs could effectively personalise student feedback, early architectures faced significant technical bottlenecks. Most notably, these included inconsistent multi-step mathematical reasoning and the tendency to 'hallucinate'—generating factually incorrect or fabricated information \cite{cobbe2021training, ji2022survey, lewis2020retrieval}.

This landscape shifted dramatically with the release of ChatGPT-3.5 in November 2022, catalysing an accelerated development phase in the artificial intelligence (AI) sector. By demonstrating functional, advanced logical reasoning and human-like response generation, the model established a new performance baseline for Large Language Models (LLMs), shifting both research priorities and market expectations. This transition prompted major technology firms (including Meta and Google) to accelerate their own development cycles to maintain competitive parity. Consequently, the intensified market pressure and high perceived valuation of generative AI have significantly compressed traditional development and deployment timelines \cite{wang2025large}.

The increased scale of these models has led to emergent capabilities beyond simple pattern recognition, including complex synthesis and structured reasoning \cite{wei2022emergent, webb2023emergent}. Early architectures already demonstrated the potential to achieve first-class grades at the university level, with subsequent iterations frequently outperforming the majority of human students in controlled examinations \cite{yeadon2022death, scarfe2024real}. Furthermore, benchmarks such as the GPQA diamond benchmark document this trend, showing a clear, consistent upward trajectory in reasoning proficiency across successive model iterations \cite{artificialanalysis2026gpqa}. As a result, LLMs have begun to approximate aspects of the analytical and problem-solving skills traditionally cultivated through formal education.

This appropriation of what was previously considered an exclusively human cognitive domain has raised significant concerns among pedagogists about its potential impact on the overall educational landscape. For example, a 2023 survey conducted in Sweden among people aged 15 to 24 years reported that 55\% had used AI in ways they believed were not permitted in their educational setting \cite{jonsson2025rapportslapp}. Furthermore, the low cost to entry of LLMs may therefore encourage a reliance on automated assistance, potentially affecting students’ development of independent problem-solving and critical thinking skills, while also exposing them to incorrect or misleading AI outputs \cite{gregorcic2023chatgpt, polverini2023how}. Finally, the lack of consistent AI detection methods prevented institutions from implementing hard policies to deter AI plagiarism, creating an institutional exigency to properly adapt to this new technology \cite{russellgroup2023ai}.

\subsection{Motivation}

%This study does not posit LLMs as a replacement for traditional instruction or independent learning. Instead, it adopts a human-in-the-loop (HITL) philosophy, analysing these models strictly as augmentative tools. This adaptation begins with defining the effective parameters of LLM usage, ensuring that students learn to leverage the technology as a cognitive supplement rather than a substitute for critical thought. 
This paper presents three complementary studies examining the capabilities of LLM architectures from mid-2024 to late-2025. These studies benchmark model performance in technical problem-solving and academic feedback generation, evaluating both solution accuracy and the pedagogical utility of automated responses when paired with human oversight.

Regarding technical problem-solving, the capacity of modern LLMs to consistently resolve advanced queries with high accuracy underscores their viability as robust educational tools. For students, these systems function as interactive learning assistants, offering a twofold pedagogical utility. First, within independent study, the conversational architecture of LLMs enables students to engage in iterative dialogue as they navigate complex topics probing nuances to resolve misconceptions through continuous questioning, thereby fostering stronger learning retention and critical thinking \cite{delic2016socratic}. Second, the ubiquity and accessibility of AI systems present significant potential to democratise high-quality instructional support. By providing personalised assistance independent of a student’s socioeconomic status or social capital, LLM technology could help narrow long-standing educational disparities \cite{halpern2005social}.

Furthermore, the ability of LLMs to provide accurate, instant feedback on student submissions represents a fundamental shift in the academic feedback cycle. For learners, mitigating feedback latency—the delay typically required to assess a solution’s correctness—enables real-time, independent evaluation. This immediacy mirrors the rapid response mechanisms of gamified learning environments, enhancing engagement and fostering more persistent learning states \cite{hamari2014does}. For educators, integrating LLM automation offers substantial gains in operational efficiency. By leveraging AI to manage repetitive administrative tasks or routine lesson preparation, instructors can redirect their focus toward higher-level pedagogical engagement \cite{paris2022instructors, chu2025llm}. A precedent for this shift is already evident; in the UK, for instance, the use of AI in assessment and grading processes is increasingly permitted. However, the burden now falls upon the academic community to outline the practical limitations and establish guidelines for such applications \cite{shearing2025teachers, dst2025teachers, govuk2026generative}.

Ultimately, while LLMs present significant pedagogical advantages, they are not without inherent vulnerabilities. The risks of cognitive over-reliance and institutional complacency remain, even when robust guidance is provided. By establishing an empirical baseline, this study provides a framework for defining the performance parameters of these models on advanced exam questions while tracking their iterative advancement over time. Furthermore, although the primary objective centres on performance benchmarking, qualitative analysis yielded notable insights into model behaviour, offering a preliminary examination of the heuristic biases stemming from underlying training methodologies and architectural design.

\section{Methodology}

\subsection{Experimental Design}
This research was structured across three distinct studies. The initial investigation, Performance Benchmark 1 (PB1), spanned October 2024 to March 2025. This was followed by two concurrent longitudinal studies—Performance Benchmark 2 (PB2) and the Multimodal (MM) study—conducted between October 2025 and January 2026.

Each study employed a three-stage methodological pipeline across the selected cohort of LLMs (\ref{tab:models}):

\begin{itemize}
\item \textbf{Stage I: Solution Generation.} The models were tasked with solving a dataset of physics problems to generate a corpus of three independent answers per question. 
\item \textbf{Stage II: Human Evaluation.} To establish an objective performance baseline, these generated solutions were manually assessed against a standardised marking rubric.
\item \textbf{Stage III: AI Evaluation.} Using a targeted subset of the solutions, the LLMs were prompted to act as graders, facilitating a direct comparison of reliability between human and AI marking.
\end{itemize}

The different AI models used for the various stages are shown in Table \ref{tab:models}. The reason why different generations of models are used is not only historical: the evolution of model performance with its generation is connected to the additional features that these generations provide, allowing a better understanding of how these features behave in real-world scenarios.

\begin{table*}[t]
    \centering
    \renewcommand{\arraystretch}{1.3}
%    \begin{tabular}{|l|p{0.28\textwidth}|p{0.28\textwidth}|p{0.28\textwidth}|}
\begin{tabular}{|l|l|l|l|}
        \hline
        \textbf{Model Family} & \textbf{PB1} & \textbf{PB2} & \textbf{MM} \\ \hline
        \textbf{OpenAI} & ChatGPT-4o, o1, o3-mini-high & ChatGPT-5.1 & ChatGPT-4o, 5.1 \\ \hline
        \textbf{Google} & Gemini 1.5 Pro, 2.0 Flash & Gemini 2.5 Flash, 3.0 Pro & Gemini 2.5 Flash, 3.0 Pro \\ \hline
        \textbf{DeepSeek} & DeepSeek-V3 & DeepSeek-V3.2 & DeepSeek-V3.2* \\ \hline
    \end{tabular}
    \caption{LLMs evaluated across all studies, categorised by model family \cite{chatgpt2025, gemini2025, deepseek2025}. For brevity, specific model variant suffixes are omitted in subsequent text. Note: *DeepSeek-V3.2 was restricted to Stage III of the MM study due to native multimodal input limitations.}
    \label{tab:models}
\end{table*}

For Stages I and II of PB1, the study employed a predefined problem set (Set A) and its corresponding marking rubric, originally developed in prior research at University College London \cite{Moketal2025}. Crucially, the complete Set A dataset was utilised exclusively during these initial two stages to establish the human evaluation baseline. The set comprised 30 questions evenly distributed across Classical Mechanics (CM), Electromagnetism (EM), and Quantum Mechanics (QM). To mirror the structural composition of a standard undergraduate physics examination, each topic included two qualitative comprehension questions and eight quantitative calculation problems. Furthermore, six of these problems incorporated visual figures, facilitating a preliminary assessment of the models' multimodal processing capabilities. The maximum available marks for each question in this set are detailed in Table \ref{tab:set-a-marks}.

\begin{table}[H] 
\centering 
\begin{tabular}{|c|c|c|c|} 
\hline 
\textbf{Question \#} & \textbf{CM} & \textbf{QM} & \textbf{EM} \\ \hline 
1 & 4 & 5 & 8 \\ \hline 
2 & 6 & 4 & 5 \\ \hline 
3 & 6 & 9 & 14 \\ \hline 
4 & 18 & 20 & 11 \\ \hline 
5 & 9 & 7 & 11 \\ \hline 
6 & 16 & 17 & 5 \\ \hline 
7 & 13 & 8 & 6 \\ \hline 
8 & 10 & 9 & 6 \\ \hline 
9 & 7 & 17 & 15 \\ \hline 
10 & 8 & 16 & 7 \\ \hline 
\end{tabular} 
\caption{Available marks for Set A.}
\label{tab:set-a-marks}
\end{table}

For Stage III (AI Evaluation) of PB1, the dataset was distilled into an 18-question subset (Set B). To maintain the benchmarking integrity established by Set A, this subset was carefully prepared to preserve a representative balance across the core subject areas and question typologies. During this phase, all selected LLMs graded responses previously generated by ChatGPT-4o. This specific model was chosen as the primary grading target because its initial outputs demonstrated the highest variance in quality, providing the broad spectrum of correct, partially correct, and flawed reasoning paths required to rigorously test the evaluative sensitivity of the AI graders. For methodological consistency, all questions retained their original Set A numerical designations. Furthermore, a table detailing the question numbers included in Set B has been included below, Table \ref{tab:q-labels-map}.

\begin{table}[H]
\centering
\begin{tabular}{|c|c|c|c|}
\hline
\textbf{Question \#} & \textbf{CM} & \textbf{QM} & \textbf{EM} \\ \hline
1 & 2 & 1 & 2 \\ \hline
2 & 5 & 2 & 4 \\ \hline
3 & 7 & 3 & 6 \\ \hline
4 & 8 & 5 & 7 \\ \hline
5 & 9 & 6 & 9 \\ \hline
6 & 10 & 9 & 10 \\ \hline
\end{tabular}
\caption{Table of questions included in Set B}
\label{tab:q-labels-map}
\end{table}

PB2 functioned as a longitudinal extension of PB1, deploying Set B exclusively across all three experimental stages. Reusing these specific questions and their corresponding grading rubrics created a strictly controlled environment. This consistency allowed for an accurate assessment of how subsequent model generations advanced in both raw problem-solving proficiency and automated assessment capabilities over time.

The MM study utilised a newly developed dataset (Set C) comprising 15 questions evenly distributed across the three previously established subject areas. Set C was designed specifically for native multimodal assessment; every problem required the direct integration and interpretation of an accompanying diagram, providing a rigorous test of spatial and multimodal reasoning. The distribution of available marks for this unique set is outlined in Table \ref{tab:set-c-marks}. Lastly, to accommodate the large volume of data associated with these studies, a comprehensive repository on GitHub has been provided as a reference \cite{physics_llm_repo}.

\begin{table}[H] 
\centering 
\begin{tabular}{|c|c|c|c|} 
\hline 
\textbf{Question \#} & \textbf{CM} & \textbf{QM} & \textbf{EM} \\ \hline 
1 & 5 & 4 & 4 \\ \hline 
2 & 6 & 8 & 3 \\ \hline 
3 & 9 & 8 & 4 \\ \hline 
4 & 10 & 9 & 5 \\ \hline 
5 & 10 & 8 & 11 \\ \hline 
\end{tabular} 
\caption{Available marks for Set C}
\label{tab:set-c-marks}
\end{table}

\subsection{Experimental Workflow}
\subsubsection{Stage I: Solution Generation}

To maintain structural and formatting consistency across all trials, each question was transcribed into LaTeX before entry into the model. This conversion process leveraged the native multimodal capabilities of ChatGPT-4o to extract text and mathematical notation directly from the source PDFs. 

\begin{figure}[H]
    \centering
    \includegraphics[width=0.7\linewidth]{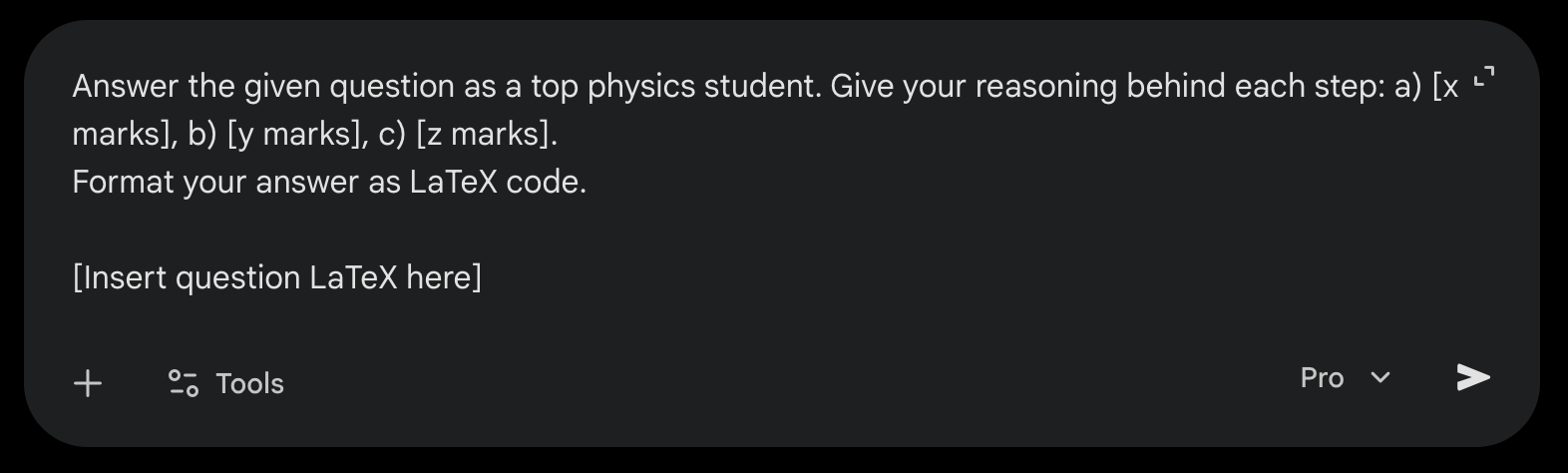}
    \caption{Template prompt inputted into the Gemini web interface, with the Pro architecture selected \cite{gemini2025}.}
    \label{fig:prompt_gem}
\end{figure}

To simulate real-world usage and facilitate cross-study analysis, all queries were submitted via the models' respective web interfaces using a predefined prompt template (Figure \ref{fig:prompt_gem}). To account for the inherent stochasticity of autoregressive generation, a consistency benchmark was established by generating three independent solutions per problem. Furthermore, to prevent context contamination, each solution was generated within a strictly isolated context window utilising zero-shot prompting.

The overarching prompt structure incorporated expert persona assignment, explicitly instructing the model to act as a “top physics student” to elicit a rigorous, mark-oriented methodology \cite{xu2023expertprompting}. Additionally, the models were directed to employ a step-by-step Chain-of-Thought (CoT) reasoning approach to bolster their problem-solving capabilities \cite{wei2022chain, wang2022self}. To appropriately calibrate the expected depth and complexity of the response, each prompt included a clear breakdown of the available marks. Finally, qualitative questions were limited to 200 words to enforce concision and prevent verbose hallucinations.

\begin{table*}[tbh]
\centering
\renewcommand{\arraystretch}{1.2}
\begin{tabular}{|l|l|l|}
\hline
\textbf{Study} & \textbf{Date Range} & \textbf{Models} \\
\hline
PB1 (Part 1) & 03/11/24 -- 20/12/24 & ChatGPT-4o, o1, Gemini 1.5 \\
PB1 (Part 2) & 28/01/25 -- 12/02/25 & ChatGPT-o3, Gemini 2.0, DeepSeek-V3 \\
PB2          & 11/11/25 -- 03/12/25 & All PB2 Models \\
MM           & 11/11/25 -- 03/12/25 & All MM Models \\
\hline
\end{tabular}
\caption{Data collection windows for Stage I: Solution Generation. Note, the break in PB1 data collection occurred because the models had not yet been released.}
\label{tab:stage1_phases}
\end{table*}

\subsubsection{Stage II: Human Evaluation}

For the human evaluation baseline (Stage II) of the PB studies, the established marking rubric from the UCL dataset \cite{Moketal2025} was utilised. Conversely, for the MM study (Set C), a bespoke marking scheme was developed. This rubric was constructed by delineating generalised solution pathways and apportioning marks across fundamental logical milestones (e.g., diagrammatic setup, variable identification, and algebraic execution).

To optimise evaluation efficiency across the extensive PB1 dataset, a tiered assessment protocol was implemented. The first generated solution for each query was evaluated and scored in full. Subsequent iterations were then scanned for critical procedural milestones and key mathematical outcomes. If a procedural deviation or error was identified in any iteration, all three independent solutions for that specific problem were retroactively marked in full. By contrast, the comparatively condensed datasets in PB2 and the MM study permitted a comprehensive, two-pass manual marking process for all outputs, comprising an initial grading phase and a secondary cross-examination to ensure evaluative consensus.

Across all evaluations, responses employing valid alternative methodologies were awarded full credit, provided sufficient deductive reasoning was demonstrated. Lastly, final quantitative answers were only credited when based on sound supporting logic; instances in which a model erroneously forced a correct final value from flawed antecedent steps received a score of zero.

\subsubsection{Stage III: AI Evaluation}

The evaluation prompt adopted a revised expert persona, positioning the models as "physics professors" to enforce a rigorous, pedagogical evaluative standard. The models were constrained to assign integer scores and provide generalised feedback. Furthermore, the input context strictly delimited the problem statements, candidate solutions, and the corresponding marking rubric.

Initial testing during Stage III of PB1 compared an "informed" condition (rubric provided) against a "blind" condition (only total marks provided). Results indicated that in the absence of a formal rubric, models exhibited severely degraded evaluative reliability. Consequently, the blind condition was omitted from all subsequent trials to isolate pure grading accuracy from the models' implicit rubric-generation capabilities.

Furthermore, PB1 initially employed a batch-grading format, wherein all three generated solutions for a given problem were evaluated simultaneously. However, this approach introduced two documented positional biases. The first was the "halo effect," in which the perceived quality of the initial solution inadvertently anchors the model's grading standard for subsequent answers \cite{zheng2023judging}. The second was the "lost in the middle" phenomenon, where an LLM's attention allocation degrades, causing it to overlook or misjudge information buried in the centre of an extended context window \cite{liu2024lost}. To mitigate these vulnerabilities, PB2 and the MM study transitioned to individualised inputs. By isolating each evaluation, the models' assessments remained strictly anchored to the objective rubric rather than the relative quality or structural position of adjacent responses.

Finally, for the MM study, Stage III (AI Evaluation) was adapted to specifically assess the models' capacity to transcribe and grade handwritten content. Using Set B as a foundation, both "perfect" and "imperfect" handwritten solutions were generated. The perfect solutions established a control for transcription accuracy, isolating errors stemming purely from visual misinterpretation. Conversely, the imperfect scripts were deliberately constructed to embed common student misconceptions, thereby testing higher-order evaluative judgment. 

Finally, preliminary testing indicated that native multimodal processing (direct image ingestion) consistently outperformed traditional optical character recognition (OCR) pipelines, such as Mathpix. This performance disparity was particularly pronounced when processing non-textual elements, such as free-body diagrams. Consequently, DeepSeek-V3.2—which lacked a native multimodal architecture and relied entirely on a supplementary text-conversion tool—was thus excluded from Stages I and II of the MM study.

\begin{table*}[tbh]
\centering
\renewcommand{\arraystretch}{1.2}
\begin{tabular}{|l|l|l|}
\hline
\textbf{Study} & \textbf{Date Range} & \textbf{Models} \\
\hline
PB1 & 14/03/25 -- 21/03/25 & All Models \\
PB2 & 13/01/26 -- 18/01/26 & All Models \\
MM  & 11/11/25 -- 03/12/25 & All Models \\
\hline
\end{tabular}
\caption{Data collection windows for Stage III: AI Evaluation.}
\label{tab:stage3_phases}
\end{table*}

\section{Results and Discussion}
\subsection{Human Evaluation}
\subsubsection{PB1}

Given the highly competitive nature of the LLM landscape during the study period, frontier models were frequently released in clustered cycles to maintain market parity. Consequently, the evaluated models have been categorised into three chronological generations:

\begin{itemize}
    \item \textbf{Generation 1} (May 2024): ChatGPT-4o and Gemini 1.5 Pro.
    \item \textbf{Generation 2} (December 2024): ChatGPT-o1 and Gemini 2.0 Flash.
    \item \textbf{Generation 3} (January 2025): DeepSeek-V3 and ChatGPT-o3-mini-high.
\end{itemize}

\begin{figure}[H]
    \centering
    \includegraphics[width=0.7\linewidth]{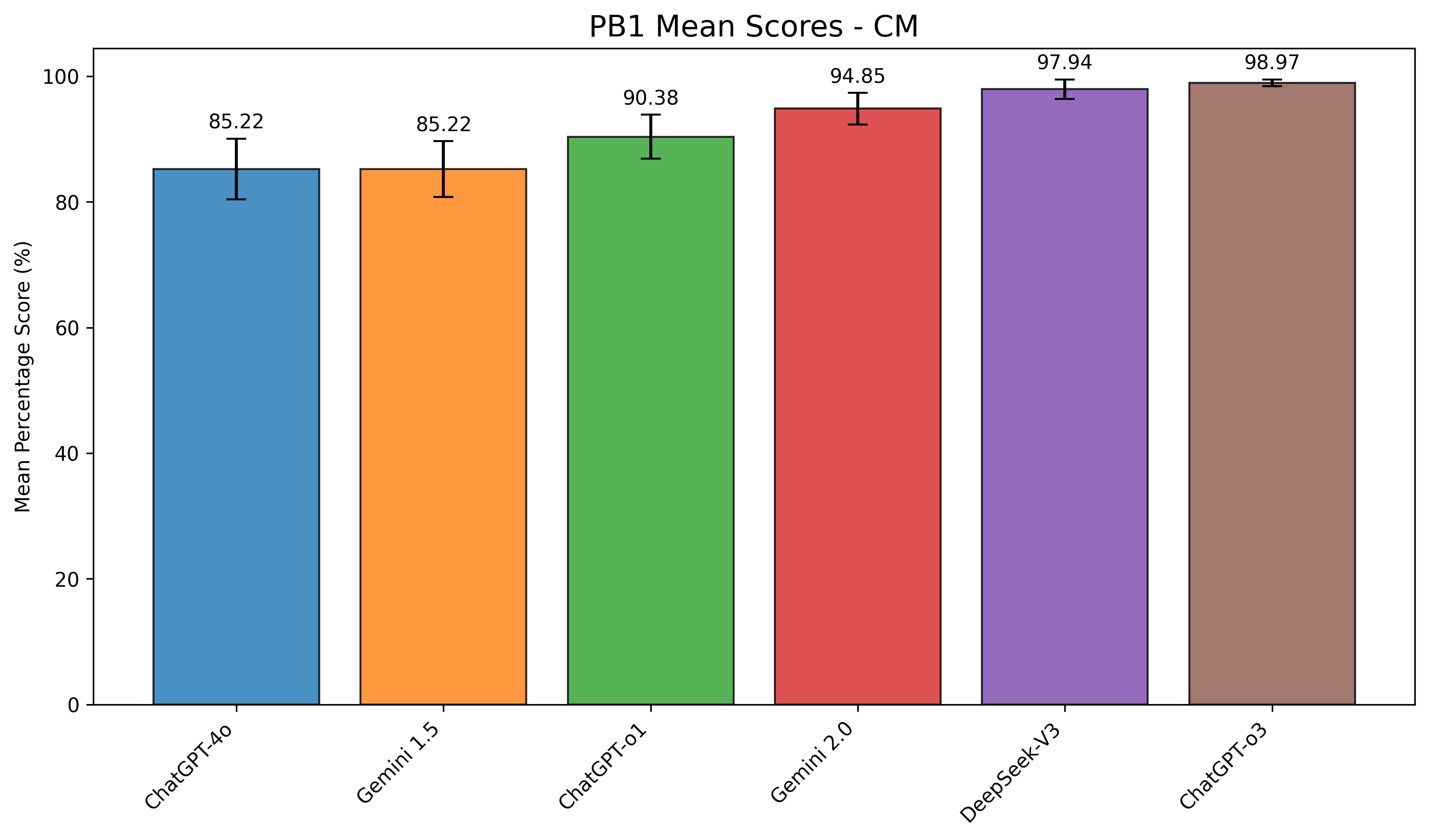}
    \caption{Mean percentage scores achieved by each model in the Classical Mechanics (CM) topic, including error bars to denote variance.}
    \label{fig:pb1_cm_bar}
\end{figure}

\begin{figure}[H]
    \centering
    \includegraphics[width=0.7\linewidth]{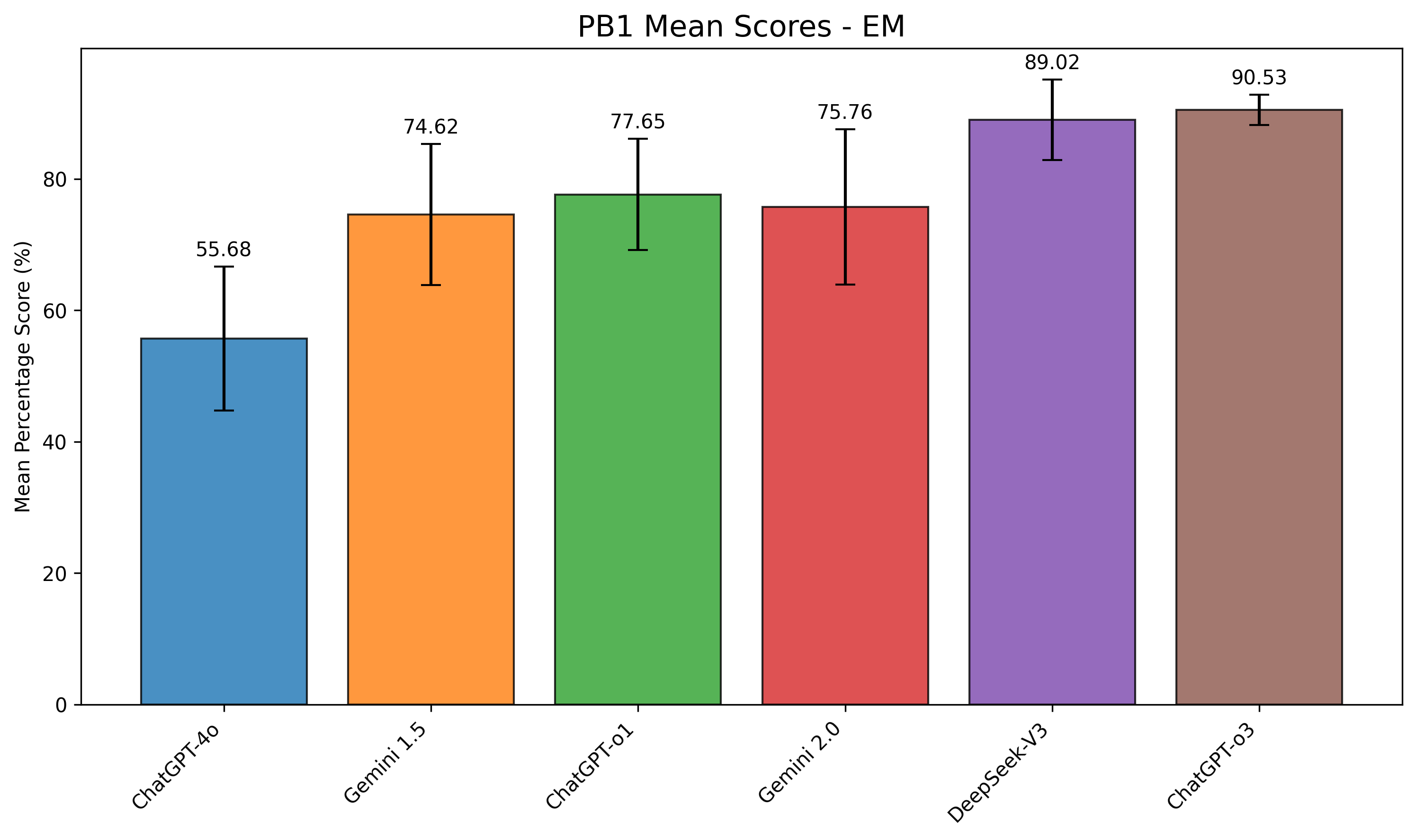}
    \caption{Mean percentage scores achieved by each model in the Electromagnetism (EM) topic, including error bars to denote variance.}
    \label{fig:pb1_em_bar}
\end{figure}

\begin{figure}[H]
    \centering
    \includegraphics[width=0.7\linewidth]{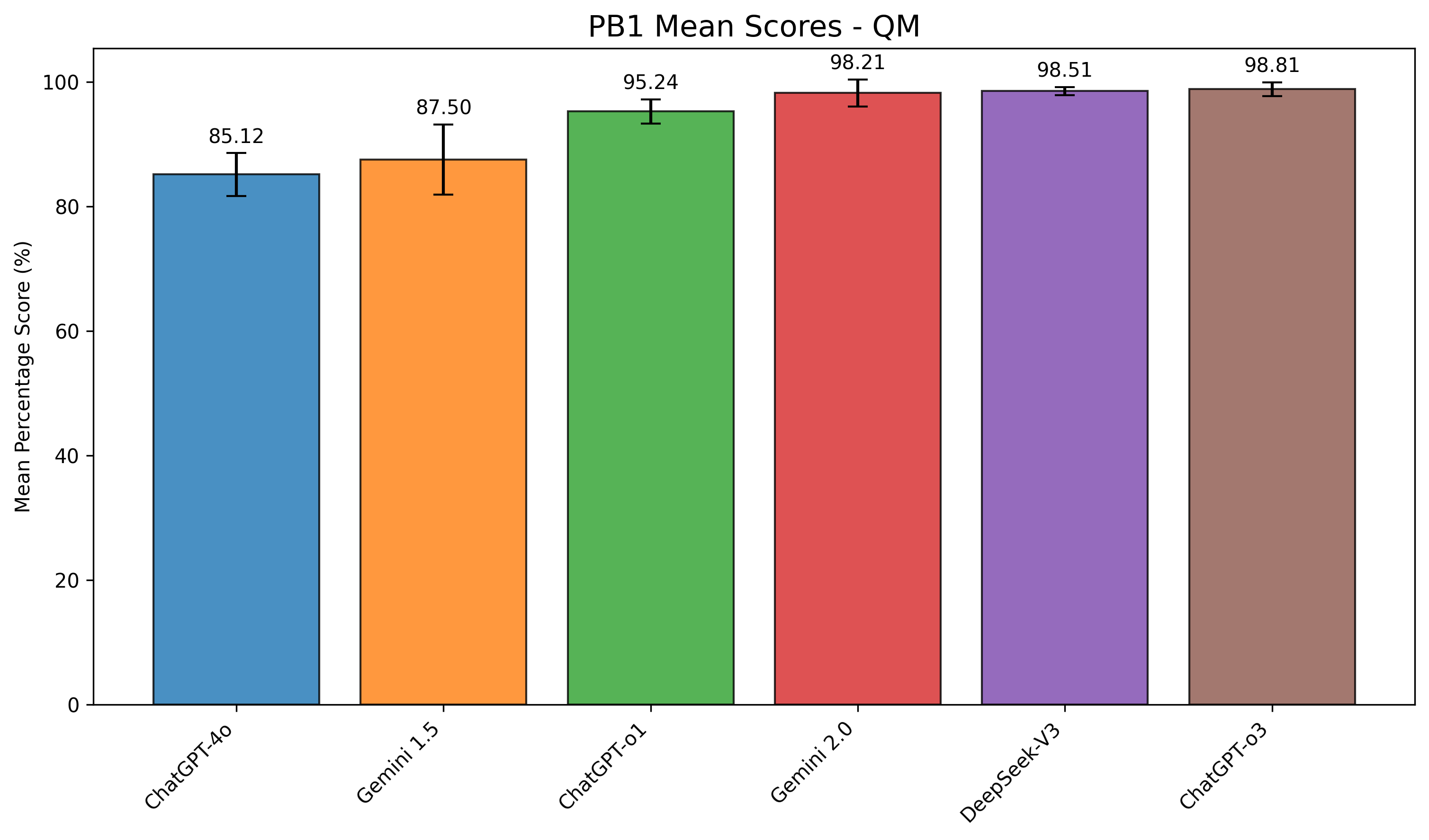}
    \caption{Mean percentage scores achieved by each model in the Quantum Mechanics (QM) topic, including error bars to denote variance.}
    \label{fig:pb1_qm_bar}
\end{figure}

\begin{figure}[H]
    \centering
    \includegraphics[width=0.7\linewidth]{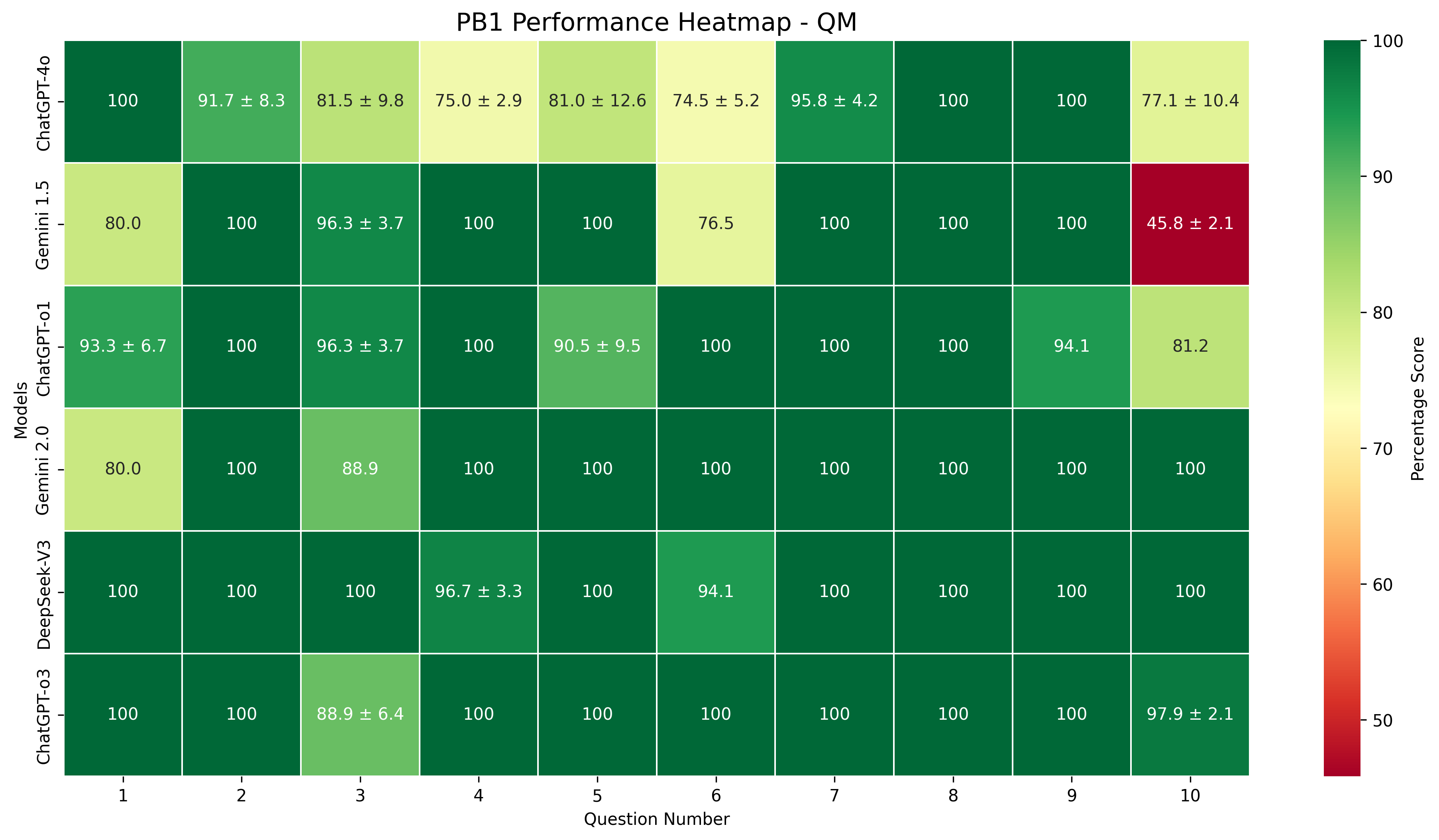}
    \caption{Heat map showing model performance in Quantum Mechanics.}
    \label{fig:pb1_qm_heat}
\end{figure}

\begin{figure}[H]
    \centering
    \includegraphics[width=0.7\linewidth]{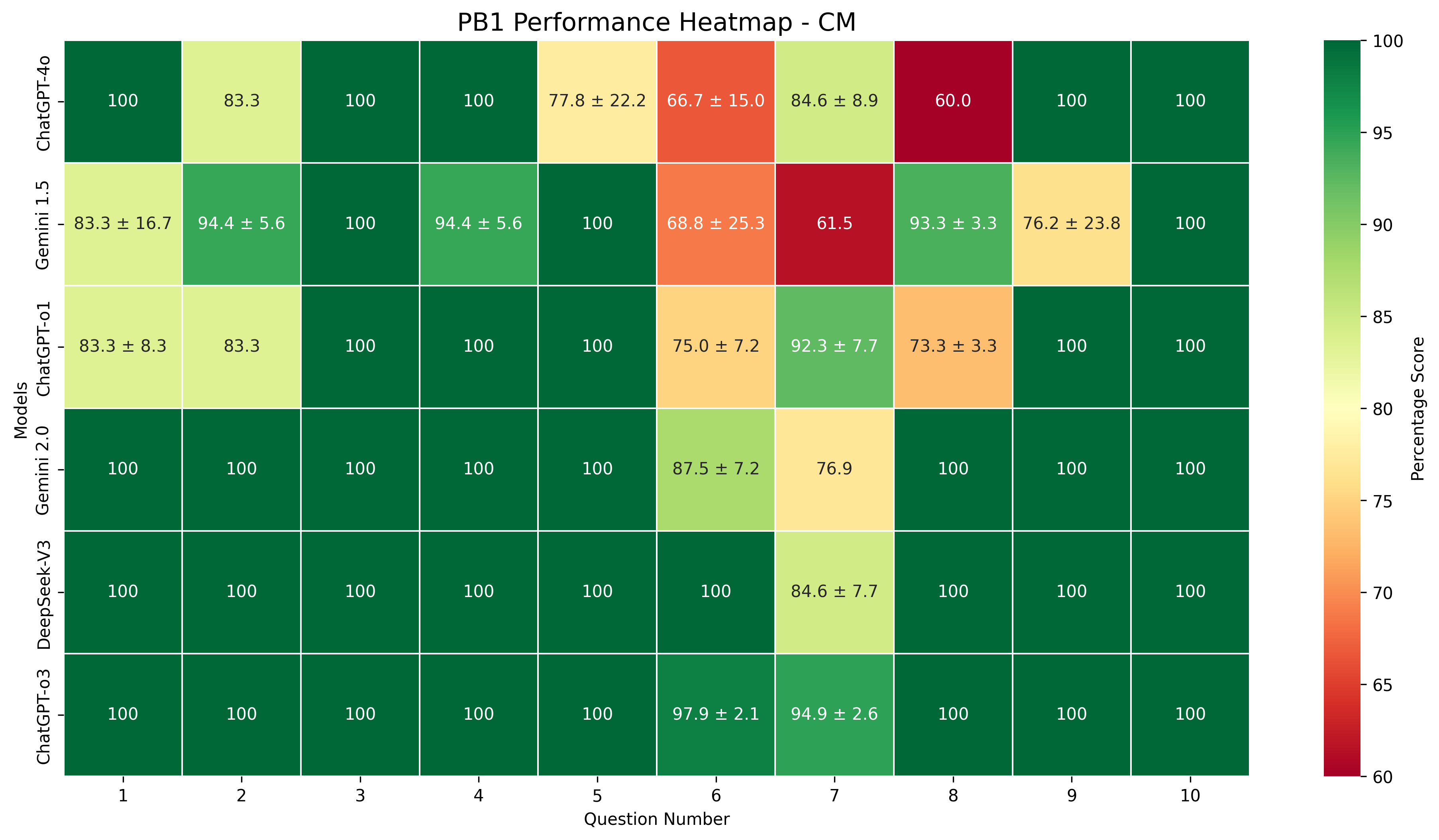}
    \caption{Heat map showing model performance in Classical Mechanics.}
    \label{fig:pb1_cm_heat}
\end{figure}

\begin{figure}[H]
    \centering
    \includegraphics[width=0.7\linewidth]{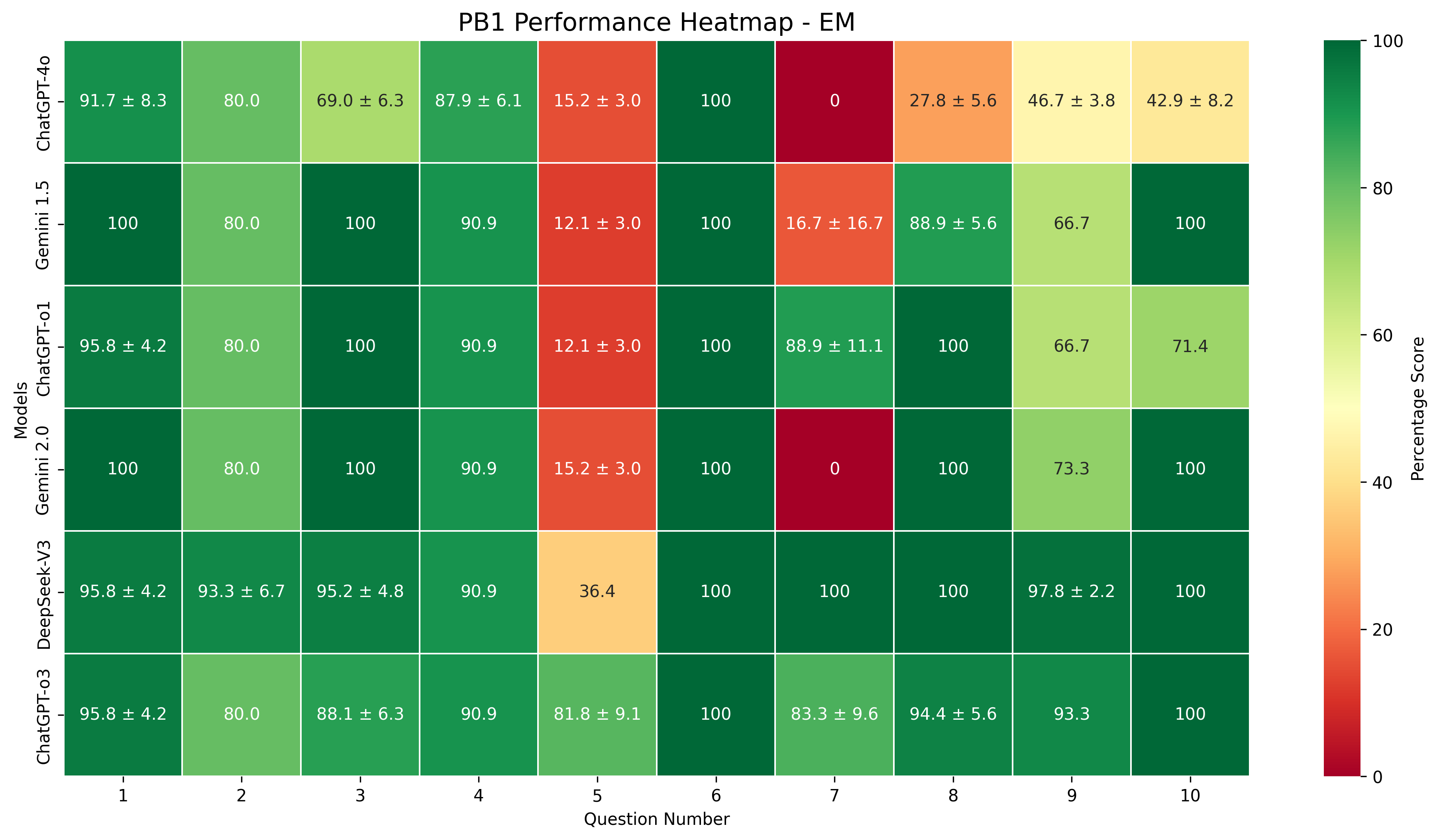}
    \caption{Heat map showing model performance in Electromagnetism.}
    \label{fig:pb1_em_heat}
\end{figure}

To best visualise the overall inter- and intra-generational performance trends, two distinct visual representations were employed: bar charts (Figures \ref{fig:pb1_cm_bar} -- \ref{fig:pb1_qm_bar}) and heat maps (Figures \ref{fig:pb1_qm_heat} -- \ref{fig:pb1_em_heat}). The bar charts provide a global view of model accuracy and precision across the various tests performed, while heat maps were integrated to isolate specific queries where the models consistently failed to achieve full marks, highlighting areas of systemic vulnerability.

\begin{table*}[t]
\centering
\label{tab:pb1_summary}
\begin{tabular}{lccc}
\toprule
\textbf{Model} & \textbf{Classical Mechanics} & \textbf{Quantum Mechanics} & \textbf{Electromagnetism} \\
\midrule
ChatGPT-4o & $85.22 \pm 4.82$ & $85.12 \pm 3.44$ & $55.68 \pm 10.97$ \\
Gemini 1.5 Pro & $85.22 \pm 4.43$ & $87.50 \pm 5.65$ & $74.62 \pm 10.75$ \\
ChatGPT-o1 & $90.38 \pm 3.48$ & $95.24 \pm 1.93$ & $77.65 \pm 8.50$ \\
Gemini 2.0 Flash & $94.85 \pm 2.50$ & $98.21 \pm 2.18$ & $75.76 \pm 11.83$ \\
DeepSeek-V3 & $97.94 \pm 1.54$ & $98.51 \pm 0.64$ & $89.02 \pm 6.15$ \\
ChatGPT-o3 & $98.97 \pm 0.53$ & $98.81 \pm 1.11$ & $90.53 \pm 2.30$ \\
\bottomrule
\end{tabular}
\caption{PB1 Bar Chart data Summary Table}
\end{table*}

The primary metric employed for this benchmarking phase was the mean percentage score for each subject area. This value was calculated by aggregating the total marks achieved from a human grader across all three independent iterations of a given problem and normalising the result against the maximum attainable score. As illustrated in Figures \ref{fig:pb1_cm_bar} -- \ref{fig:pb1_qm_bar}, the most modern versions of the LLMs not only have consistently higher scores, but also a decrease in performance variance. For instance, ChatGPT-o3 emerged as the premier benchmark for stability, maintaining standard errors of 2.30\% or lower across all topics. While DeepSeek-V3 matched this top-tier accuracy, it consistently trailed ChatGPT-o3 in precision, with the sole exception of Quantum Mechanics (QM), where it achieved the lowest standard error of the cohort (0.64\%).

Overall, the gap between the lowest- and highest-scoring models—varied significantly by subject: 13.75\% for Classical Mechanics (CM), 13.69\% for QM, and a striking 34.85\% for Electromagnetism (EM). This disproportionate variance in EM was driven almost exclusively by ChatGPT-4o's initial poor performance (55.68\%). If this Generation 1 outlier is excluded, this gap for EM tightens to 15.91\%, aligning closely with the one observed in CM and QM, suggesting a relatively consistent field of competition among the more recent architectures.

The behaviour of the performance over model generations depends on the specific subject. In CM and QM, baseline performance was robust from the outset, with Generation 1 models comfortably clearing the 85\% threshold. Notably, QM almost reached benchmark saturation as early as Generation 2, with Gemini 2.0 Flash surpassing the 98\% mark. By the third generation, both subjects were effectively saturated; DeepSeek-V3 and ChatGPT-o3 achieved near-parity with scores approaching 99\%. Conversely, Electromagnetism had a notably lower baseline, consistently floating around 75-80\% until Generation 3,  but even the top-performing model, ChatGPT-o3, peaking at only 90.53\%. This subject also induced unique evaluative instability. For example, while Gemini 2.0 strongly improved version 1.5 in CM and QM, its EM performance stagnated and its standard error actually increased to 11.8\%.

While Quantum Mechanics (QM) is conceptually the most abstract domain, it exhibited the most consistent generational improvement. This is because QM's rigorous, algebraic formalism aligns highly with LLM architectures optimised for syntactic pattern recognition and symbolic fidelity. Given this structural compatibility, raw model scaling alone is highly effective in mitigating logical hallucinations. Furthermore, because QM carried the highest allocation of raw marks, it provided the greatest evaluative granularity, establishing it as the most direct and reliable benchmark for testing a model's sustained computational robustness.

Within Classical Mechanics, a similar scaling pattern to QM emerged; however, CM progressed more slowly, failing to reach benchmark saturation until Generation 3. This slower trajectory indicates the presence of structural reasoning bottlenecks that Generation 2 models could not overcome. For example, CM Q6 ( (Figure \ref{fig:CM_Q6})) required the explicit handling and balancing of multiple independent forces. This systemic balancing act introduced computational overhead that reliably defeated models until the Gen 3 architectures.

Furthermore, qualitative analysis of CM responses revealed systemic errors driven by training corpora heuristics. When autoregressive models encounter a logical divergence, they frequently default to statistically common sequences rather than maintaining rigorous physical logic. Figure \ref{fig:CM_Q7})), for instance, early models consistently applied the standard Lorentz transformation instead of the required inverse transformation, likely because the standard formulation is vastly overrepresented in physics texts \cite{song2026large, khalid2025large}. This reliance on heuristics is compounded when models emulate human cognitive shortcuts—such as skipping vector calculus derivations in favour of implicitly stating the right-hand rule. By making these intuitive leaps, models deviate from a strict, step-by-step Chain-of-Thought (CoT) approach. This artificially widens the logical gaps between generated tokens, overtaxing the model's autoregressive reasoning capacity and increasing the probability of hallucinations.

Lastly, Electromagnetism (EM) exhibited a distinct performance profile, acting as the persistent frontier subject for LLMs. While some marks were lost to the rigid qualitative demands of the mark scheme---such as in EM Q2, where models were frequently penalised for failing to explicitly state assumed foundational facts (e.g., that light is an EM wave)---the primary difficulty stemmed from a core architectural constraint: the inability to effectively reason within spatial geometry. This limitation was most evident in the nearly universal failure mode of EM Q4 (Figure \ref{fig:EM_Q4}). Across Generations 1 through 3, almost every model scored exactly 90.9\% with zero variance. The models successfully performed the overarching flux calculations but failed the final qualitative step required to deduce the direction of the induced current using the right-hand rule. Because concepts such as "handedness" and physical rotation are notoriously difficult to represent via text tokens, the models lacked the three-dimensional spatial grounding required to complete the problem, defaulting instead to an inverted statistical heuristic.

This disconnect between syntactic processing and visual grounding culminated in the catastrophic failure observed in EM Q5 (Figure \ref{fig:EM_Q5})), where Generation 1 and 2 performance languished between 12.1\% and 15.2\%. This failure highlighted a crucial structural quirk in the Set A benchmark: of the six questions featuring accompanying visual diagrams, five could be reliably solved through text inference alone. EM Q5 was the sole exception, making the direct extraction of topological information from the circuit diagram an absolute prerequisite for success. The models did not fail this question because the underlying Kirchhoff's laws were too complex; rather, they hit a fundamental visual reasoning cliff. Forced to rely on text representations, they routinely hallucinated incorrect junction points or placed resistors on the wrong parallel branches, rendering all subsequent calculations physically incorrect. It was not until ChatGPT-o3 that a model demonstrated the capacity-- albeit inconsistently-- to correctly perceive and reason with the provided diagrammatic topology, a limitation that strongly justified the inclusion of the dedicated Multimodal (MM) study.

Ultimately, these failures underscore a critical ceiling in early LLM proficiency, reinforcing the necessity established in Section 1 to rigorously evaluate these models as robust physical problem solvers rather than mere syntactic engines. While architectures up to Generation 3 achieved near-total mastery of the ``language'' of physics---excelling in the abstract algebraic derivations of Quantum Mechanics---they demonstrated limited utility in dealing with spatial reality and diagrammatic comprehension. Overall, the indication is that as of early 2025, models demonstrated remarkable proficiency but showed significant areas for improvement. 

\subsubsection{PB2}

Before detailing the results for recent architectures, a methodological shift must be noted. All subsequent non-multimodal evaluations utilised the condensed Set B dataset. Again, to maintain cross-study continuity, all questions retain their original Set A numerical designations (as mapped in Table \ref{tab:q-labels-map}). Furthermore, to account for how this dataset truncation inherently affects average topic scores, an updated summary table (Table \ref{app:pb1b_summary}) and corresponding bar charts (Figures \ref{fig:pb1b_cm_bar_chart}--\ref{fig:pb1b_qm_heat_map}) for the PB1 cohort on Set B are provided in Appendix B

To evaluate the upper performance limits of contemporary LLM architectures, the testing suite was expanded to include four frontier models, categorised into two subsequent generational cohorts:

\begin{itemize} 
    \item \textbf{Generation 4 (June 2025):} Gemini 2.5 Flash.
    \item \textbf{Generation 5 (Nov/Dec 2025):} ChatGPT-5.1, Gemini 3.0 Pro, DeepSeek-V3.2.
\end{itemize}

\begin{table*}[htbp]
\centering
\begin{tabular}{lccc}
\toprule
\textbf{Model} & \textbf{Classical Mechanics} & \textbf{Quantum Mechanics} & \textbf{Electromagnetism} \\
\midrule
Gemini 2.5 Flash & $96.23 \pm 2.10$ & $100.00 \pm 0.00$ & $94.56 \pm 2.34$ \\
ChatGPT-5.1 & $98.74 \pm 0.85$ & $100.00 \pm 0.00$ & $97.96 \pm 1.34$ \\
Gemini 3.0 Pro & $99.37 \pm 0.93$ & $98.31 \pm 0.98$ & $96.60 \pm 3.28$ \\
DeepSeek-V3.2 & $100.00 \pm 0.00$ & $97.74 \pm 0.83$ & $96.60 \pm 2.43$ \\
\bottomrule
\end{tabular}
\caption{PB2 Summary Table: Mean Percentage Score $\pm$ Standard Error (\%)}
\label{tab:pb2_summary}
\end{table*}

\begin{figure}[H]
    \centering
    \includegraphics[width=0.7\linewidth]{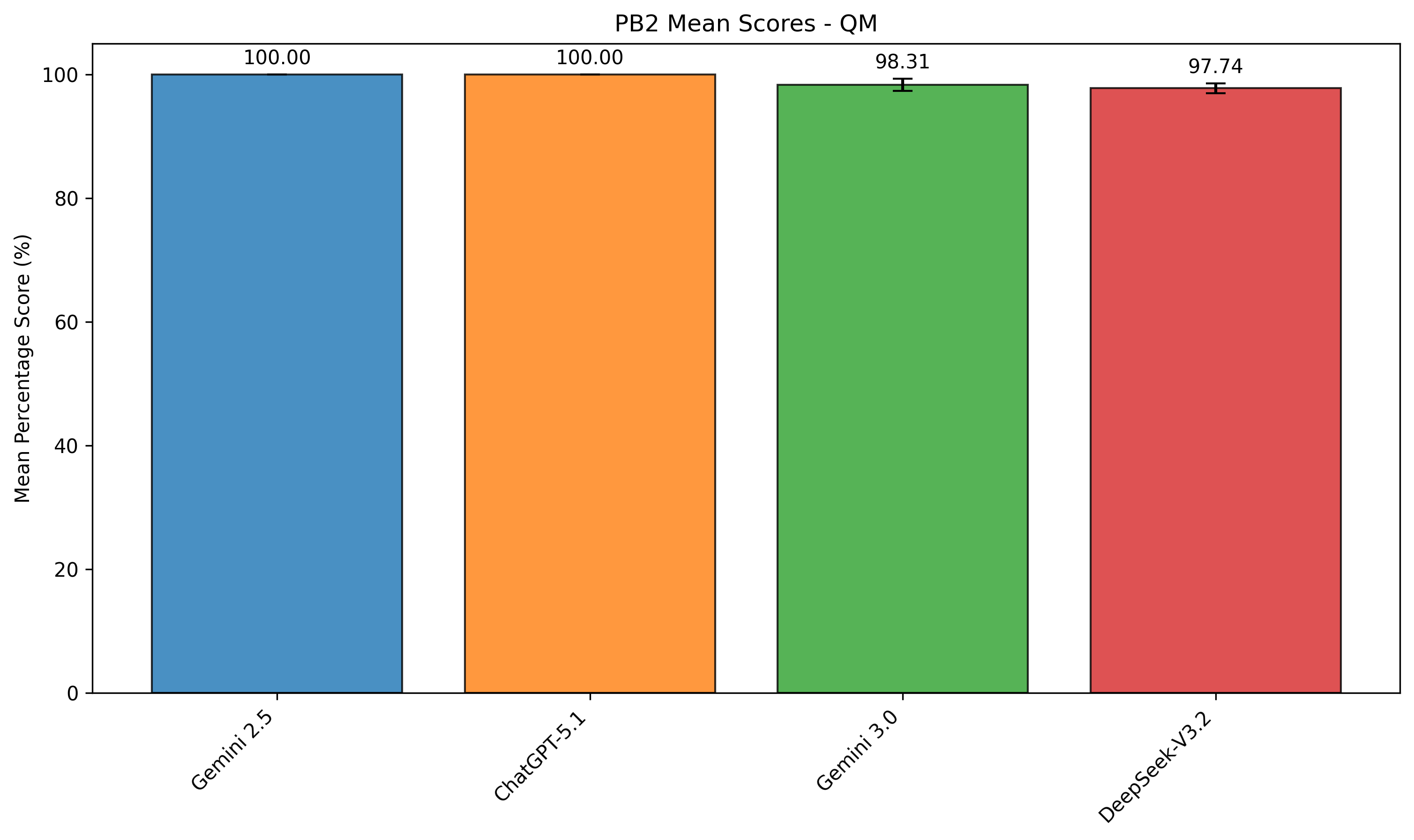}
    \caption{Averaged percentage scores for Quantum Mechanics across Gen 4 and Gen 5 models.}
    \label{fig:pb2_qm_bar}
\end{figure}

\begin{figure}[H]
    \centering
    \includegraphics[width=0.7\linewidth]{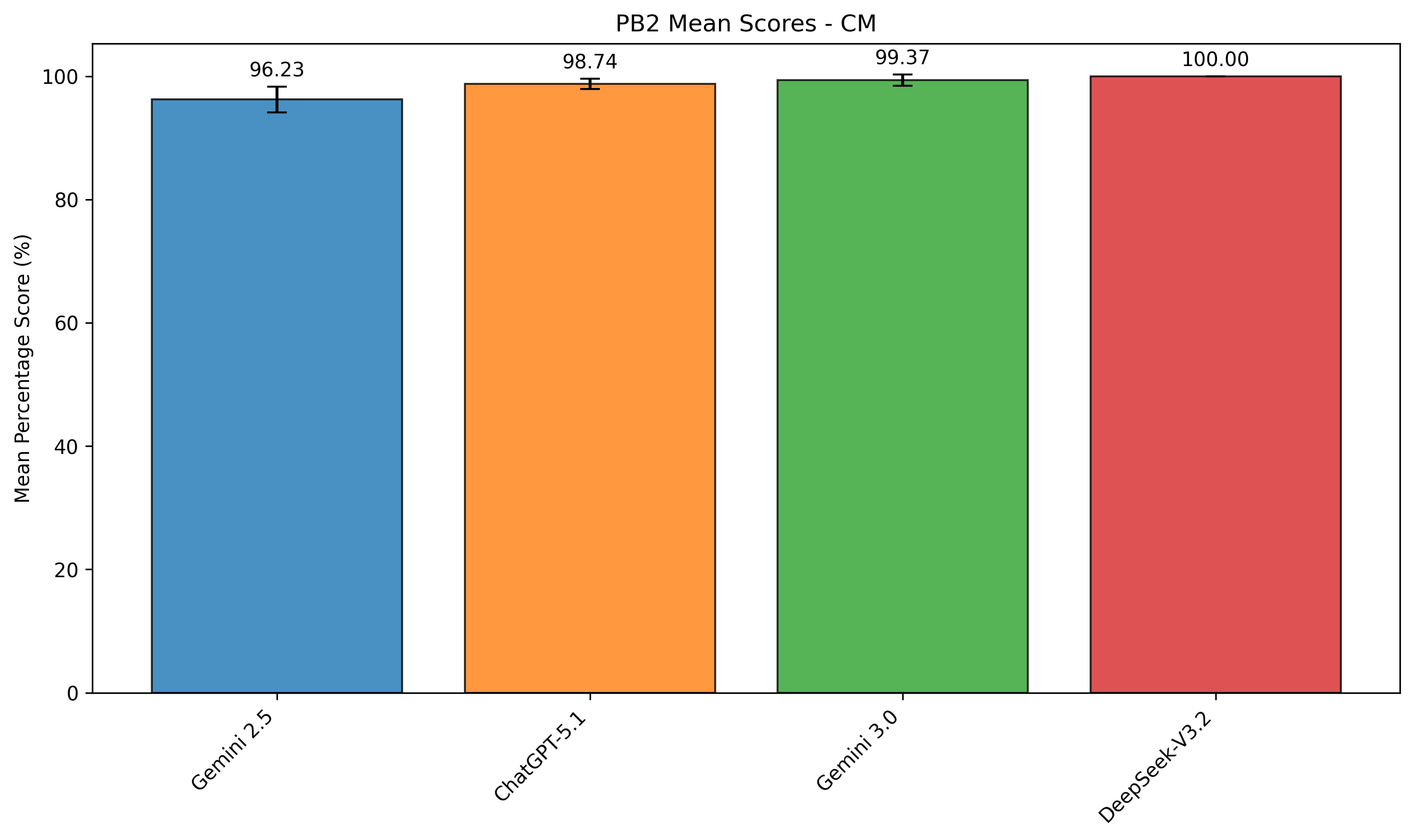}
    \caption{Averaged percentage scores for Classical Mechanics across Gen 4 and Gen 5 models.}
    \label{fig:pb2_cm_bar}
\end{figure}

\begin{figure}[H]
    \centering
    \includegraphics[width=0.7\linewidth]{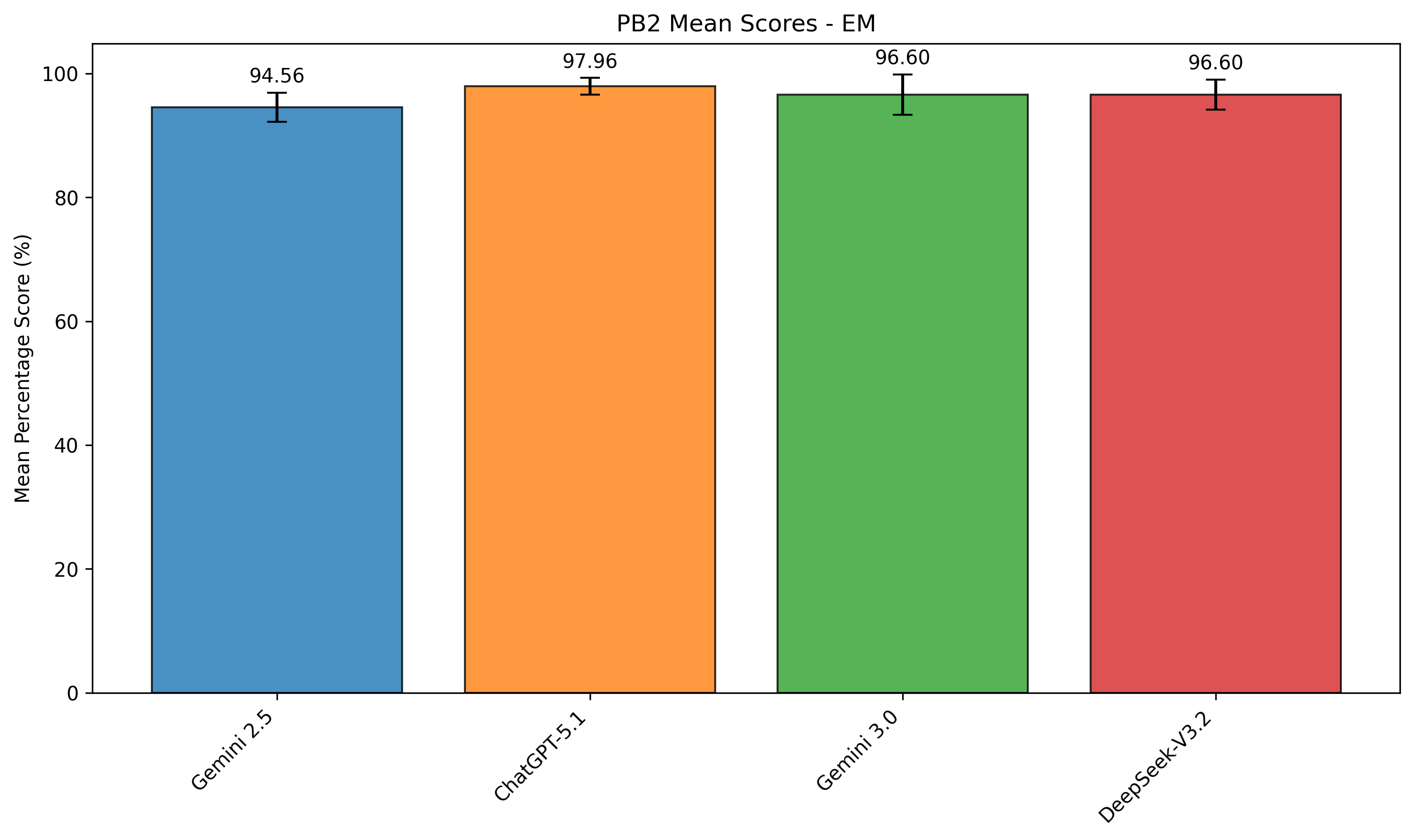}
    \caption{Averaged percentage scores for Electromagnetism across Gen 4 and Gen 5 models.}
    \label{fig:pb2_em_bar}
\end{figure}

As illustrated by the accompanying bar charts (Figures \ref{fig:pb2_qm_bar}--\ref{fig:pb2_em_bar}) and summary table (Table \ref{tab:pb2_summary}), the performance data indicates a definitive trajectory toward benchmark saturation. Standard errors (SE) across the subjects decreased to near-zero margins.

In Quantum Mechanics (QM), the models effectively reached the dataset ceiling. ChatGPT-5.1 achieved a flawless 100\% mean score with zero standard error, while Gemini 3.0 Pro (98.31\%) and DeepSeek-V3.2 (97.74\%) clustered tightly just behind this absolute upper bound. Notably, Gemini 2.5 Flash also achieved a perfect 100\% score, despite its earlier release date and distilled architecture. Crucially, however, while Gemini 2.5 Flash matched or outperformed Generation 5 models in QM, it fell slightly behind the leading Generation 3 models in both CM and EM (when comparing against the adjusted Set B baselines). 

This anomaly underscores an important caveat regarding benchmark saturation: the ceiling effect. Because performance reached the dataset's maximum threshold, the Generation 5 models lacked the evaluative headroom to demonstrate their expanded reasoning capabilities relative to their predecessors. Furthermore, as established during the PB1 analysis, QM's rigorous, symbolic formalism makes it the most naturally suited domain for LLM syntactic processing. This structural alignment allowed the distilled architecture of Gemini 2.5 Flash to achieve full parity in QM, even as its reduced parameter count constrained its performance in the more spatially and sequentially demanding domains of CM and EM.  

Classical Mechanics (CM) results demonstrate a significant performance improvement. Previously, models were severely constrained by the spatial and system setup overhead required to track shifting coordinate axes and free-body diagrams. The Generation 4 and 5 cohorts substantially mitigated this spatial bottleneck. DeepSeek-V3.2 achieved a 100\% mean across all questions, while ChatGPT-5.1 (98.74\%) and Gemini 3.0 Pro (99.37\%) demonstrated robust stability with standard errors dropping below 1\%. The emergence of these flawless and near-flawless scores indicates that these models are not only performing exceptionally well but have effectively hit the evaluative limit of the question set. The near-zero standard errors observed across CM and QM suggest that the benchmarks from Sets A and B are approaching saturation and can no longer reliably measure or differentiate the true upper bounds of these frontier architectures.

\begin{figure}[H]
    \centering
    \includegraphics[width=0.7\linewidth]{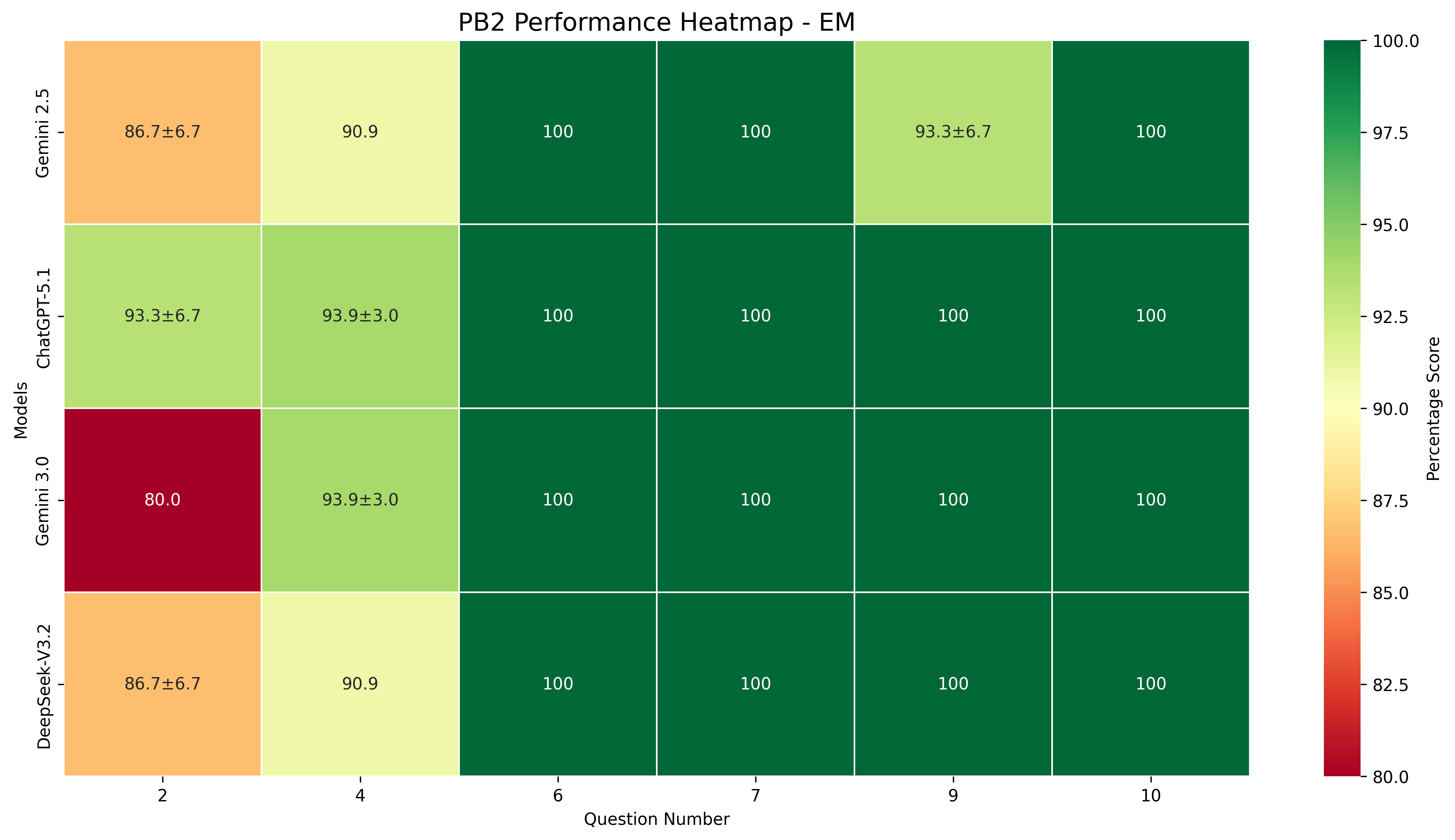}
    \caption{Heat map of Electromagnetism performance highlighting remaining spatial cold spots.}
    \label{fig:pb2_em_heat}
\end{figure}

Despite the high overall performance, Electromagnetism (EM) remained the most challenging subject. While structural scaling pushed total mean scores into the mid-to-high 90s across the board, a granular look at the EM heat map (Figure \ref{fig:pb2_em_heat}) reveals persistent difficulty with specific questions, notably EM Q2 and Q4. However, the recurring errors in Q2 were largely an artefact of a mark scheme peculiarity rather than a misunderstanding of the underlying physics, in which models frequently glossed over and failed to explicitly state the foundational deduction that light is an EM wave. In contrast, the difficulties observed in EM Q4 reflect a genuine, albeit diminishing, architectural limitation. While Generation 5 models occasionally achieved the final mark by correctly inferring the current direction, this inconsistency indicates that 3D spatial orientation tasks — such as applying the right-hand rule — still pose friction when processed purely as text tokens, often leading models to default to the mean, as discussed in PB1.

\begin{figure}[H]
    \centering
    \includegraphics[width=0.7\linewidth]{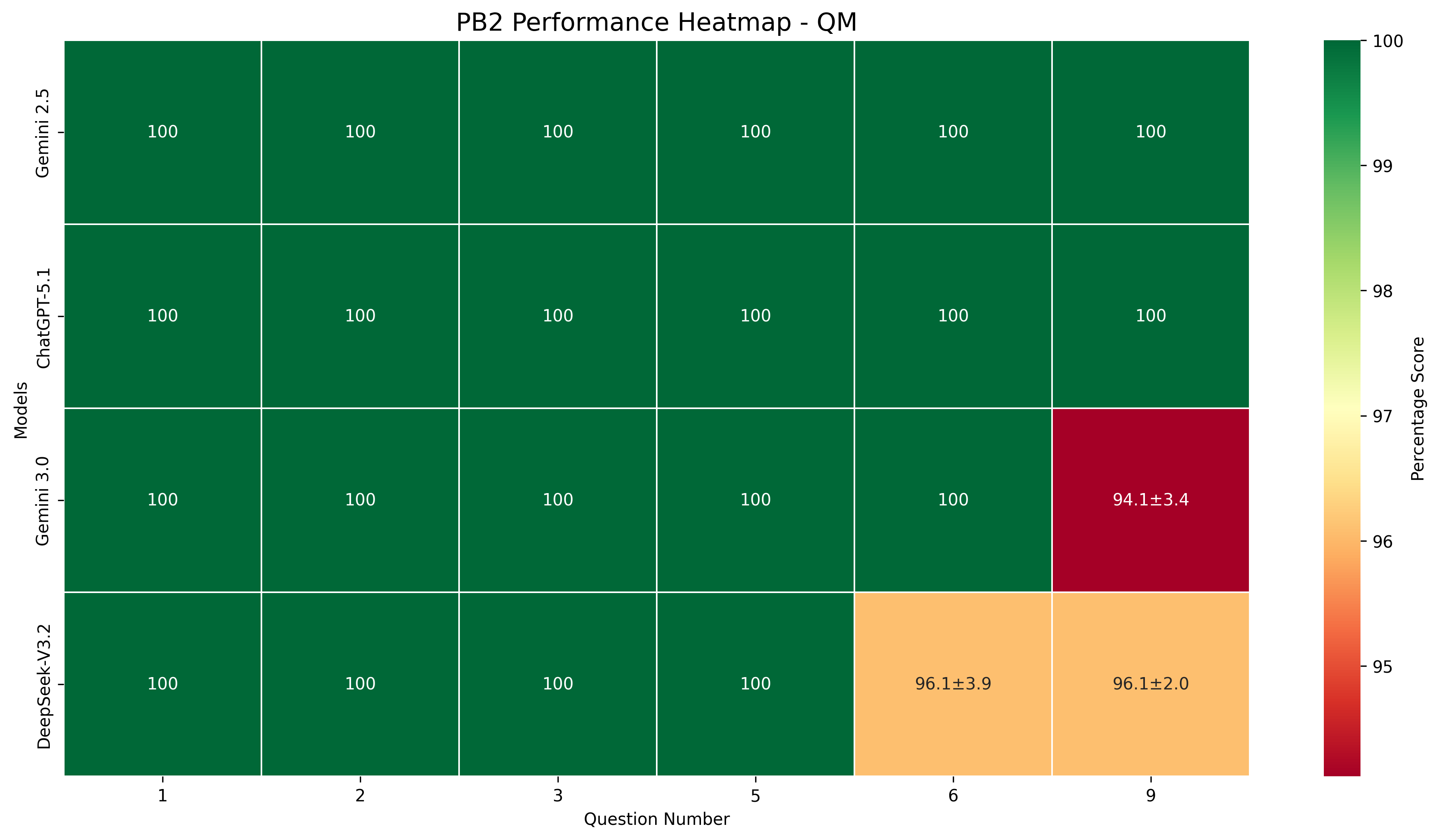}
    \caption{Heat map of Quantum Mechanics performance showing near-total saturation.}
    \label{fig:pb2_qm_heat}
\end{figure}

\begin{figure}[H]
    \centering
    \includegraphics[width=0.7\linewidth]{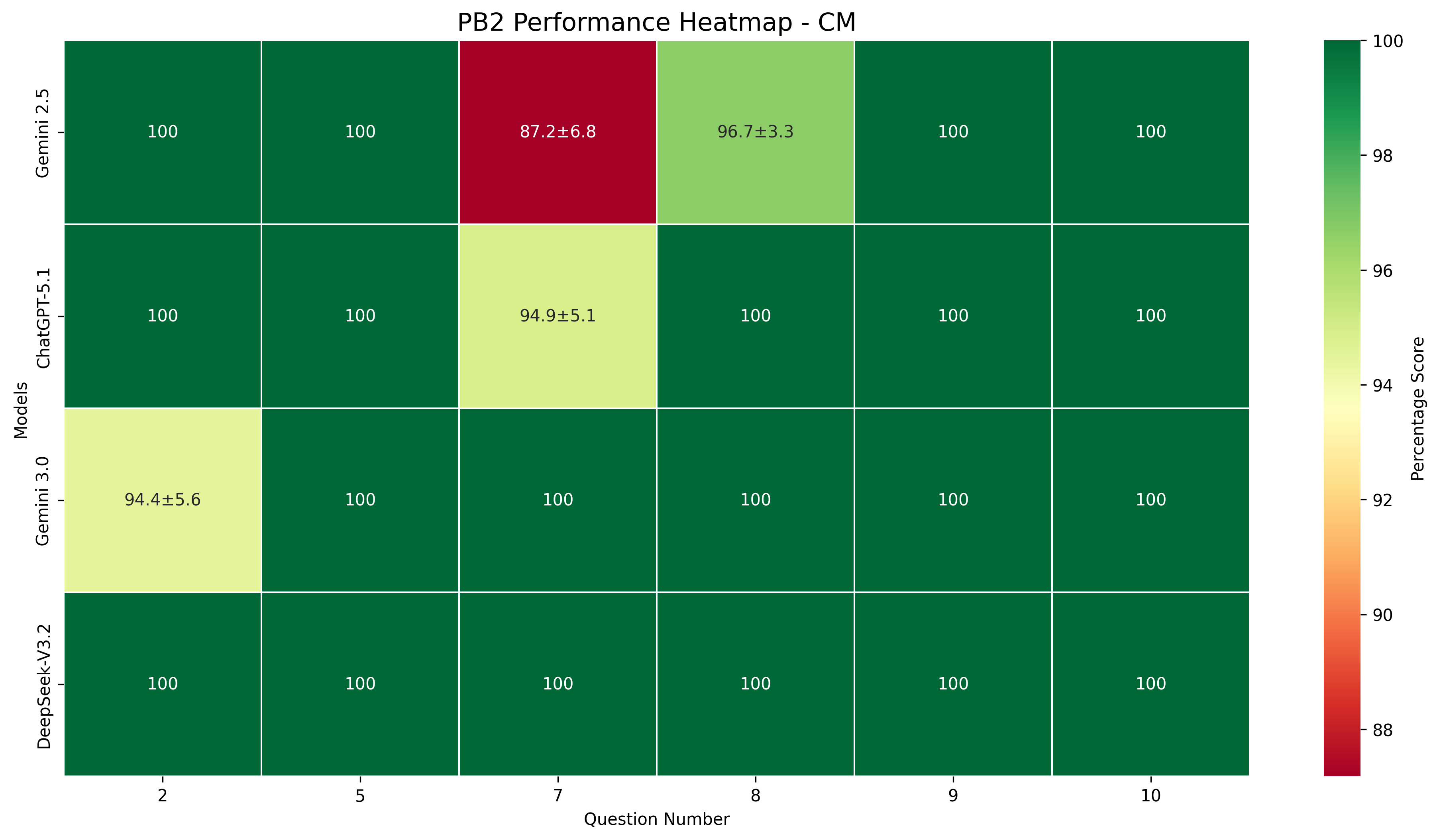}
    \caption{Heat map of Classical Mechanics performance showing the elimination of setup bottlenecks.}
    \label{fig:pb2_cm_heat}
\end{figure}

Ultimately, these results underscore performance differences driven by both architectural design and chronological development. While Gemini 2.5 Flash defined the lower bound of this cohort in the more spatially demanding subjects of CM and EM, this reflects its earlier Generation 4 release date and its structural optimisation for inference speed and cost-efficiency over exhaustive, multi-step reasoning. Yet, compared with the Generation 3 baselines, it still performed at a highly competitive level despite these architectural trade-offs. Overall, the PB2 phase establishes that, as of late 2025, frontier LLMs possess the rigorous logic to serve as highly accurate, independent problem solvers at the undergraduate physics level. While minor caveats remain regarding abstract 3D spatial orientation tasks, their near-perfect performance across these benchmarks confirms that modern architectures can be widely trusted to consistently and accurately resolve complex physical derivations.

\subsubsection{MM}

The preceding phases established that while Generation 5 models possess robust text-based reasoning, earlier generations suffered from a disconnect between syntactic processing and visual grounding. To test whether modern architectures have bridged this gap, the models were evaluated on Set C: a dedicated, natively multimodal problem set comprised of questions that could only be solved by extracting geometric and topological data directly from the provided diagrams. The cohort for this phase included the PB2 models---excluding DeepSeek-V3.2, which lacked a native multimodal implementation---alongside ChatGPT-4o as a generative baseline. 

\begin{figure}[H]
    \centering
    \includegraphics[width=0.7\linewidth]{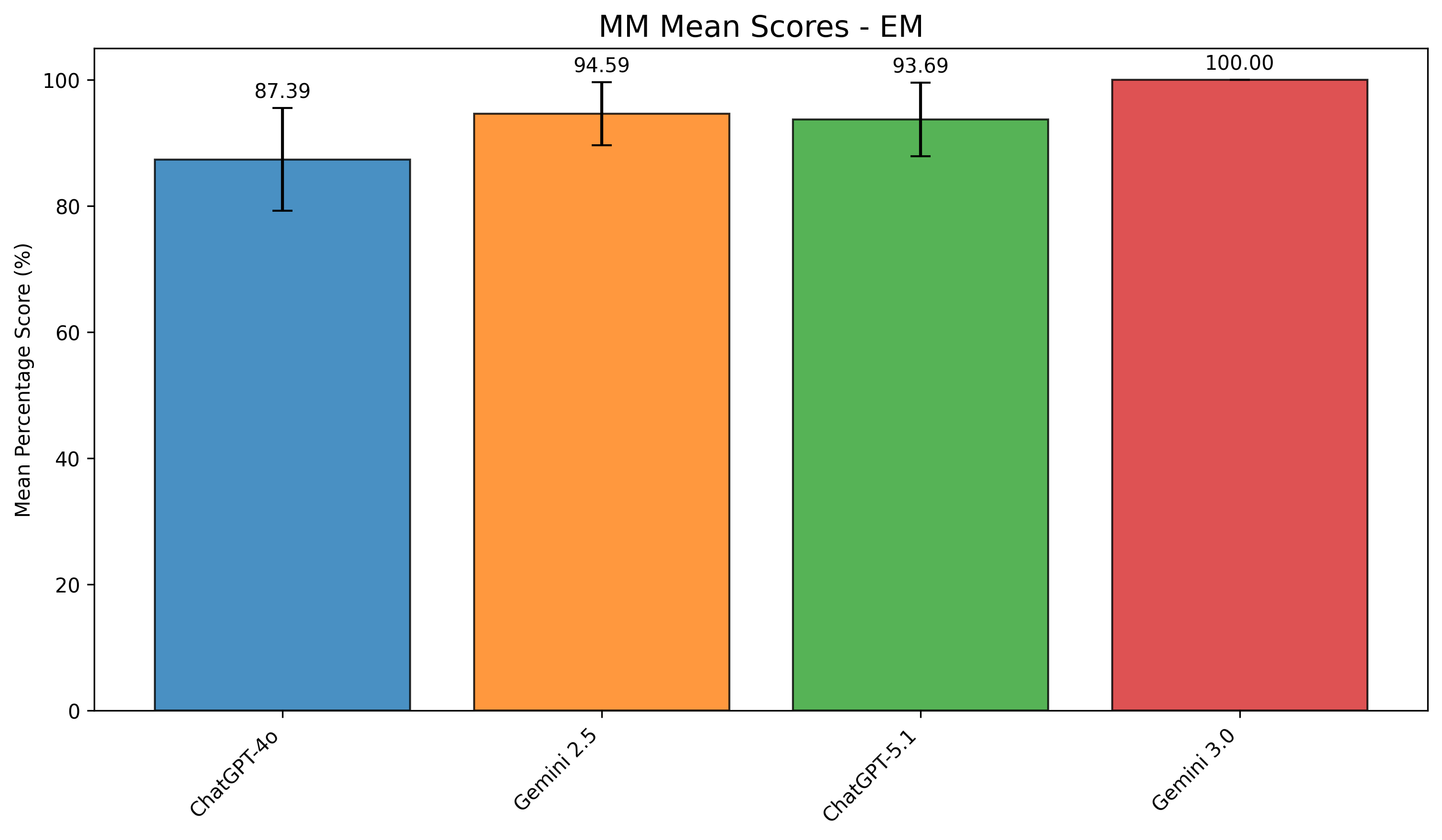}
    \caption{Multimodal performance in Electromagnetism: Transitioning from text-hallucination to visual topography.}
    \label{fig:mm_em_bar}
\end{figure}

\begin{figure}[H]
    \centering
    \includegraphics[width=0.7\linewidth]{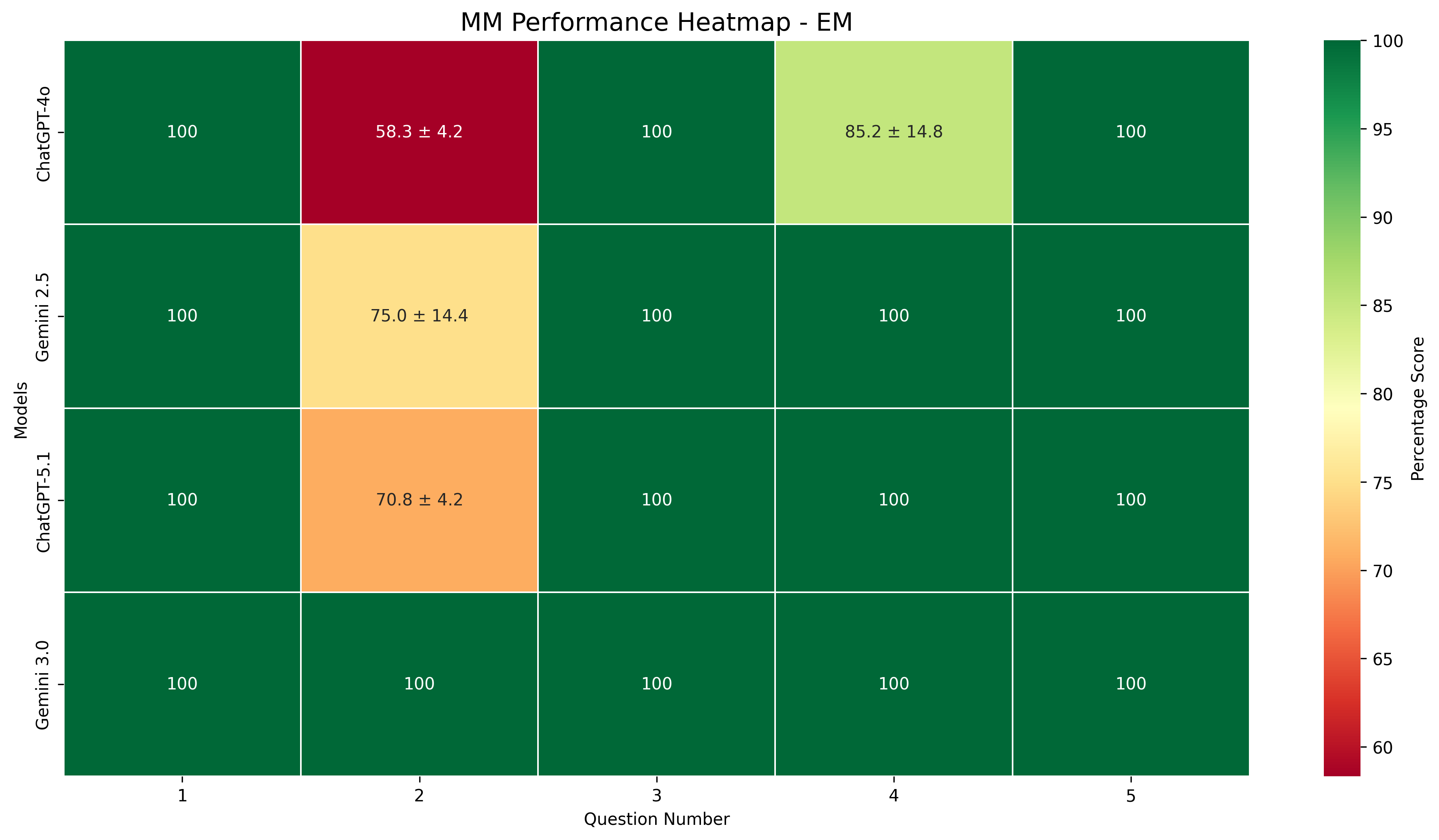}
    \caption{Multimodal performance in Electromagnetism: Heat-Map.}
    \label{fig:mm_em_heat}
\end{figure}

\begin{figure}[H]
    \centering
    \includegraphics[width=0.7\linewidth]{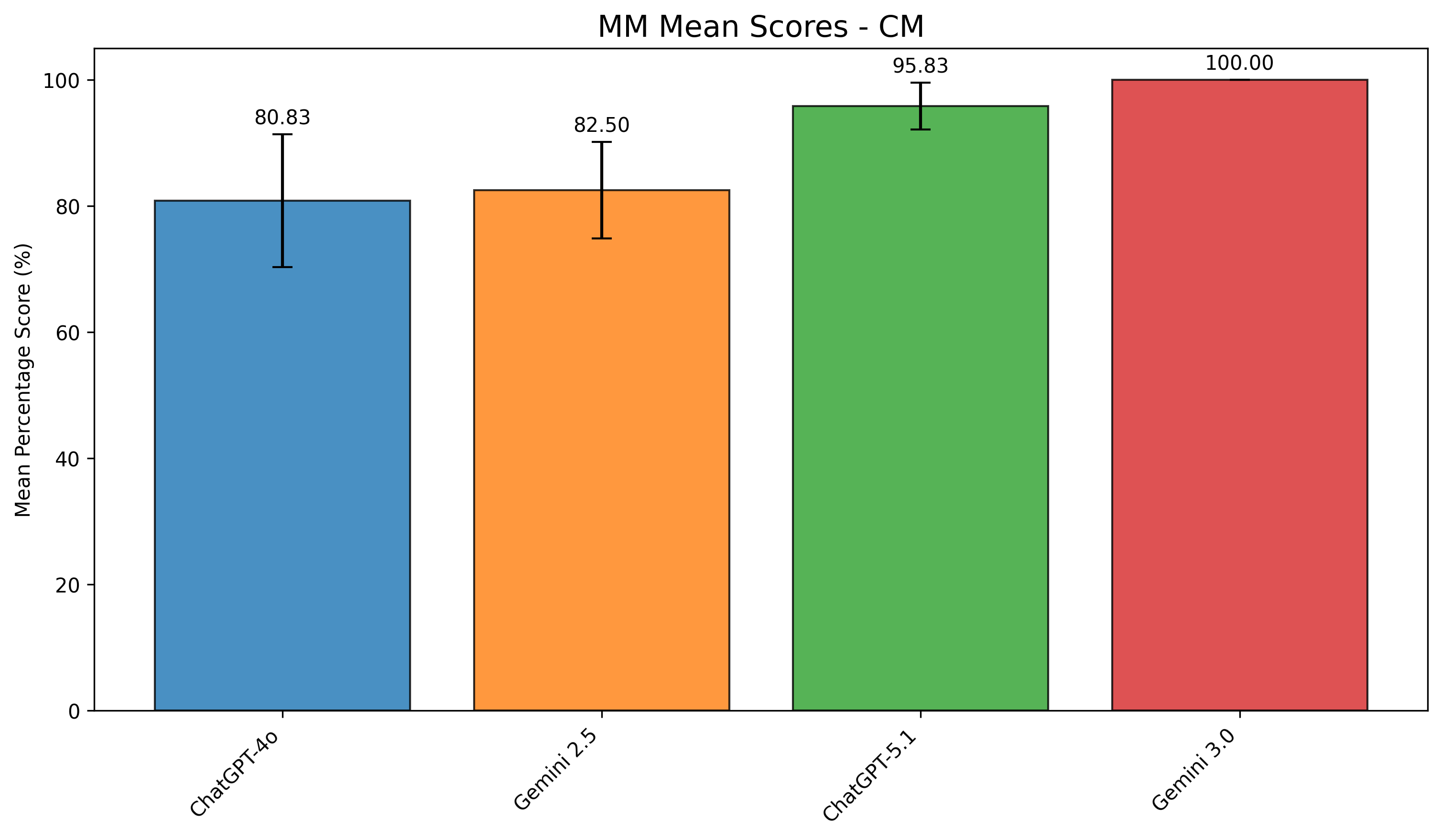}
    \caption{Multimodal performance in Classical Mechanics: Visual grounding of free-body diagrams.}
    \label{fig:mm_cm_bar}
\end{figure}

\begin{figure}[H]
    \centering
    \includegraphics[width=0.7\linewidth]{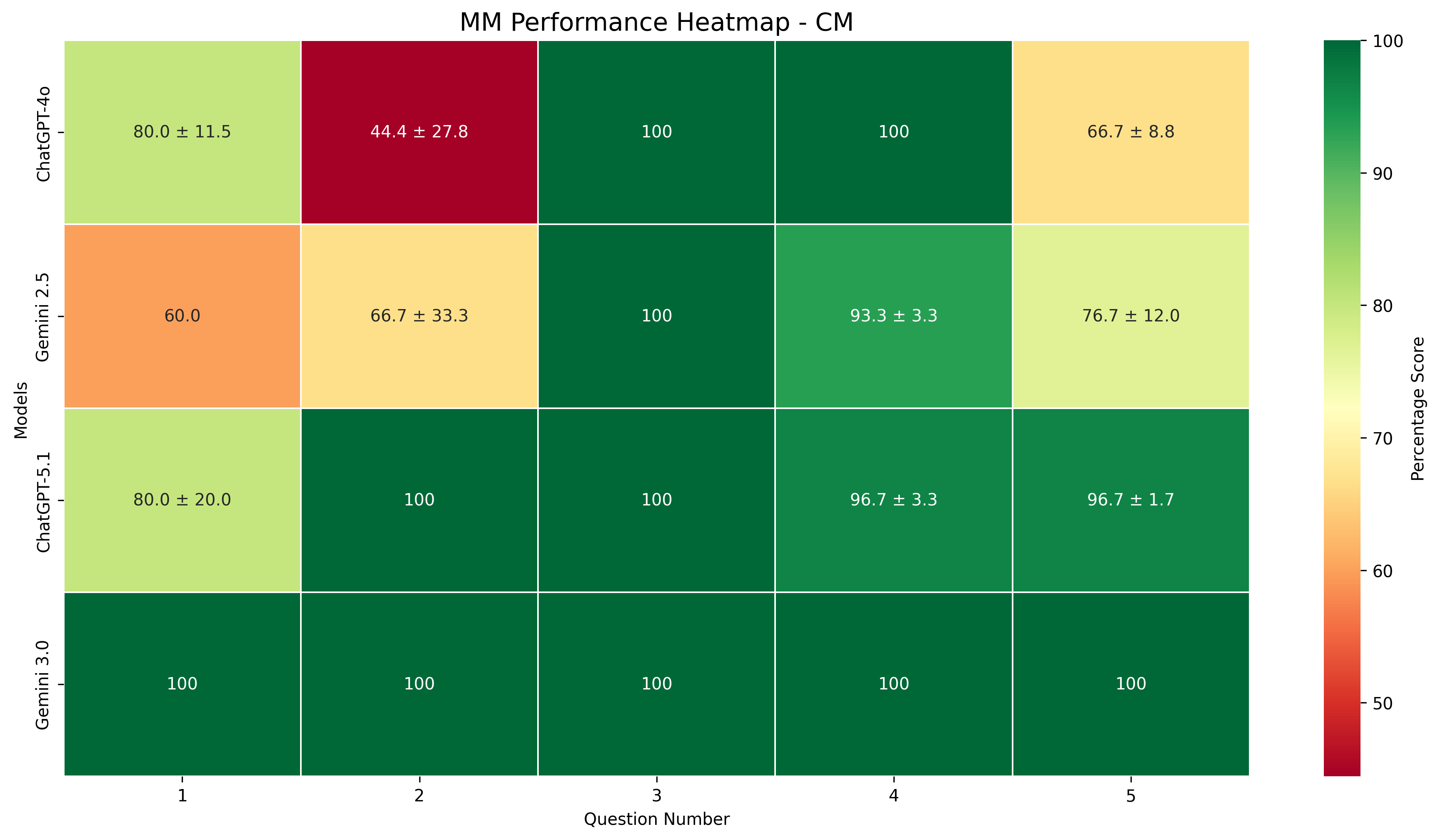}
    \caption{Multimodal performance in Electromagnetism: Transitioning from text-hallucination to visual topography.}
    \label{fig:mm_cm_heat}
\end{figure}

\begin{figure}[H]
    \centering
    \includegraphics[width=0.7\linewidth]{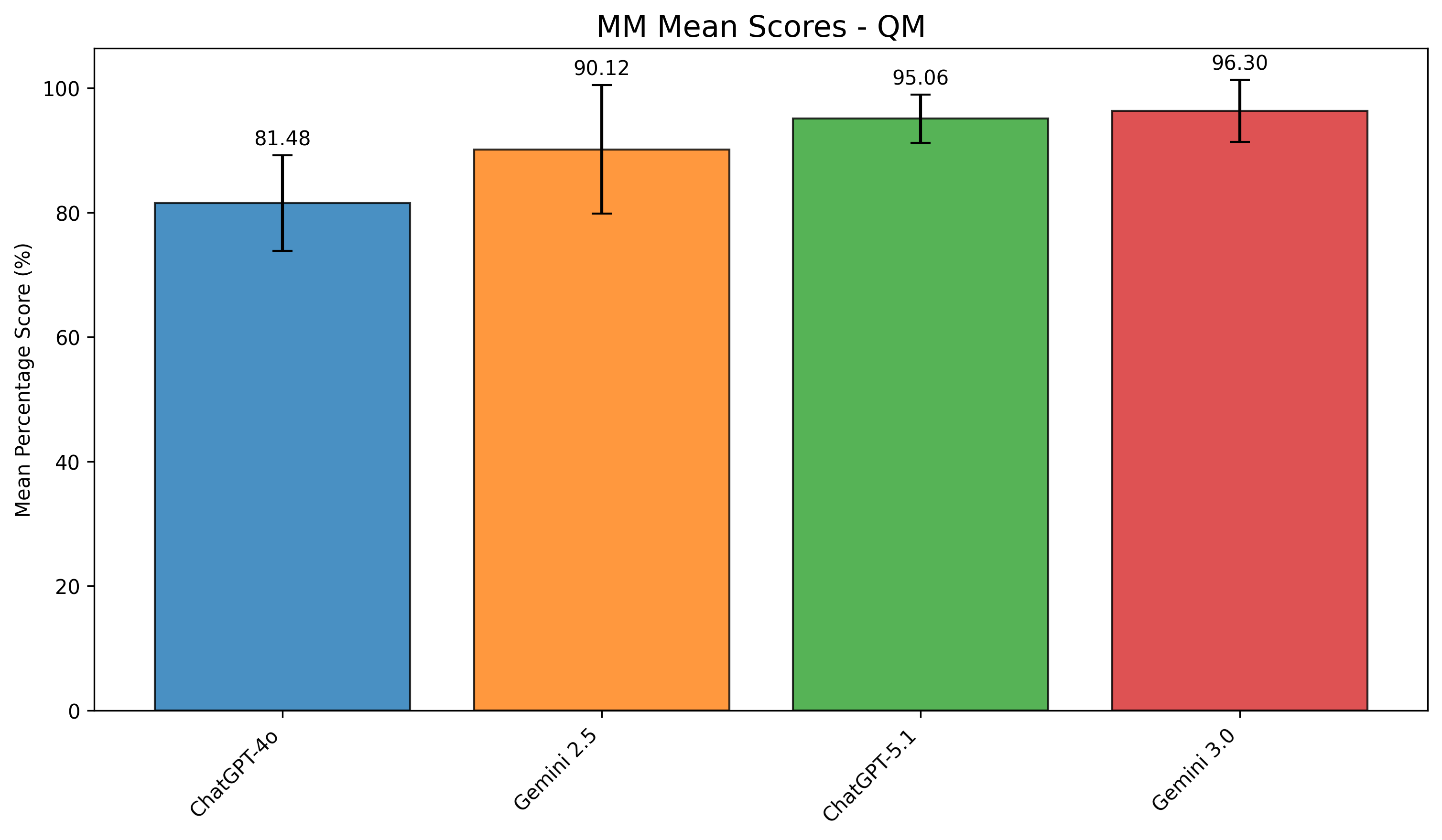}
    \caption{Multimodal performance in Quantum Mechanics: Topological mapping of Feynman diagrams.}
    \label{fig:mm_qm_bar}
\end{figure}

\begin{figure}[H]
    \centering
    \includegraphics[width=0.7\linewidth]{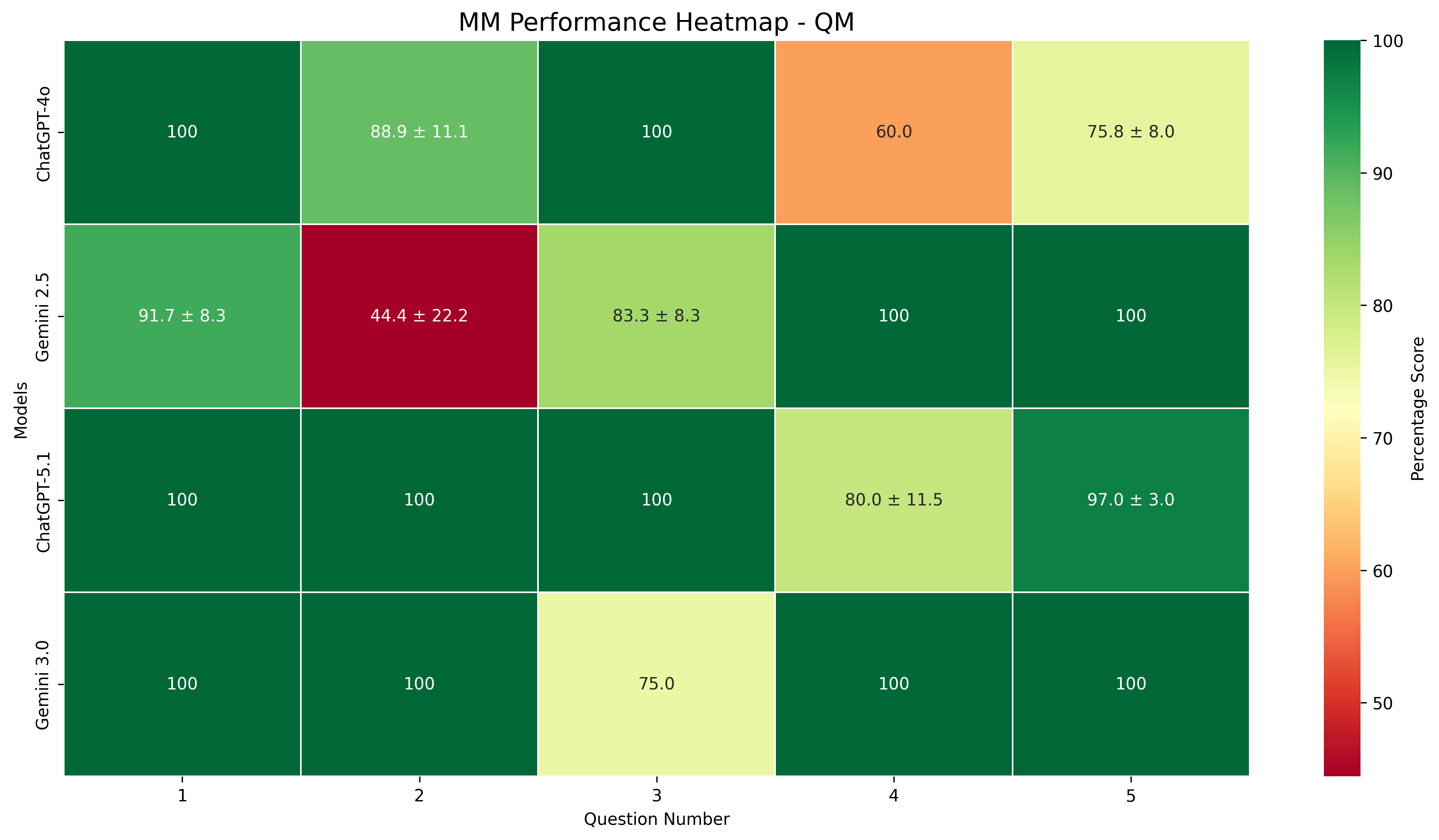}
    \caption{Multimodal performance in Electromagnetism: Transitioning from text-hallucination to visual topography.}
    \label{fig:mm_qm_heat}
\end{figure}

The resulting performance data (Figures \ref{fig:mm_em_bar}--\ref{fig:mm_qm_heat} and Table \ref{tab:mm_summary}) reveals a definitive step-change in spatial problem-solving capabilities. On these natively multimodal tasks, the top performer, Gemini 3.0 Pro, achieved a flawless 100.00\% mean score in both Classical Mechanics and Electromagnetism, demonstrating a high level of accuracy and consistency when extracting essential data from diagrams. 

The Electromagnetism (EM) benchmark provides strong evidence of this architectural progression, particularly regarding Set C EM Q2. As a direct visual translation of Set A EM Q5---the complex Kirchhoff matrix that induced catastrophic reasoning cliffs in PB1---it reprised its role as primary point of friction across the cohort (Figure \ref{fig:mm_em_heat}). When tasked with directly processing the circuit's visual topography, Gemini 3.0 Pro performed exceptionally well, achieving a flawless 100\% score. Notably, despite its distilled Generation 4 architecture, Gemini 2.5 performed on par with the Generation 5 flagship, ChatGPT-5.1.

A closer review of the performance heat map reveals the specific mechanical limitations separating these models. While ChatGPT-5.1 and Gemini 2.5 Flash successfully applied Kirchhoff's laws and correctly deduced portions of the circuit architecture, both models still committed minor topological errors. Given Gemini 3.0 Pro's absolute success on this problem while its contemporary lagged slightly behind, it can be inferred that the Gemini vision-language architecture was better equipped at the time to handle this specific brand of diagrammatic and topographical challenge. Furthermore, the baseline performance of ChatGPT-4o is notable: unlike its complete failure in the text-only PB1 environment, the model was not outright blocked by the visual reasoning cliff, likely due to OpenAI's rolling updates that continuously refine the architecture's multimodal processing behind the scenes.

Classical Mechanics (CM) results exhibit a clear generational divide in multimodal capabilities. The Generation 5 models—ChatGPT-5.1 and Gemini 3.0 Pro—performed exceptionally well, dropping only a negligible number of marks across the entire set. In contrast, the baseline ChatGPT-4o and the distilled Gemini 2.5 Flash recorded very similar performance profiles, lagging behind the flagship models despite the significant gap in their respective release dates. As illustrated by the performance heat-map, question-by-question scaling steadily improved across generations, a trend most clearly evident in the progressive score increases in Q2 and Q5. 

However, in Q2 (Figure \ref{fig:CMc_Q2}), earlier architectures exhibited large standard errors (±27.8\% for ChatGPT-4o and ±33.3\% for Gemini 2.5 Flash). This large variance stemmed from failing to visually determine that the applied force was horizontal, rather than parallel to the incline. When ChatGPT-4o and Gemini 2.5 Flash failed to extract this specific spatial orientation, they defaulted to the standard text-based heuristic of treating the force as parallel to the incline. This initial visual misinterpretation rendered their subsequent mathematical derivations deeply flawed, cascading into substantial mark penalties. However, the success of the Generation 5 models on this specific problem again underscores their ability to extract more information from visual media, thereby reducing the need to default to a statistical heuristic. 

Finally, the Quantum Mechanics (QM) module in Set C diverged from the pure abstract algebra of earlier sets, presenting the models with a collection of Feynman diagrams. Models were required to analyse these diagrams to identify missing exchange particles and vertices, and to determine which interactions are allowed or forbidden, by verifying charge and baryon-number conservation across the visual nodes. While ChatGPT-4o struggled with the visual-topological mapping of these quantum states (falling to 81.48\%), the Gen 5 models excelled. Both ChatGPT-5.1 (95.06\%) and Gemini 3.0 (96.30\%) demonstrated a high capacity to translate visual nodes and directional arrows into strict conservation laws.

The strong performance of Generation 5 architectures in these natively multimodal tasks is characterised by a systematic deconstruction of visual inputs. Whether natively embedded or emergently learned, these frontier models effectively mimic Visual Chain-of-Thought (VCoT) processing \cite{zhao2025cotvla}. Instead of extracting the necessary data directly from the visuals, the model constructs a preliminary semantic representation. By translating raw imagery into structured text, this intermediate step effectively grounds the model's logic prior to any algebraic calculation.

Ultimately, the findings from this multimodal phase provide a definitive resolution to the spatial reasoning bottlenecks initially identified in PB1. By successfully extracting and applying geometric configurations directly from diagrams, Generation 5 architectures have demonstrated that visual complexity is no longer a fundamental barrier to physical derivation. When this multimodal capability is integrated with the near-perfect syntactic and algebraic proficiency established in PB2, a clear conclusion emerges regarding the overall utility of frontier LLMs in this domain. As of late 2025, these models function not merely as advanced text calculators; they possess the capacity to autonomously interpret complex, multi-modal system states and integrate them seamlessly into their reasoning processes. However, a critical caveat remains: when a model fails to accurately extract information from a diagram, it becomes highly susceptible to defaulting to text-based statistical patterns derived from its training corpus. Consequently, while their analytical utility is formidable, their initial diagrammatic interpretations still warrant careful verification to ensure foundational physical premises are correctly established.

\begin{table*}[htbp]
\centering
\begin{tabular}{lccc}
\toprule
\textbf{Model} & \textbf{Classical Mechanics} & \textbf{Quantum Mechanics} & \textbf{Electromagnetism} \\
\midrule
ChatGPT-4o & $80.83 \pm 10.55$ & $81.48 \pm 7.67$ & $87.39 \pm 8.12$ \\
Gemini 2.5 & $82.50 \pm 7.63$ & $90.12 \pm 10.33$ & $94.59 \pm 5.00$ \\
ChatGPT-5.1 & $95.83 \pm 3.74$ & $95.06 \pm 3.89$ & $93.69 \pm 5.83$ \\
Gemini 3.0 & $100.00 \pm 0.00$ & $96.30 \pm 5.00$ & $100.00 \pm 0.00$ \\
\bottomrule
\end{tabular}
\caption{MM Data Summary Table}
\label{tab:mm_summary}
\end{table*}

\subsection{AI Evaluation}
\subsubsection{PB1}

\begin{figure}[H]
\centering
\includegraphics[width=0.7\linewidth]{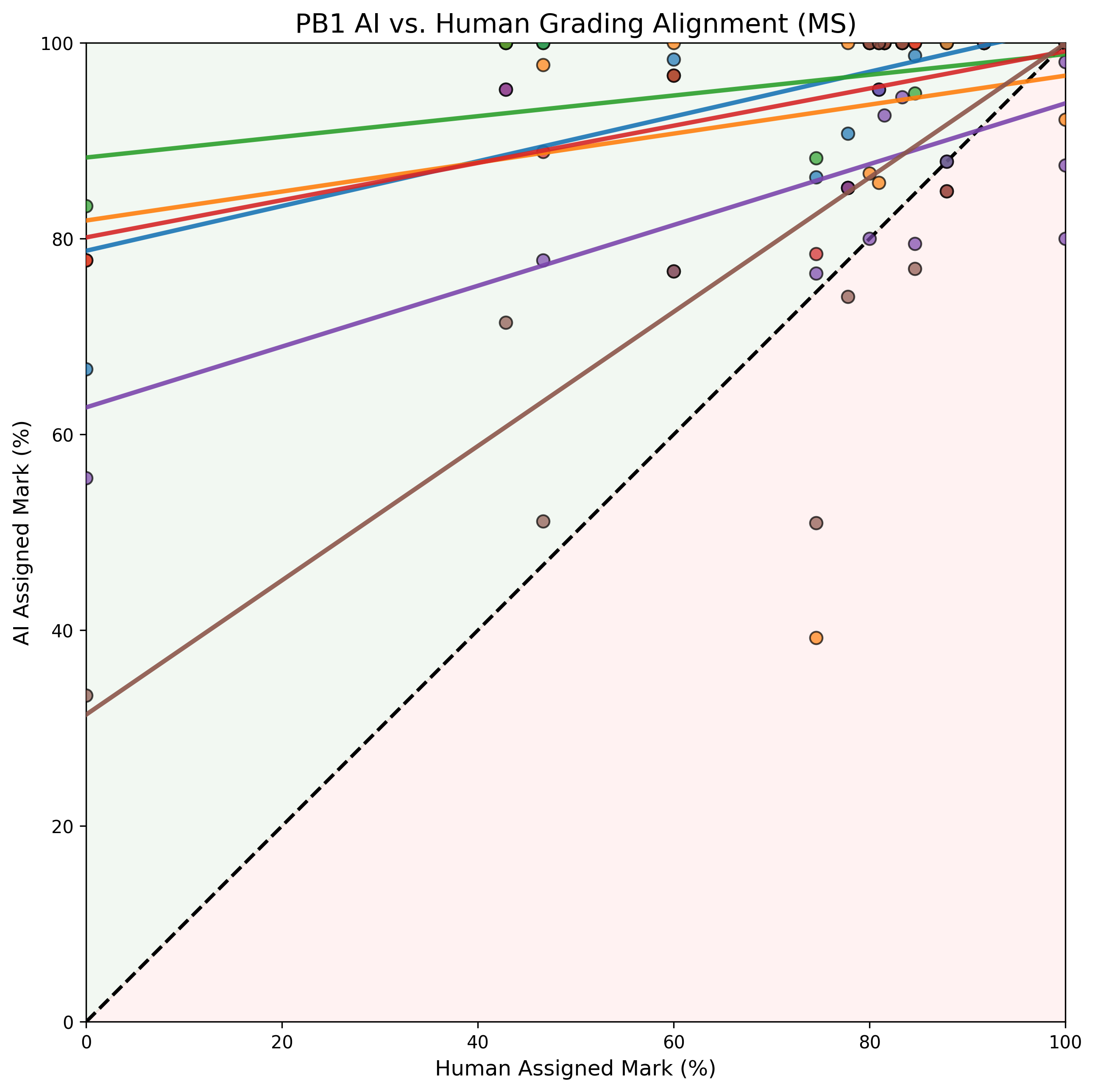}
\caption{Linear Regression of Model Alignment (With Mark Scheme)}
\label{fig:ms-regression}
\end{figure}

\begin{figure}[H]
\centering
\includegraphics[width=0.7\linewidth]{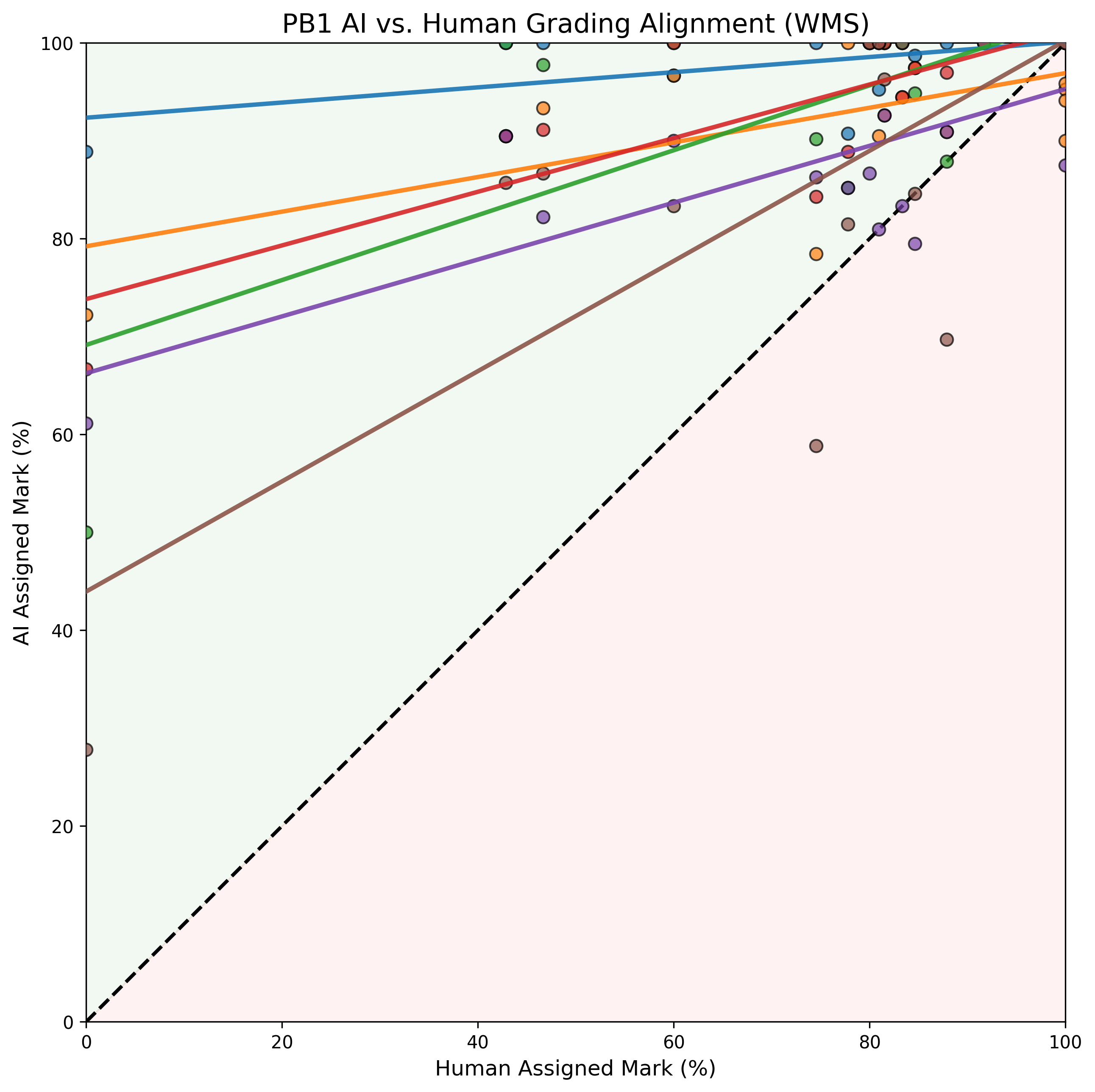}
\caption{Linear Regression of Model Alignment (Without Mark Scheme)}
\label{fig:wms-regression}
\end{figure}

After having probed the LLM's ability to solve problems, this chapter evaluates the reliability of using LLMs as markers. 
Figures \ref{fig:ms-regression} and \ref{fig:wms-regression} present the relation between grades awarded by the six LLMs and humans across the three core topics, comparing the cases where the LLMs were either provided with, or deprived of, the mark schemes. To mirror the Phase 2 human-marking baseline, the LLMs graded three ChatGPT-4o solutions for each of the 18 questions in Set B. The resulting statistical profiles are detailed in Tables \ref{tab:alignment-ms} and \ref{tab:alignment-wms}.

The analysis of the mark-scheme-tethered (MS) data confirms a systemic bias toward grade inflation and compression. Across the cohort, high Y-intercepts—ranging from a minimum of 31.38 (ChatGPT-o3) to 88.29 (ChatGPT-o1)—reveal a "floor effect" where early LLMs exhibit a tendency to avoid assigning low marks. Even when a human evaluator would score a response near zero, these models default to a high baseline. This is further corroborated by shallow gradients (0.10 to 0.31), indicating a lack of sensitivity to the rubric's technical nuances. The result is a narrow, upwardly shifted distribution that fails to differentiate between rigorous solution paths and incorrect attempts.

A potential methodological factor driving this tight clustering of scores is the batch-grading approach employed during this phase. Because the LLMs evaluated all three generated solutions for a given problem simultaneously within a single prompt, they became susceptible to positional biases. The most prominent is the "halo effect," where the perceived quality of the first solution anchors the model’s internal grading standard, pulling subsequent evaluations toward the initial score \cite{zheng2023judging}. 

Stacking multiple dense physics derivations into one context window exacerbates the "lost in the middle" phenomenon \cite{liu2024lost}. As the context expands, LLM attention degrades for data located in the middle of the prompt, exhibiting a U-shaped attention curve where the front and back are prioritised. Thus, unable to maintain an independent, granular focus across all three solutions, the models default to a homogenised score, blurring the evaluations between distinct mathematical attempts.

The performance data reveals a clear distinction between legacy generalist models and newer reasoning architectures. ChatGPT-o3 emerges as the outlier, showing the highest alignment with human graders. With the highest Pearson correlation (0.8379) and an $R^2$ of 0.7022, it accounts for over 70\% of the variance in human marks. Its relatively steep slope (0.6858) and lowest Mean Absolute Error (11.3\%) mark it as the only evaluator in the PB1 cohort capable of proportionally scaling its marks to mirror human judgment.

Gemini 1.5 Pro and ChatGPT-o1 define the lower bound of grading reliability. Gemini 1.5’s $R^2$ of 0.0654 suggests its marks are effectively decoupled from the human baseline. Meanwhile, ChatGPT-o1’s Y-intercept of 88.29 and slope of 0.1056 reveal a model restricted to a high-scoring band. These models demonstrate a tendency to credit perceived intent rather than the technical accuracy required in QM or EM. DeepSeek-V3 and ChatGPT-4o occupy a middle ground, showing moderate correlation ($\approx$ 0.70) but still exhibiting systemic over-marking.

\begin{table*}[tbh]
\centering
\label{tab:alignment-ms}
\begin{tabular}{|l|c|c|c|c|c|c|}
\hline
\textbf{Model} & \textbf{Slope} & \textbf{Y-Int} & \textbf{Pearson $r$} & \textbf{$R^2$} & \textbf{MAE (\%)} & \textbf{RMSE (\%)} \\ \hline
ChatGPT-4o     & 0.2287         & 78.76        & 0.7103               & 0.5046         & 19.12           & 27.85            \\ \hline
Gemini 1.5     & 0.1480         & 81.85        & 0.2557               & 0.0654         & 20.76           & 30.24            \\ \hline
ChatGPT-o1     & 0.1056         & 88.28        & 0.4634               & 0.2148         & 19.13           & 30.00            \\ \hline
Gemini 2.0     & 0.1902         & 80.12        & 0.6174               & 0.3812         & 17.85           & 27.57            \\ \hline
DeepSeek-V3    & 0.3106         & 62.76        & 0.6924               & 0.4794         & 13.86           & 21.39            \\ \hline
ChatGPT-o3     & 0.6858         & 31.38        & 0.8379               & 0.7022         & 11.30           & 15.48            \\ \hline
\end{tabular}
\caption{Alignment and IRR Statistics: With Mark Scheme (MS)}
\end{table*}

The analysis of marking performed  Without Mark Scheme (WMS) shown in Table \ref{tab:alignment-wms}  reveals a shift in evaluative behaviour. When stripped of a formal rubric, the systemic over-marking identified in the MS phase evolves into a broader collapse of grade differentiation. ChatGPT-4o represents the extreme of this trend; with a Y-intercept of 92.35 and a slope of 0.0775, it assigns near-uniform high grades. Without a rubric to anchor its critique, it defaults to awarding high distinctions regardless of the solution's actual quality.

In contrast, Gemini 2.0 demonstrates strong relative alignment in the WMS cohort. It achieved the highest Pearson correlation (0.8288) and $R^2$ (0.6869), suggesting that its internal evaluative baseline for physics solutions is consistent with human judgment. While it over-marks in absolute terms (Y-intercept of 73.83), it successfully identifies the relative quality of solutions even without the provided key.

Reasoning-heavy models demonstrate greater resilience in this rubric-free environment. ChatGPT-o3 remains the most consistent, maintaining the lowest Y-intercept (43.95) and the steepest gradient (0.5628). Although its $R^2$ dropped relative to the MS phase, it remains the only model to utilise the full breadth of the marking scale. DeepSeek-V3 also remains competitive, posting the lowest MAE (13.35\%), indicating its internal knowledge base is robust enough to provide a stable baseline in the absence of a key.

Finally, the RMSE vs. Human percentages in the WMS phase reveal an anomaly: for models like Gemini 1.5, performance improved (dropping from 30.25\% to 26.53\%) when the mark scheme was removed. This implies that for certain architectures, a formal rubric may oversaturate the model's context window, leading to higher accuracy when the model focuses solely on distribution marks, rather than comparing to a mark scheme. 

\begin{table*}[tbh]
\centering
\caption{Alignment and IRR Statistics: Without Mark Scheme (WMS)}
\label{tab:alignment-wms}
\begin{tabular}{|l|c|c|c|c|c|c|}
\hline
\textbf{Model} & \textbf{Slope} & \textbf{Y-Int} & \textbf{Pearson $r$} & \textbf{$R^2$} & \textbf{MAE (\%)} & \textbf{RMSE (\%)} \\ \hline
ChatGPT-4o     & 0.0775         & 92.35        & 0.5927               & 0.3513         & 21.02           & 31.45            \\ \hline
Gemini 1.5     & 0.1769         & 79.21        & 0.6251               & 0.3908         & 18.33           & 26.53            \\ \hline
ChatGPT-o1     & 0.3315         & 69.13        & 0.7086               & 0.5021         & 17.45           & 25.64            \\ \hline
Gemini 2.0     & 0.2741         & 73.82        & 0.8288               & 0.6869         & 17.69           & 25.87            \\ \hline
DeepSeek-V3    & 0.2904         & 66.25        & 0.7612               & 0.5794         & 13.34           & 22.09            \\ \hline
ChatGPT-o3     & 0.5628         & 43.94        & 0.7608               & 0.5788         & 13.91           & 19.25            \\ \hline
\end{tabular}
\end{table*}

\begin{figure}[H]
\centering
\includegraphics[width=0.7\linewidth]{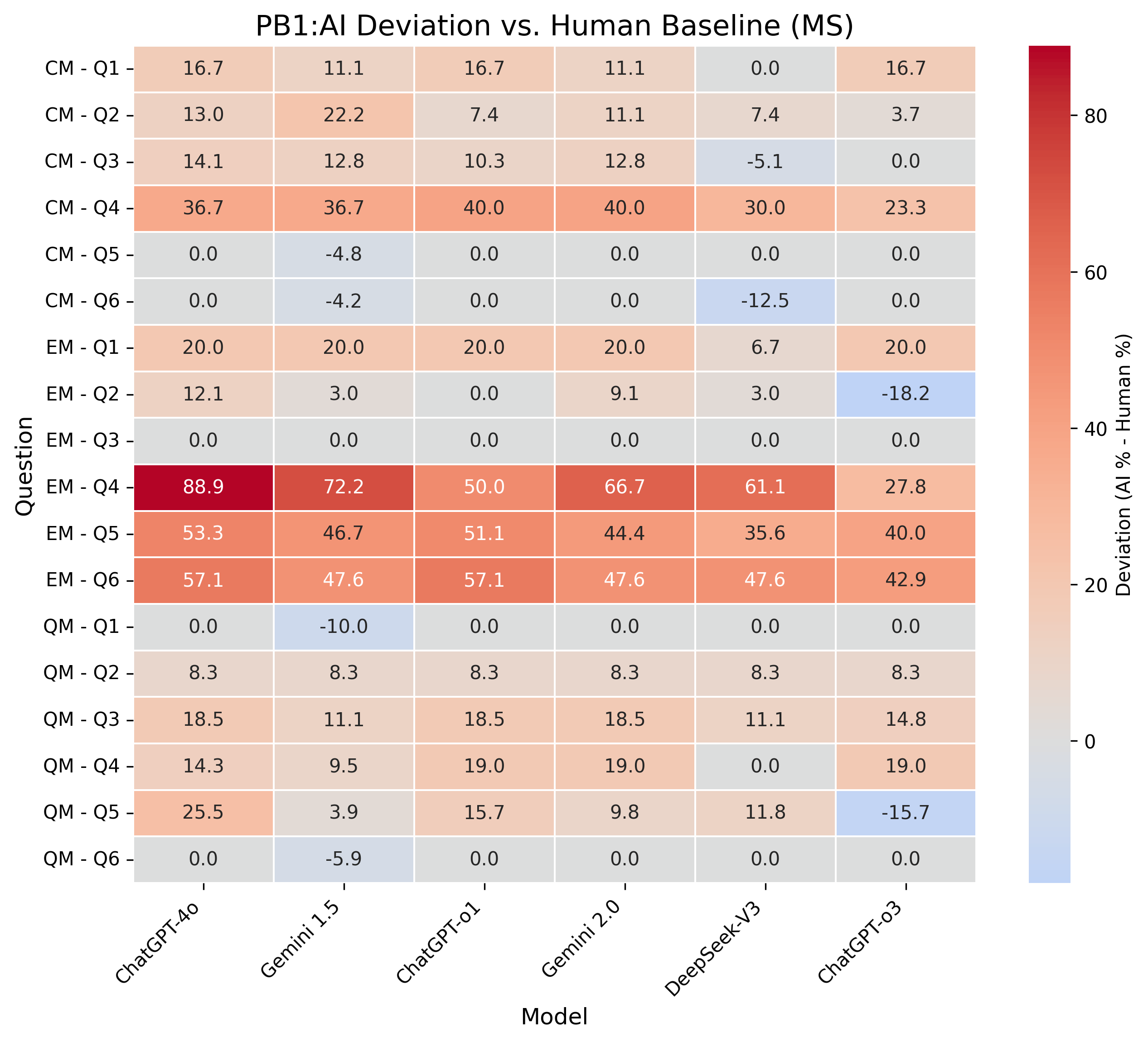}
\caption{Deviation heat-map Mapping Absolute Percentage Difference (AI vs. Human Marking)}
\label{fig:deviation-heatmap}
\end{figure}

To further pinpoint the locus of this systemic over-marking, a deviation heat-map (Figure \ref{fig:deviation-heatmap}) was generated, mapping the absolute percentage difference between AI and human grading across Set B (MS). While the regression analysis established the macro-trend of grade inflation, the heat-map reveals its highly uneven distribution. 

The most striking feature is the intense clustering of positive deviation within the Electromagnetism module—specifically EM-Q4, EM-Q5, and EM-Q6. In EM-Q4, for instance, ChatGPT-4o deviated from the human baseline by +88.9\%, effectively awarding near-perfect scores to solutions that human graders deemed fundamentally flawed. This indicates that grading leniency is not applied uniformly; rather, models exhibit the highest leniency in domains where their own underlying architectural constraints (such as spatial reasoning) are weakest. When unable to parse the physical geometry of a student's answer, legacy models default to leniency. Conversely, the vertical profile for ChatGPT-o3 visually reinforces the regression data, displaying the lowest deviations and confirming its stability across all three subject areas.

\subsubsection{PB2} 
Having established that Gen 4 and Gen 5 models possess the foundational reasoning to solve the benchmark entirely, the final stage of analysis evaluates their utility as automated assessors. To conduct this, the methodology was refined: rather than batching the three generated ChatGPT-4o solutions together within a single prompt—as was done in PB1—the Gen 4 and Gen 5 models evaluated each solution individually.

\begin{figure}[H]
\centering
\includegraphics[width=0.8\linewidth]{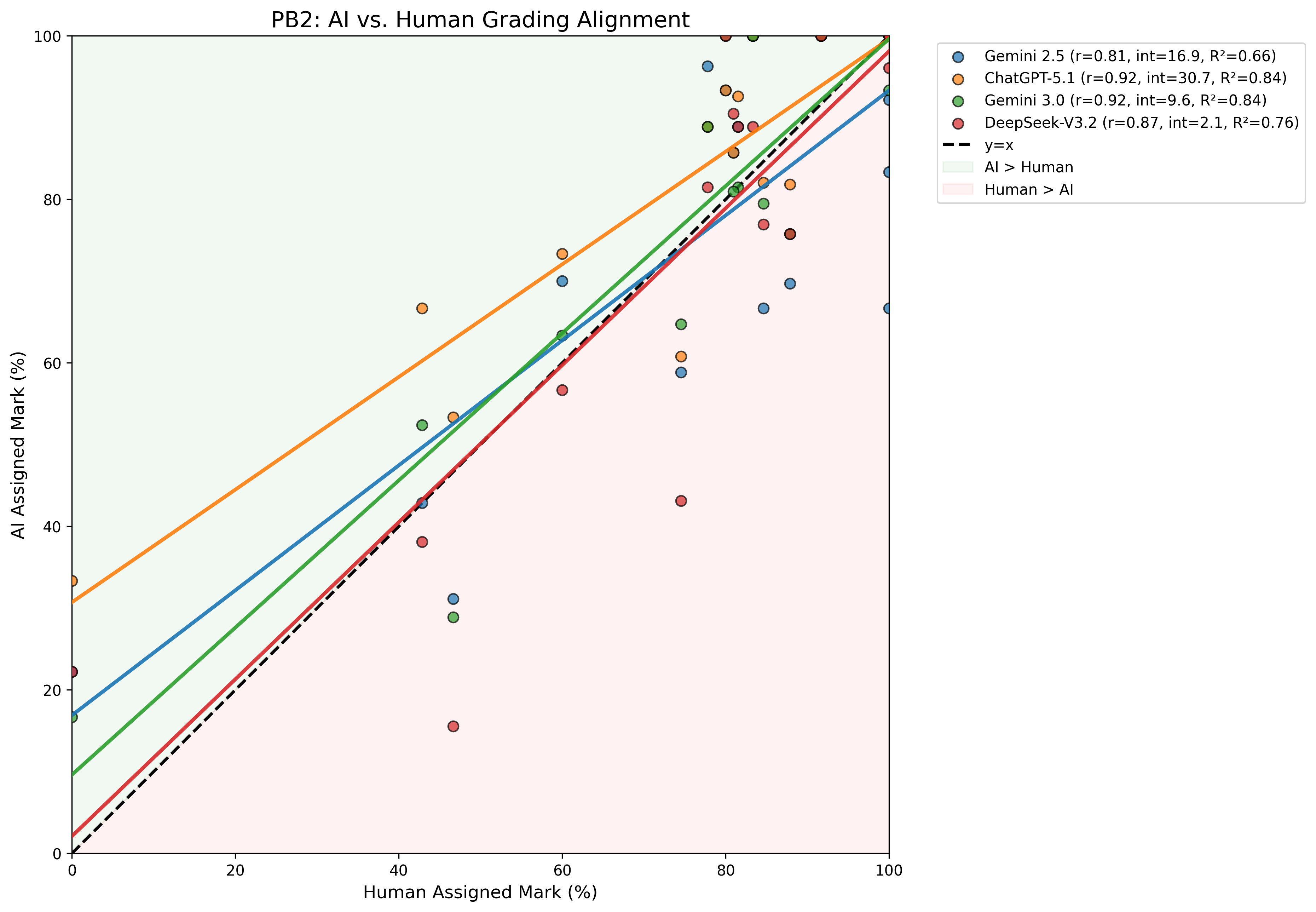}
\caption{Linear Regression of Model Alignment (Individual Grading, P2)}
\label{fig:p2-regression}
\end{figure}

\begin{figure}[H]
\centering
\includegraphics[width=0.8\linewidth]{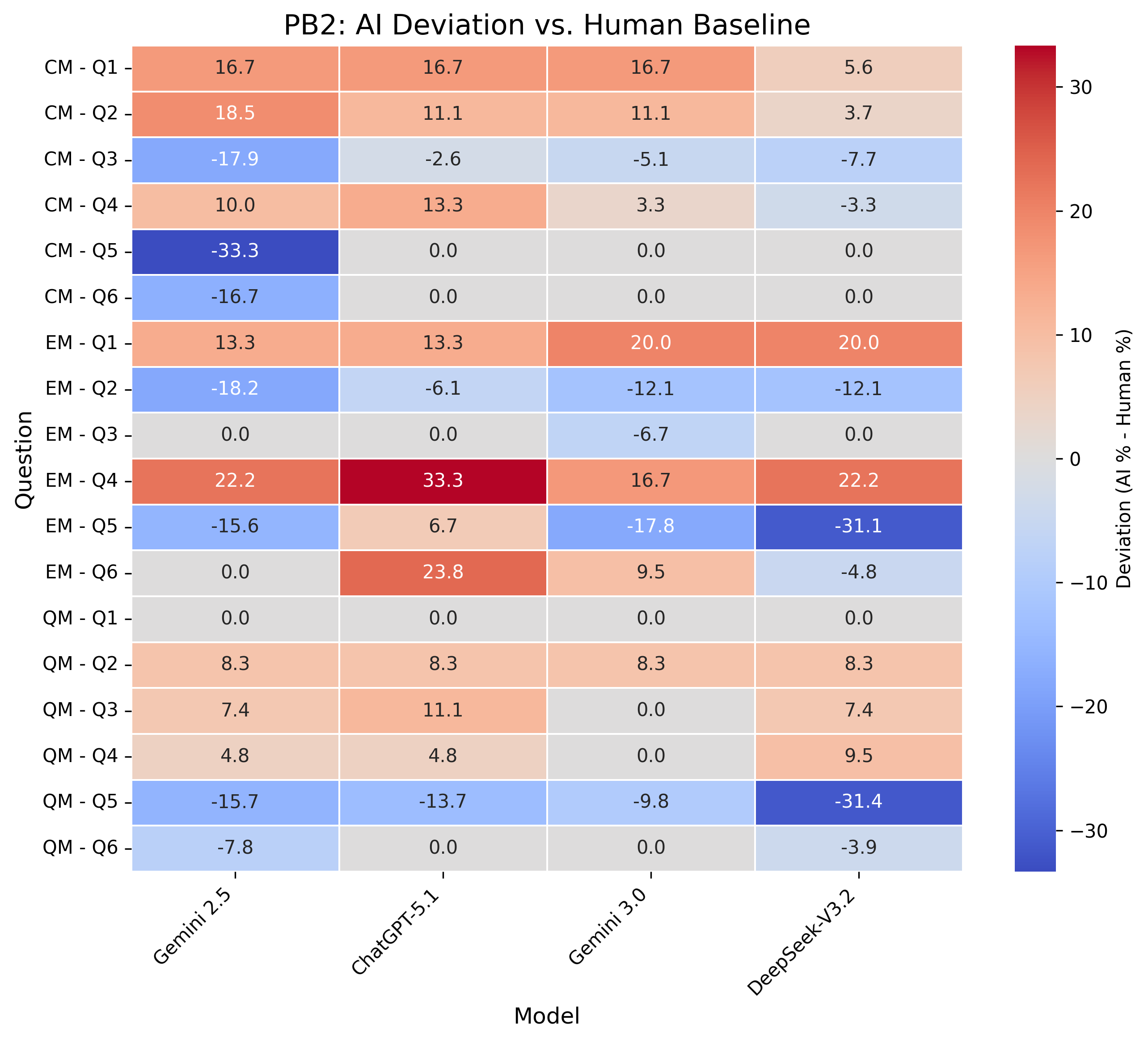}
\caption{AI Deviation vs. Human Baseline per Question (P2)}
\label{fig:p2-heatmap}
\end{figure}

This methodological shift isolates the evaluation task to mitigate contextual interference. The resulting regression statistics (Table \ref{tab:p2-alignment} and Figure \ref{fig:p2-regression}) and deviation heat-map (Figure \ref{fig:p2-heatmap}) provide a more precise representation of each model's internal grading standard by preventing the haloing and U-shaped attention curve effects observed in batched evaluation.

\begin{table*}[tbh]
\centering
\label{tab:p2-alignment}
\begin{tabular}{|l|c|c|c|c|c|c|}
\hline
\textbf{Model} & \textbf{Slope} & \textbf{Y-Int} & \textbf{Pearson $r$} & \textbf{$R^2$} & \textbf{MAE (\%)} & \textbf{RMSE (\%)} \\ \hline
Gemini 2.5     & 0.7644         & 16.88        & 0.8115               & 0.6585         & 12.58           & 15.14            \\ \hline
ChatGPT-5.1    & 0.6892         & 30.70        & 0.9175               & 0.8417         & 9.15            & 12.75            \\ \hline
Gemini 3.0     & 0.9002         & 9.60         & 0.9191               & 0.8448         & 7.61            & 10.22            \\ \hline
DeepSeek-V3.2  & 0.9603         & 2.09         & 0.8719               & 0.7602         & 9.50            & 13.66            \\ \hline
\end{tabular}
\caption{Alignment and IRR Statistics for Gen 4 \& 5 (P2)}
\end{table*}

The regression data demonstrates a substantial improvement in alignment, indicating a reduction in the systemic over-marking seen in earlier generations. DeepSeek-V3.2 emerges as a highly aligned evaluator. With a Y-intercept of 2.09 and a gradient of 0.9603, its regression line closely approximates the ideal $y=x$ baseline. It does not artificially inflate poor answers or compress the top end, successfully utilising the entire spectrum of the grading rubric.

However, Gemini 3.0 Pro achieved the strongest overall statistical alignment. It posted the highest Pearson correlation ($r = 0.9191$), an $R^2$ of 0.8448, and a low Mean Absolute Error (MAE) of 7.62\%. This indicates that Gemini 3.0 Pro accurately accounts for nearly 85\% of the variance in human grading, demonstrating a highly nuanced sensitivity to the structural integrity of a physics derivation.

ChatGPT-5.1, despite having an exceptionally high correlation ($r = 0.9175$), still exhibits a milder version of the "floor effect" observed in earlier OpenAI models. Its relatively high Y-intercept (30.71) and shallower gradient (0.6892) indicate that while it scales its grades consistently with human evaluators, it rarely assigns scores near zero, maintaining a persistent baseline of leniency.

This alignment is further contextualised by the deviation heat-map. The most immediate visual distinction from PB1 is the introduction of blue (under-marking) zones. Gen 1 through Gen 3 models are almost universally over-marked. In contrast, Gen 4 and Gen 5 models demonstrated a tendency to assign negative deviations relative to the human baseline. 

DeepSeek-V3.2, for example, shows substantial negative deviations in EM-Q5 (-31.1\%) and QM-Q5 (-31.4\%). Models like DeepSeek-V3.2 and Gemini 2.5 Flash penalise logical missteps more heavily, occasionally marking more strictly than the human baseline when confronted with flawed circuit topologies or incorrect boundary conditions. 

Ultimately, this phase confirms that, when provided with isolated context windows, Gen 5 architectures demonstrate sufficient consistency, rigour, and variance tracking to serve as reliable automated assessors in first-pass approaches.

\subsubsection{MM Handwritten Grading}
Across all models, grading accuracy was consistently higher for perfect handwritten solutions than for imperfect ones. Mean absolute error (MAE) analysis shows near-zero deviation from human grading for perfect scripts for all models, indicating reliable recognition of canonical solution structures and correct reasoning patterns. In contrast, imperfect solutions produced substantially higher MAE across all architectures, demonstrating that partial-credit evaluation remains a significantly more challenging task than recognition of fully correct work. This suggests that the core limitation lies not in visual transcription, but in reasoning-based assessment of incomplete or flawed logic.

Model-level comparisons reveal distinct behavioural differences. ChatGPT-4o achieved the lowest MAE for perfect handwritten solutions, consistently assigning exact human-equivalent grades. However, its performance degraded sharply for imperfect scripts, where it exhibited the largest discrepancy between perfect and imperfect grading accuracy. In contrast, ChatGPT-5.1, Gemini 2.5 Flash and Gemini 3.0 Pro demonstrated greater accuracy across imperfect solutions, maintaining lower MAE and indicating improved stability in evaluating partial correctness. ChatGPT-4o is the oldest model assessed and exhibits the highest error for imperfect solutions, aside from DeepSeek, which exhibited consistently elevated MAE and high variance across both perfect and imperfect solutions. This confirms that multimodal LLMs outperform OCR-mediated LLMs for handwritten physics assessment. Overall, the evidence suggests that newer multimodal architectures are less dependent on surface-level pattern matching and better able to trace multi-step reasoning when solutions deviate from canonical forms.

\begin{figure}[H]
    \centering
    \includegraphics[width=0.8\linewidth]{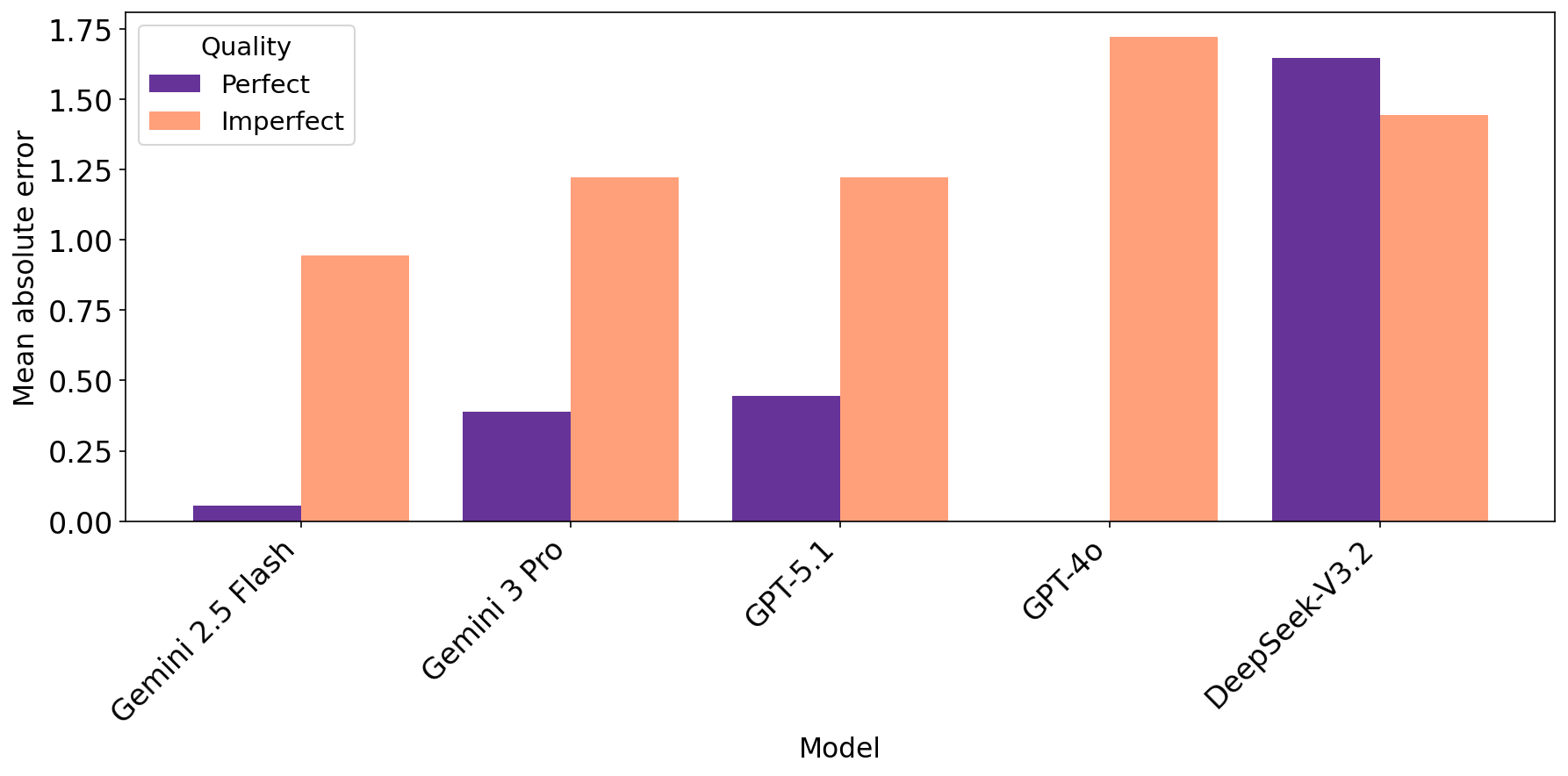}
    \label{fig:placeholder}
    \caption{Mean absolute grading error across models for perfect and imperfect handwritten physics solutions. While grading of perfect scripts shows near-zero average error for several models, imperfect solutions lead to significantly higher average errors, highlighting the difficulty of reasoning-based partial-credit evaluation.}
\end{figure}

To further examine whether grading discrepancies arise primarily from perceptual limitations or from reasoning constraints, the relationship between each model’s solution-generation accuracy and its grading error on imperfect handwritten scripts was analysed.

Across models, a clear negative correlation emerges between solution accuracy and mean absolute grading error. When the solution accuracy for diagram-based questions is considered, a moderate inverse relationship is observed ($R^2 = 0.248$, $r = -0.498$), indicating that models with stronger diagram interpretation ability tend to exhibit lower grading error. However, this association remains relatively weak.

When the solution accuracy for text-based questions is considered, the relationship strengthens considerably ($R^2 = 0.459$, $r = -0.677$). Here, approximately 46\% of the variance in grading error can be associated with differences in reasoning performance. Models with higher problem-solving proficiency demonstrate lower grading deviation on imperfect scripts.

This pattern reinforces the earlier interpretation: grading instability in handwritten contexts is more strongly linked to reasoning capacity than to perceptual transcription ability. While multimodal perception is necessary for decoding handwritten inputs, the dominant bottleneck in partial-credit allocation appears to be reasoning-based assessment. 

Although the number of models analysed is limited, the monotonic downward trend across successive model generations suggests a structural relationship rather than random variation. As reasoning proficiency improves across architectures, grading reliability correspondingly increases.

\begin{figure}[H]
    \centering
    \includegraphics[width=0.7\linewidth]{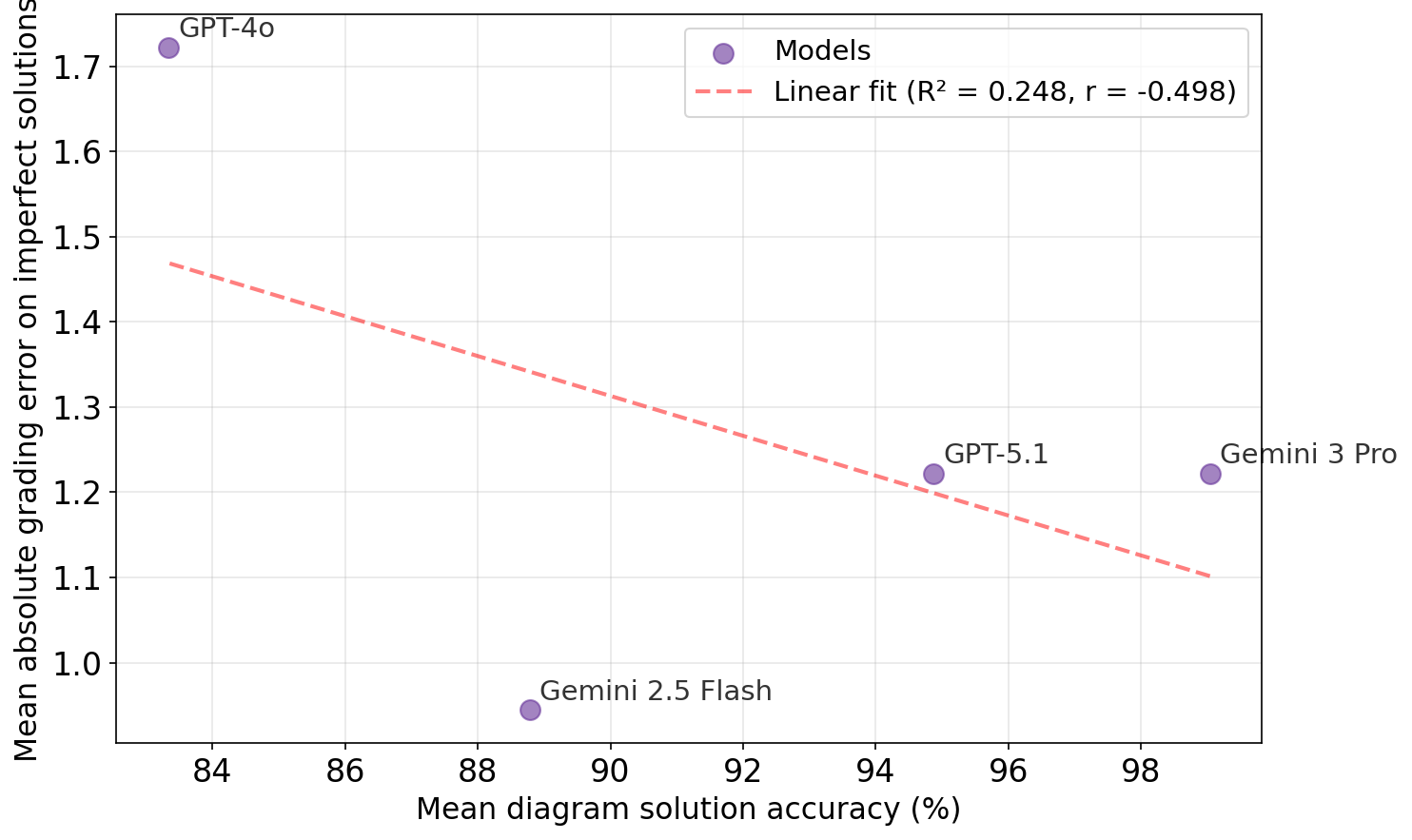}
    \label{fig:placeholder}
    \caption{Relationship between model solution accuracy on diagram-based physics problems and mean absolute grading error on imperfect handwritten solutions. Each point represents a model, and the dashed line indicates a linear regression fit. A moderate negative correlation is observed, suggesting that models with stronger diagram interpretation ability tend to produce slightly lower grading errors.}
\end{figure}

\begin{figure}[H]
    \centering
    \includegraphics[width=0.7\linewidth]{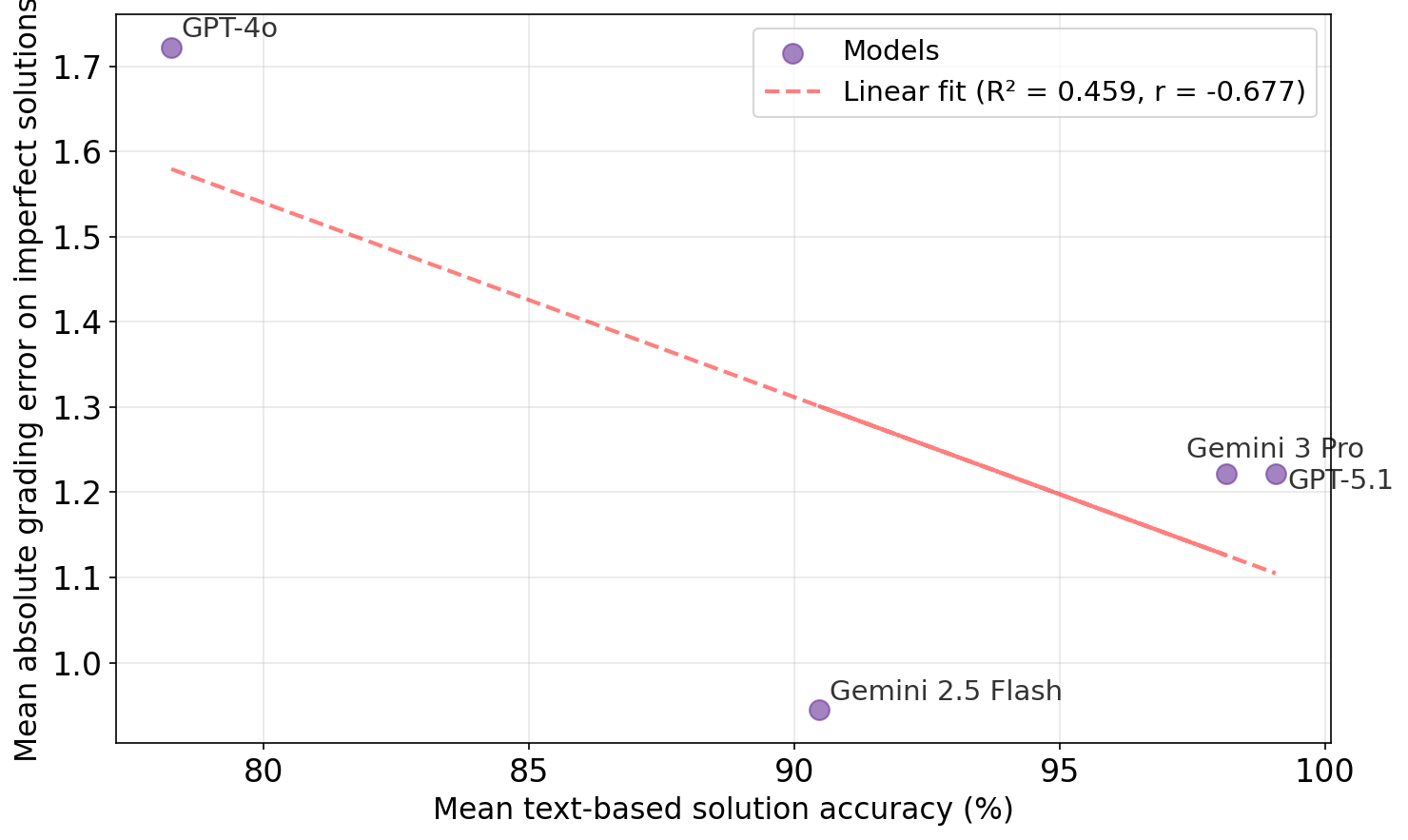}
    \label{fig:placeholder}
    \caption{Relationship between model solution accuracy on text-based physics problems and mean absolute grading error on imperfect handwritten solutions. Each point represents a model, with the dashed line showing the linear regression fit. A moderate negative correlation is observed, indicating that grading reliability is associated with reasoning performance.}
\end{figure}

Module-level analysis further highlights structural influences on grading reliability. For perfect solutions, all models achieved near-zero error across Electromagnetism (EM), Quantum Mechanics (QM), and Classical Mechanics (CM), further indicating that handwriting recognition itself is not a limiting factor for multimodal models. For imperfect solutions, however, systematic differences emerged. 

EM exhibited the largest error variability, reflecting the highly procedural nature of electromagnetic solutions, where missing or incorrect intermediate steps create ambiguity in mark allocation. A representative example occurred in an electromagnetic field calculation problem in which the student incorrectly substituted $0.16\,\mathrm{m}$ for the given plate separation of $0.15\,\mathrm{m}$. Gemini 2.5 Flash interpreted this as a minor numerical error and awarded $4/5$ with error-carried-forward credit, whereas the human marker awarded $1/5$, judging the incorrect substitution to invalidate the field magnitude and subsequent acceleration calculation. This divergence illustrates how electromagnetic problems, which rely on tightly coupled procedural steps and precise numerical substitution, create ambiguity in partial-credit allocation. This example also shows that variability in human marking can be large, since other human markers may have used different criteria. It is realistic to think that LLMs could be used for grading once the variance of the marks between the various models, and those with respect to a human marker, is comparable to that of various independent human markers.

QM showed broadly similar error distributions across models, consistent with the abstract and symbolic nature of quantum formalism, where partial correctness is harder to localise. A representative quantum mechanics example involved a potential step scattering derivation graded $17/17$ by ChatGPT-4o and $15/17$ by the human marker. The student’s derivational structure, boundary conditions, and final expressions for $T$ and $R$ were correct, with only minor formal imprecision in the definitions of $k_1$ and $k_2$ and compact presentation of intermediate algebra. Unlike the electromagnetism example, there was no single localised procedural failure that invalidated subsequent reasoning. Instead, correctness was distributed across the symbolic structure of the derivation, resulting in only modest inter-grader discrepancy. This supports the observation that quantum mechanics exhibited narrower error variability across models.

CM displayed the smallest and most stable error distributions for imperfect solutions, suggesting that mechanically traceable errors in classical dynamics are more easily identified and penalised consistently. A representative example involved a special relativity question on energy-momentum 4-vectors, where the student correctly completed part (a) and the rest-frame component of part (b), but failed to perform the required Lorentz transformation to the lab frame. Every model awarded 7/13 with identical sub-part allocations. The error was structurally unambiguous — a missing transformation step with no partial working to interpret — leaving no room for the kind of partial-credit disagreement that drove variability in EM and QM.

%Or figs 7 and 8
\begin{figure}[H]
    \centering
    \includegraphics[width=0.6\linewidth]{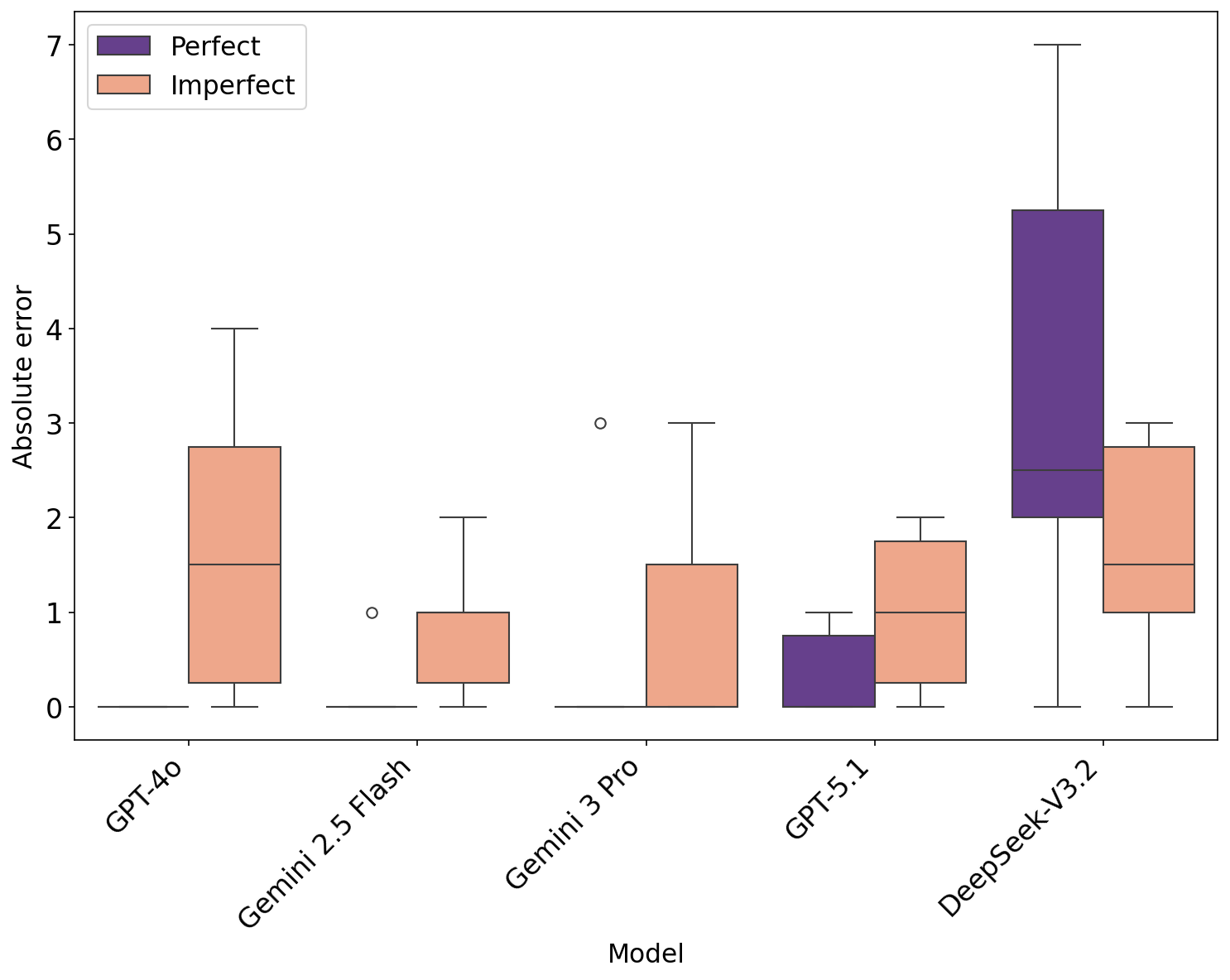}
    \label{fig:placeholder}
    \caption{Absolute grading error distributions for each model when evaluating handwritten Classical Mechanics (CM) solutions. Results are separated for perfect and imperfect scripts. Perfect solutions exhibit near-zero error across models, whereas imperfect scripts show broader distributions due to the challenge of assigning partial credit for incomplete or flawed reasoning.}
\end{figure}

\begin{figure}[H]
    \centering
    \includegraphics[width=0.6\linewidth]{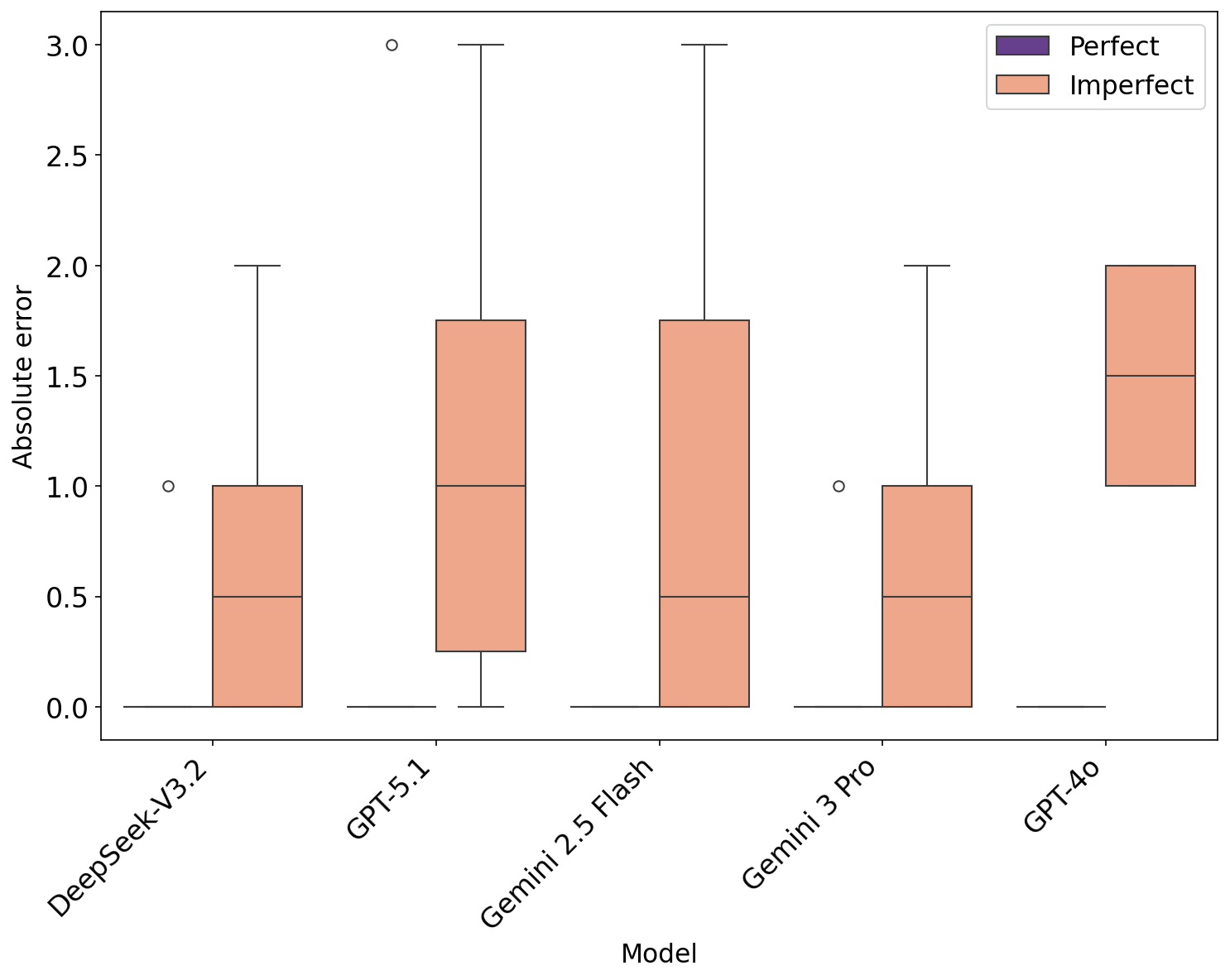}
    \label{fig:placeholder}
    \caption{Absolute grading error distributions for each model when evaluating handwritten Quantum Mechanics (QM) solutions. Perfect solutions again show minimal error, while imperfect scripts produce moderate variability reflecting the distributed and symbolic nature of correctness in quantum derivations.}
\end{figure}

\begin{figure}[H]
    \centering
    \includegraphics[width=0.6\linewidth]{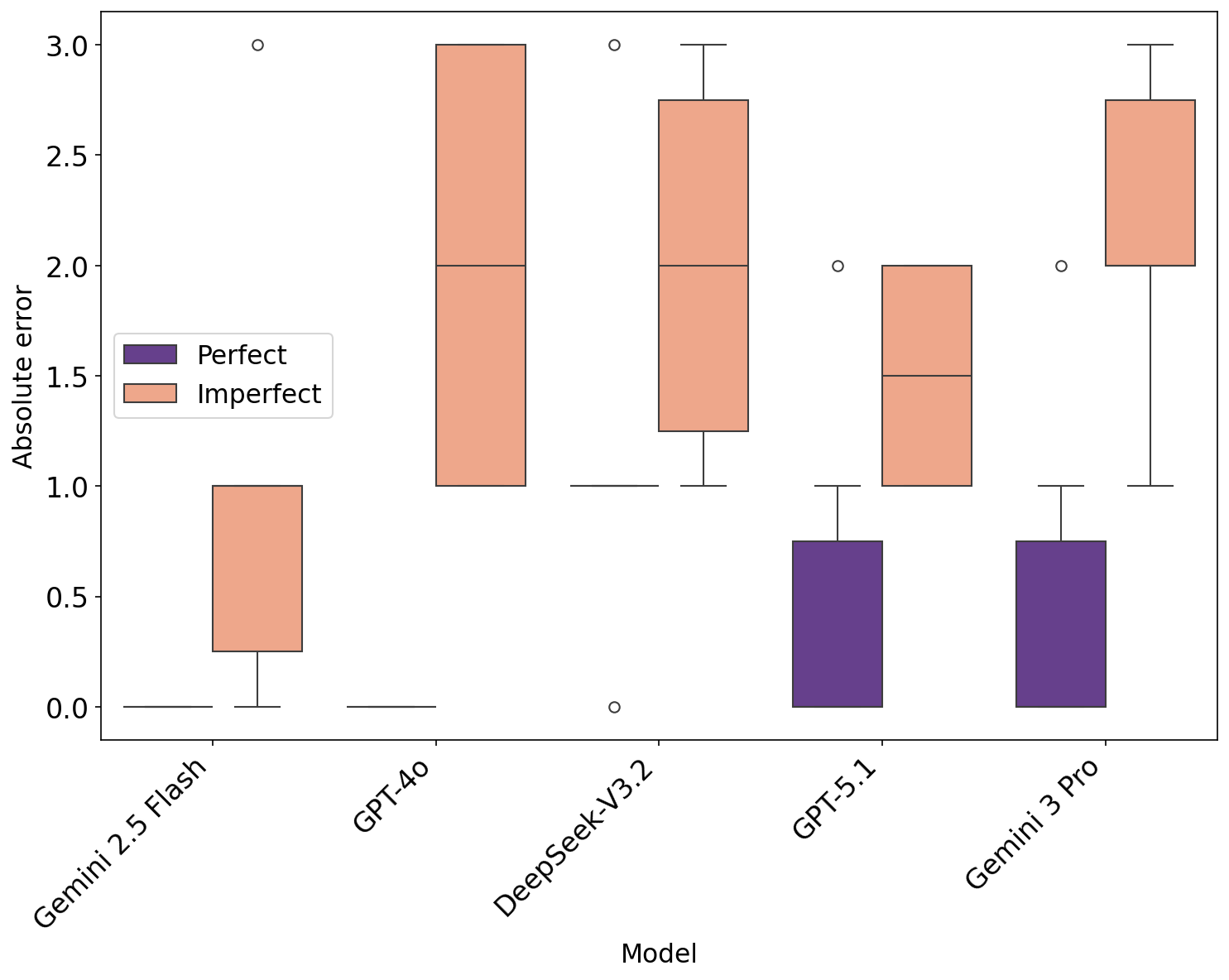}
    \label{fig:placeholder}
    \caption{Absolute grading error distributions for each model when evaluating handwritten Electromagnetism (EM) solutions. Imperfect scripts exhibit the largest variability across models, highlighting the procedural structure of electromagnetic problems where early numerical or substitution errors propagate through subsequent steps.}
\end{figure}

Signed error analysis indicates that LLM grading of handwritten solutions is not symmetrically distributed around human marks. For imperfect solutions, distributions are shifted slightly toward positive values, indicating a mild but consistent tendency toward over-marking. This effect is most pronounced at lower human grades, where small absolute discrepancies translate into large percentage differences.

The absence of a monotonic relationship between grading error and total available marks further supports this interpretation. Signed error shows no clear systematic relationship with question size, suggesting that grading discrepancies occur at the level of raw marks rather than scaling with total mark allocation. Consequently, the apparent bias in percentage-based analyses arises primarily from proportional scaling rather than intrinsic model behaviour.

\begin{figure}[H]
    \centering
    \includegraphics[width=0.7\linewidth]{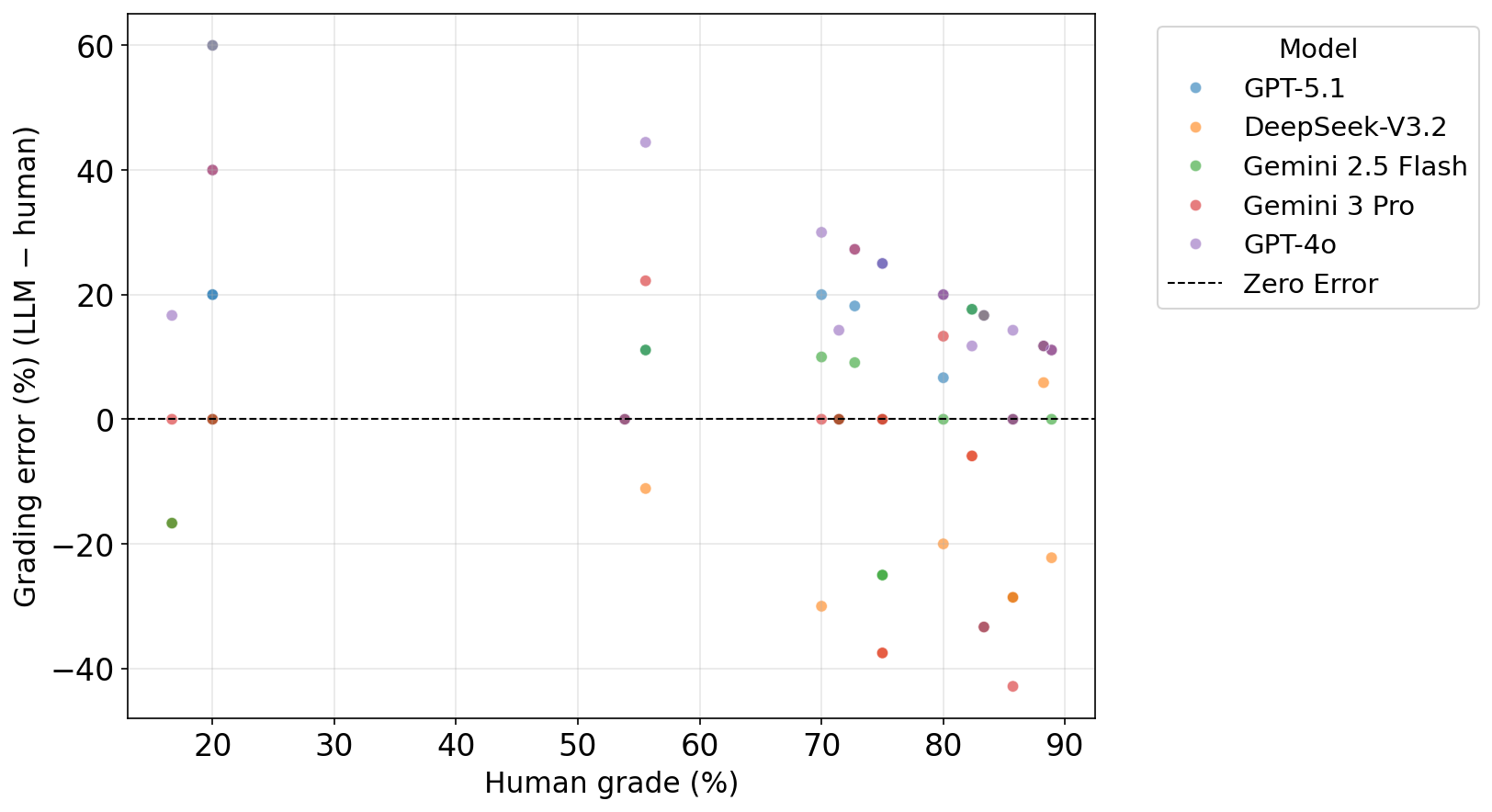}
    \caption{Percentage grading error for imperfect handwritten physics solutions, plotted against the human-assigned grade. Error is defined as the percentage difference between the LLM and human marks. Several models show a mild tendency toward positive error (over-marking), particularly at lower human grades, where small raw mark discrepancies correspond to larger percentage deviations.}
    \label{fig:percent_error_vs_grade}
\end{figure}

\begin{figure}[H]
    \centering
    \includegraphics[width=0.7\linewidth]{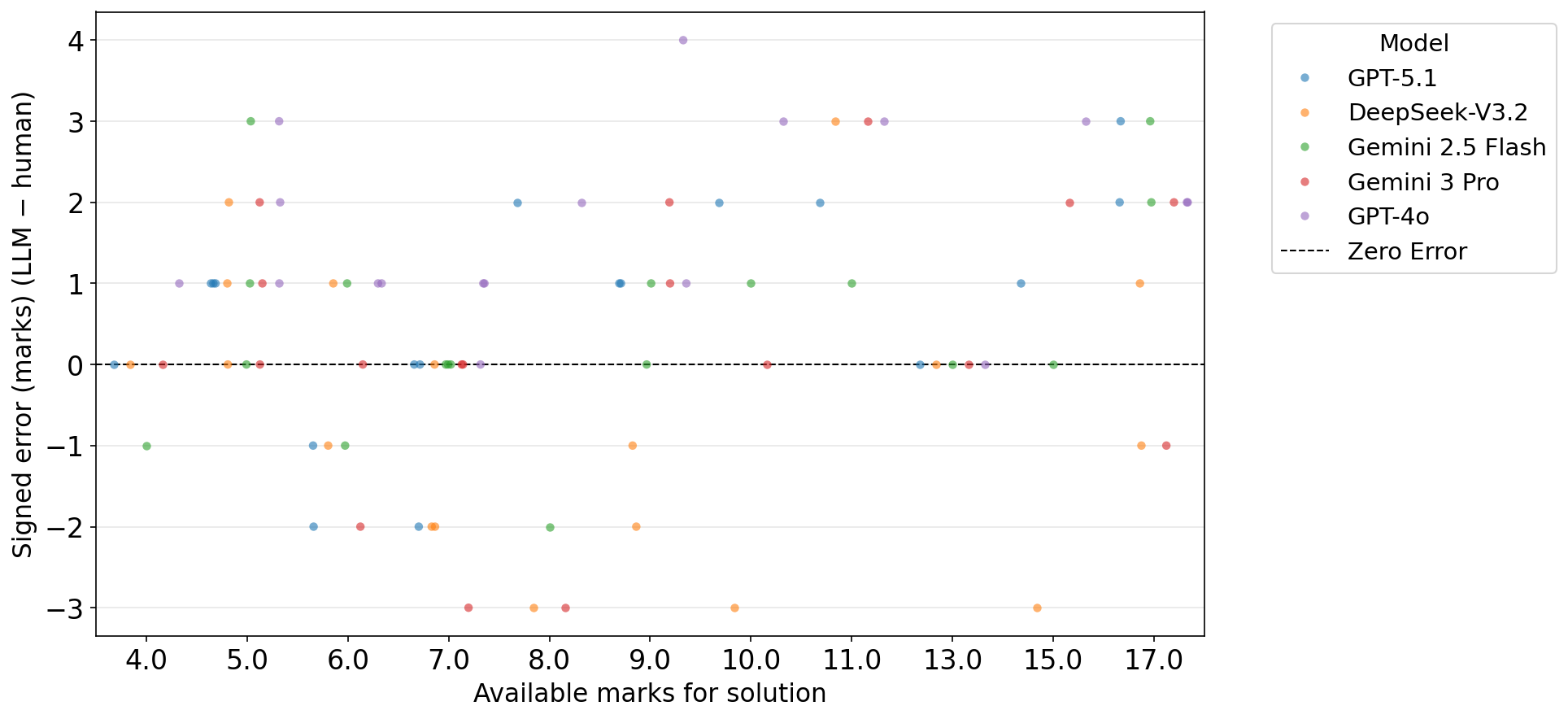}
    \caption{Signed grading error (LLM marks minus human marks) for imperfect handwritten physics solutions, plotted against the total available marks for each question. The spread of errors shows no clear systematic relationship with question size, indicating that grading discrepancies occur primarily at the level of raw marks rather than scaling with total mark allocation. This suggests that the apparent percentage-based bias seen in Figure 8 is largely a consequence of proportional scaling.}
    \label{fig:signed_error_vs_marks}
\end{figure}

Collectively, these findings indicate that current multimodal LLMs can recognise and grade fully correct handwritten physics solutions fairly reliably, but exhibit increased uncertainty and mild generosity when allocating partial credit. The core limitation is not handwriting transcription, but evaluative reasoning under uncertainty. While newer models show clear improvements in accuracy, partial-credit grading of handwritten work remains a fundamental challenge for multimodal LLMs.

\section{Conclusions}
This study evaluated Large Language Model (LLM) proficiency within undergraduate university physics education, assessing their utility both as independent problem-solvers and automated grading assistants. Across three distinct testing phases (PB1, PB2, and MM), the data reveals a fundamental shift in model capabilities, driven by architectural scaling and the integration of native multimodal processing.

Initial benchmarking (PB1) established that while early architectures (Gen 1 to Gen 3) demonstrated high competency in abstract, syntactically driven domains like Quantum Mechanics, they were severely constrained by tasks requiring spatial reasoning or system setup. This limitation manifested as a "reasoning cliff" in Classical Mechanics and Electromagnetism, where models frequently relied on training heuristics—resulting in setup hallucinations rather than generating robust physical representations. Furthermore, early models proved to be unreliable evaluators; when tasked with grading solutions, they exhibited a systemic bias toward grade inflation, heavily influenced by positional biases and an inability to penalise flawed logic consistently.

Subsequent evaluation of Gen 4 and Gen 5 models (PB2) demonstrated a clear trajectory toward benchmark saturation in text-based problem-solving. Flagship models, including ChatGPT-5.1, Gemini 3.0 Pro, and DeepSeek-V3.2, approached the theoretical ceiling of the dataset, achieving near-perfect mean scores and substantially mitigating earlier setup bottlenecks. This phase also highlighted significant improvements in evaluative reliability. By isolating the grading context to prevent haloing effects, Gen 5 models successfully aligned with human marking standards. Notably, the latest models demonstrate the capacity to rigorously penalise incorrect reasoning rather than defaulting to leniency.

The Multimodal (MM) investigation definitively proved that the persistent spatial errors observed in earlier generations were primarily input bottlenecks rather than algebraic deficits. By employing systematic visual deconstruction—a behaviour highly consistent with Visual Chain-of-Thought (VCoT) processing—Generation 5 models successfully extracted topological logic directly from diagrams, significantly improving their ability to decipher complex physical diagrams. However, the evaluation of handwritten solutions revealed a lingering limitation. While contemporary multimodal LLMs can reliably transcribe and grade fully correct handwritten physics solutions, they exhibit increased error and variability when allocating partial credit to imperfect scripts. The abstract, procedural nature of physics derivations means that models struggle to maintain evaluative consistency when student logic diverges from canonical pathways.

However, empirical evidence also shows that human graders are not immune to these divergences, and the evaluation of an imperfect solution maintains a certain degree of variability.

Ultimately, these findings confirm that modern architectures demonstrate the consistency and accuracy required to serve as highly effective, automated Teaching Assistants. Their interactive nature can be a valuable tool to help students in their learning, and their assessment capabilities can speed up feedback for students. However, their integration into formal assessment pipelines must remain supervised, particularly in scenarios requiring nuanced, partial-credit evaluation of flawed human reasoning.

\newpage

\bibliography{refs}

\clearpage

\appendix

% Tie figure and table numbering to the appendix sections
\numberwithin{figure}{section}
\numberwithin{table}{section}
% Remove the default dot between the letter and number
\renewcommand{\thefigure}{\thesection\arabic{figure}}
\renewcommand{\thetable}{\thesection\arabic{table}}

\section{Referenced Questions} 

\begin{figure}[htbp] 
    \centering
    \includegraphics[width=0.7\linewidth]{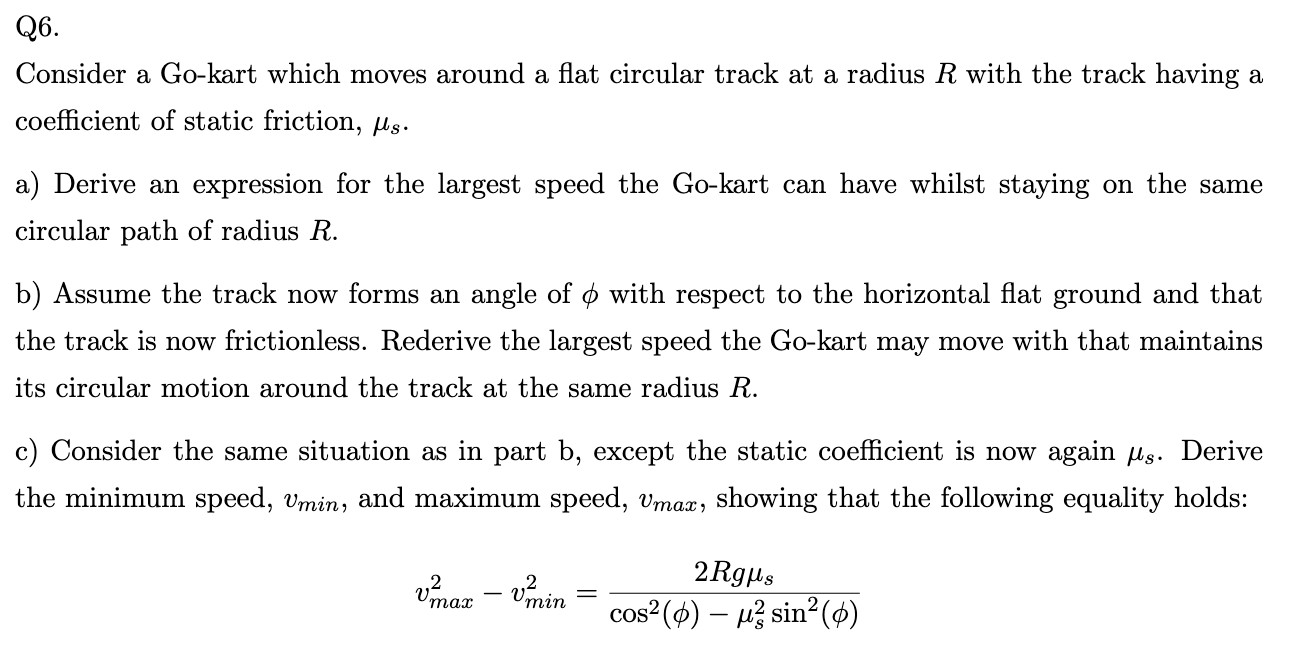}
    \caption{Classical Mechanics Question 6 From Set A}
    \label{fig:CM_Q6}
\end{figure}

\begin{figure}[htbp] 
    \centering
    \includegraphics[width=0.7\linewidth]{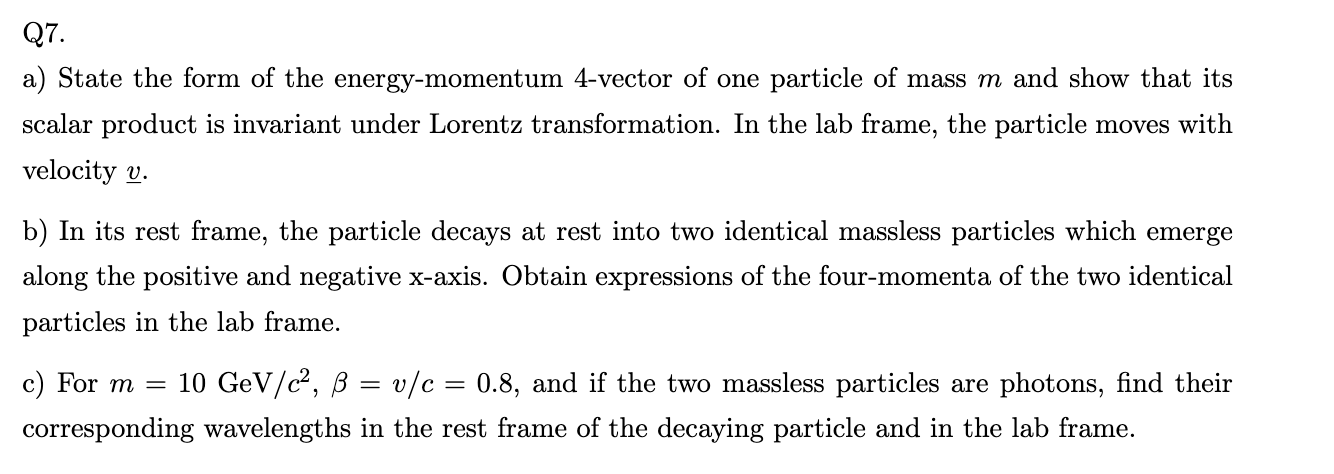}
    \caption{Classical Mechanics Question 7 From Set A}
    \label{fig:CM_Q7} 
\end{figure}

\begin{figure}[htbp] 
    \centering
    \includegraphics[width=0.7\linewidth]{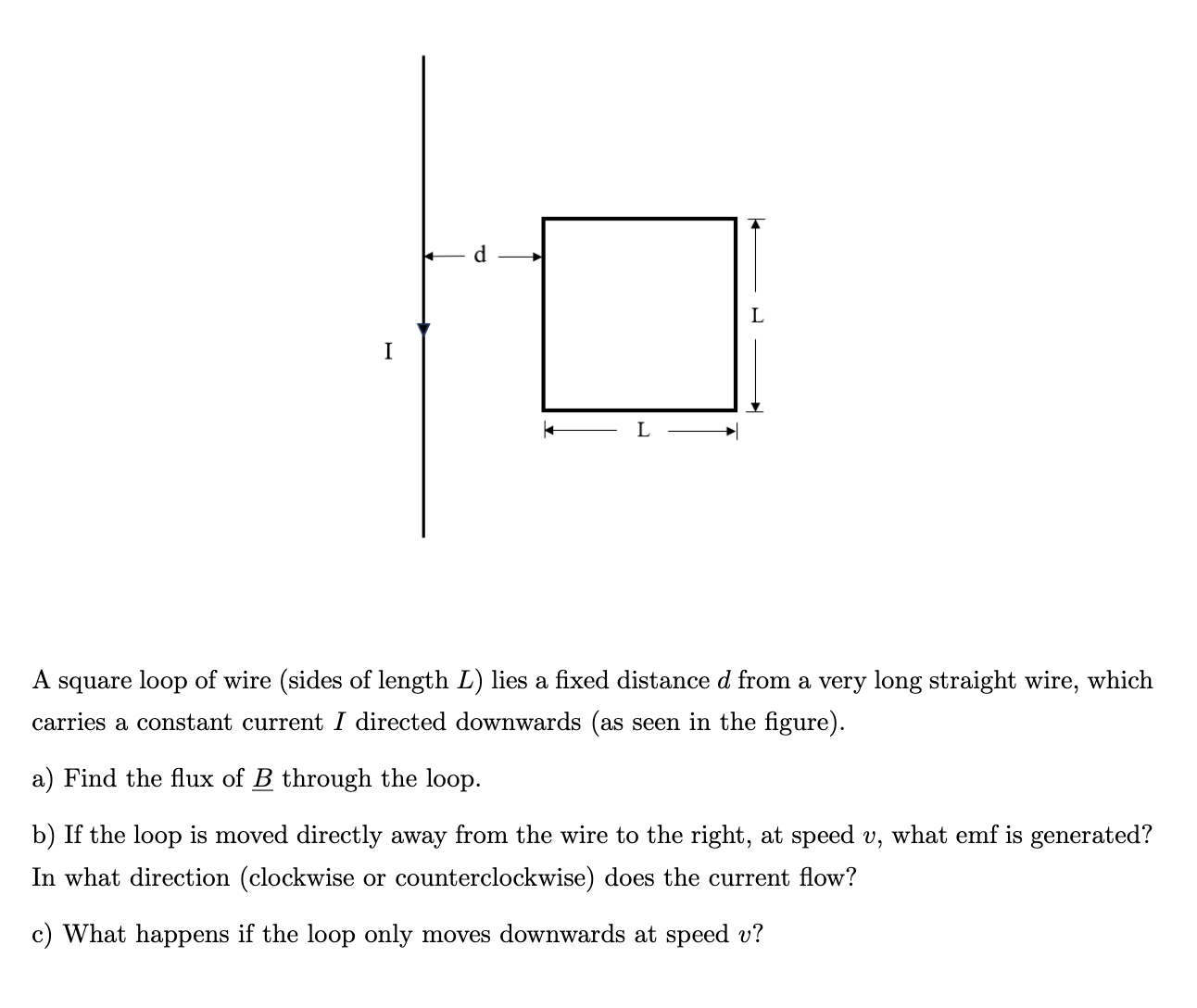}
    \caption{Electromagnetism Question 4 From Set A}
    \label{fig:EM_Q4} 
\end{figure}

\begin{figure}[htbp] 
    \centering
    \includegraphics[width=0.7\linewidth]{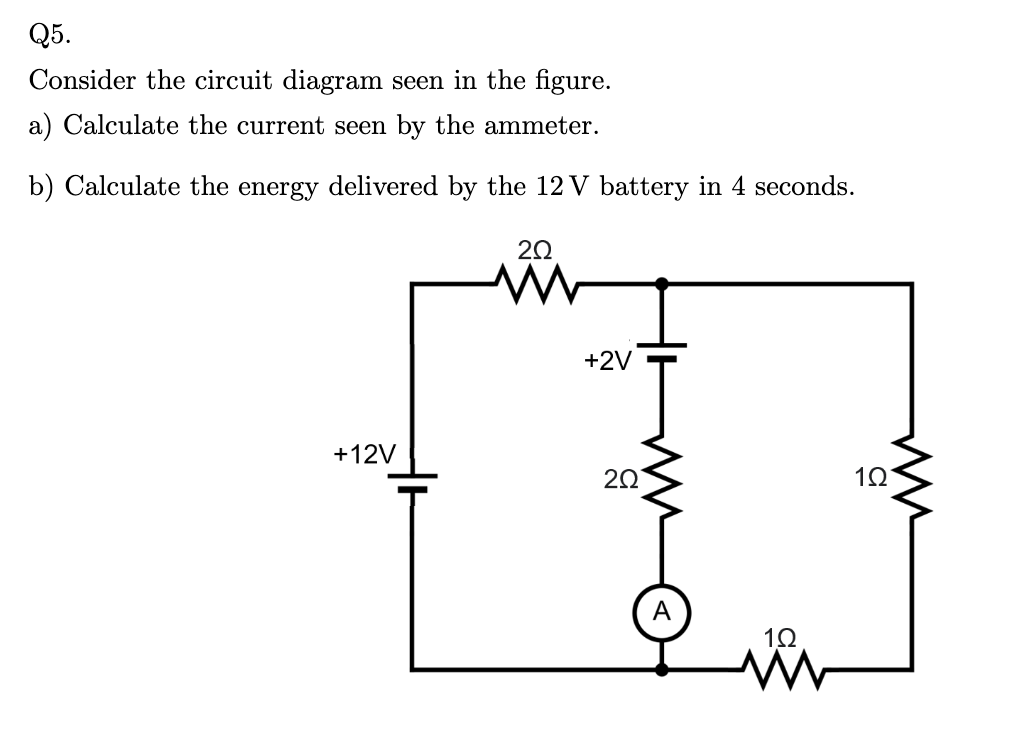}
    \caption{Electromagnetism Question 4 From Set A}
    \label{fig:EM_Q5} 
\end{figure}

\begin{figure}[htbp] 
    \centering
    \includegraphics[width=0.7\linewidth]{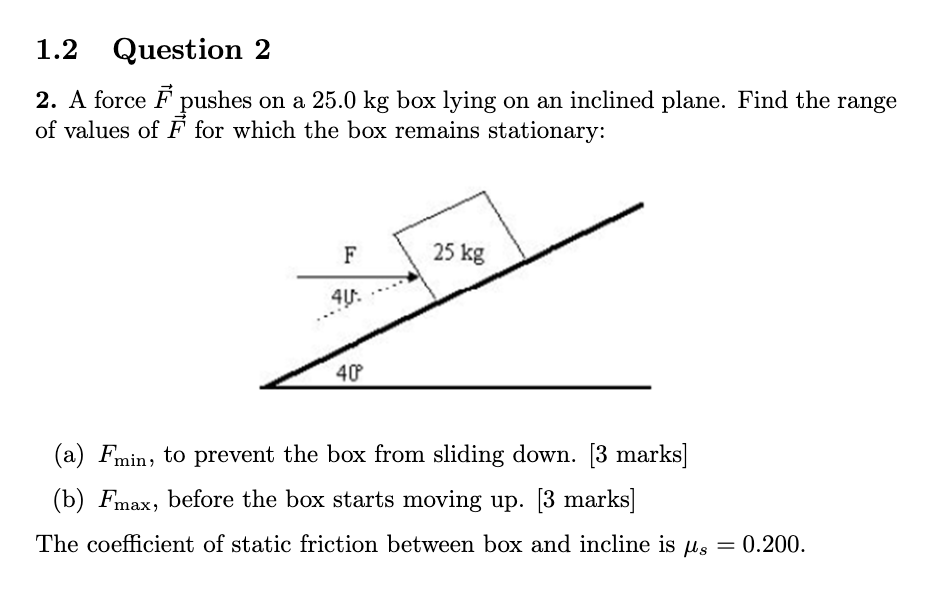}
    \caption{Classical Mechanics Q2 from Set C. Note that the minor rendering artefact over the $40^{\circ}$ label did not interfere with model comprehension or subsequent derivations.}
    \label{fig:CMc_Q2} 
\end{figure}

\clearpage
\section{Supplementary Figures and Tables}

\begin{figure}[htbp] 
    \centering
    \includegraphics[width=0.7\linewidth]{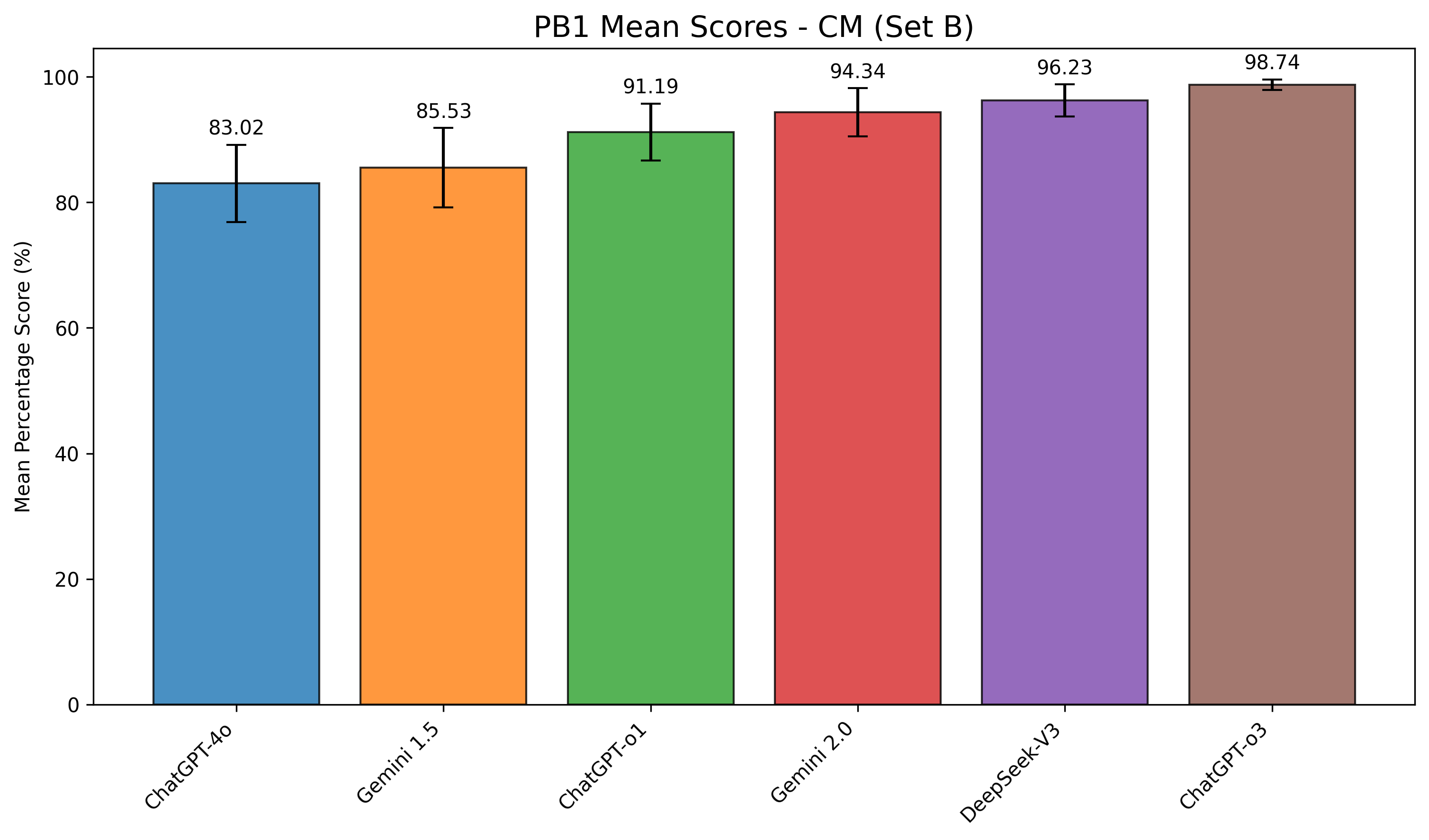}
    \caption{Bar chart showing PB1 model performance in Classical Mechanics (Set B)}
    \label{fig:pb1b_cm_bar_chart}
\end{figure}

\begin{figure}[htbp] 
    \centering
    \includegraphics[width=0.7\linewidth]{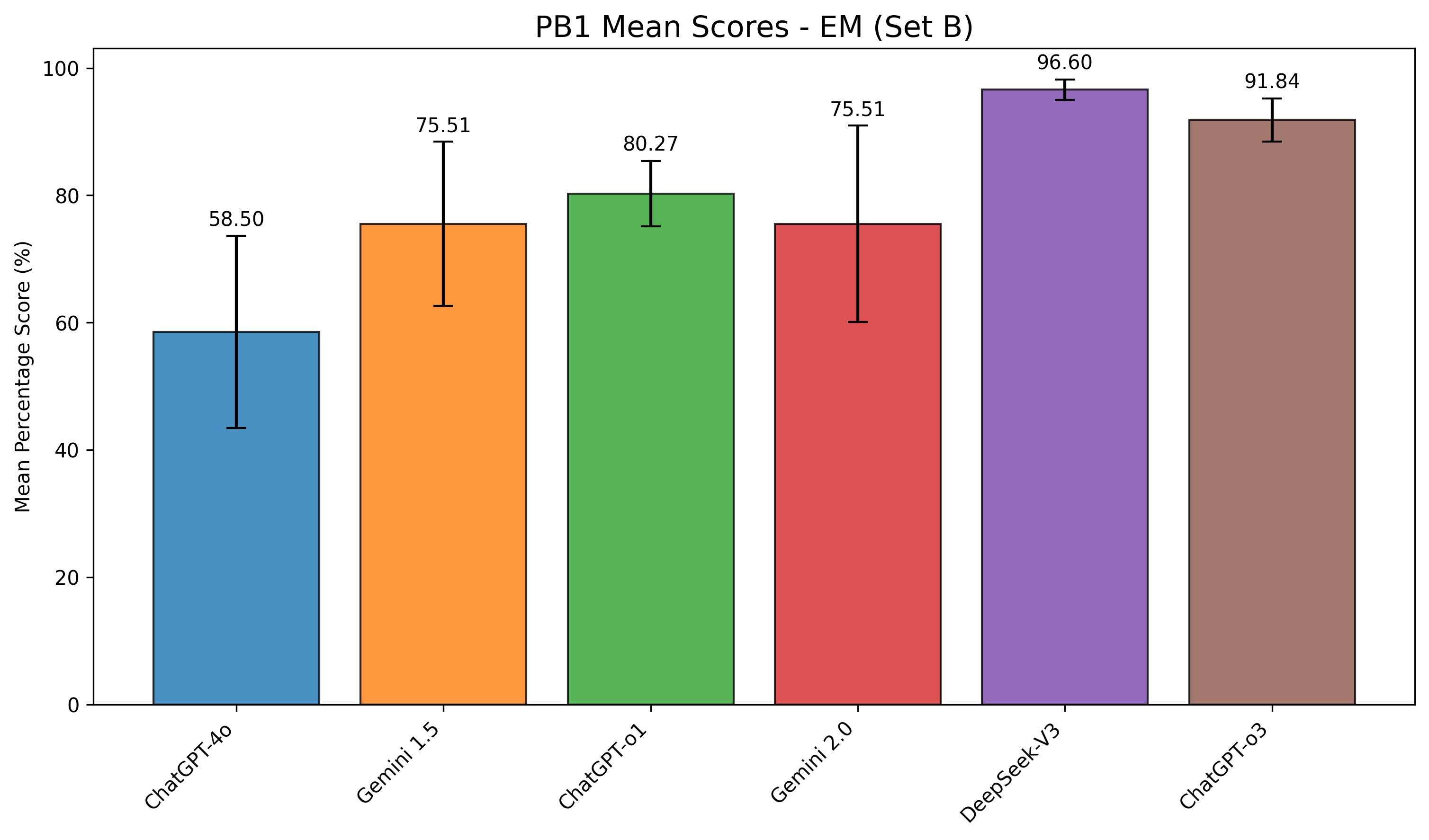}
    \caption{Bar chart showing PB1 model performance in Electromagnetism (Set B)}
    \label{fig:pb1b_em_bar_chart}
\end{figure}

\begin{figure}[htbp] 
    \centering
    \includegraphics[width=0.7\linewidth]{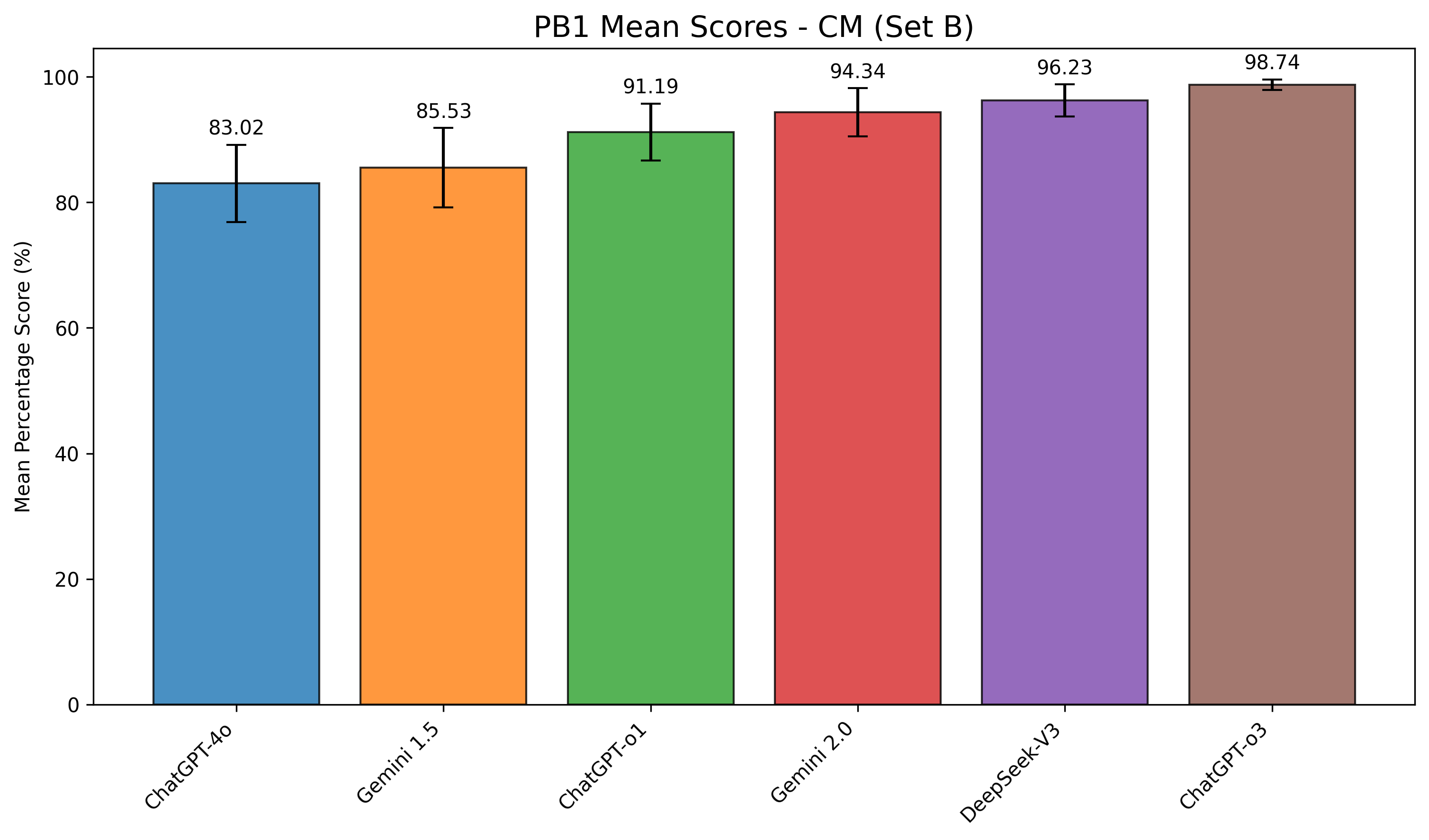}
    \caption{Bar chart showing PB1 model performance in Quantum Mechanics (Set B)}
    \label{fig:pb1b_qm_bar_chart}
\end{figure}

\begin{figure}[htbp] 
    \centering
    \includegraphics[width=0.7\linewidth]{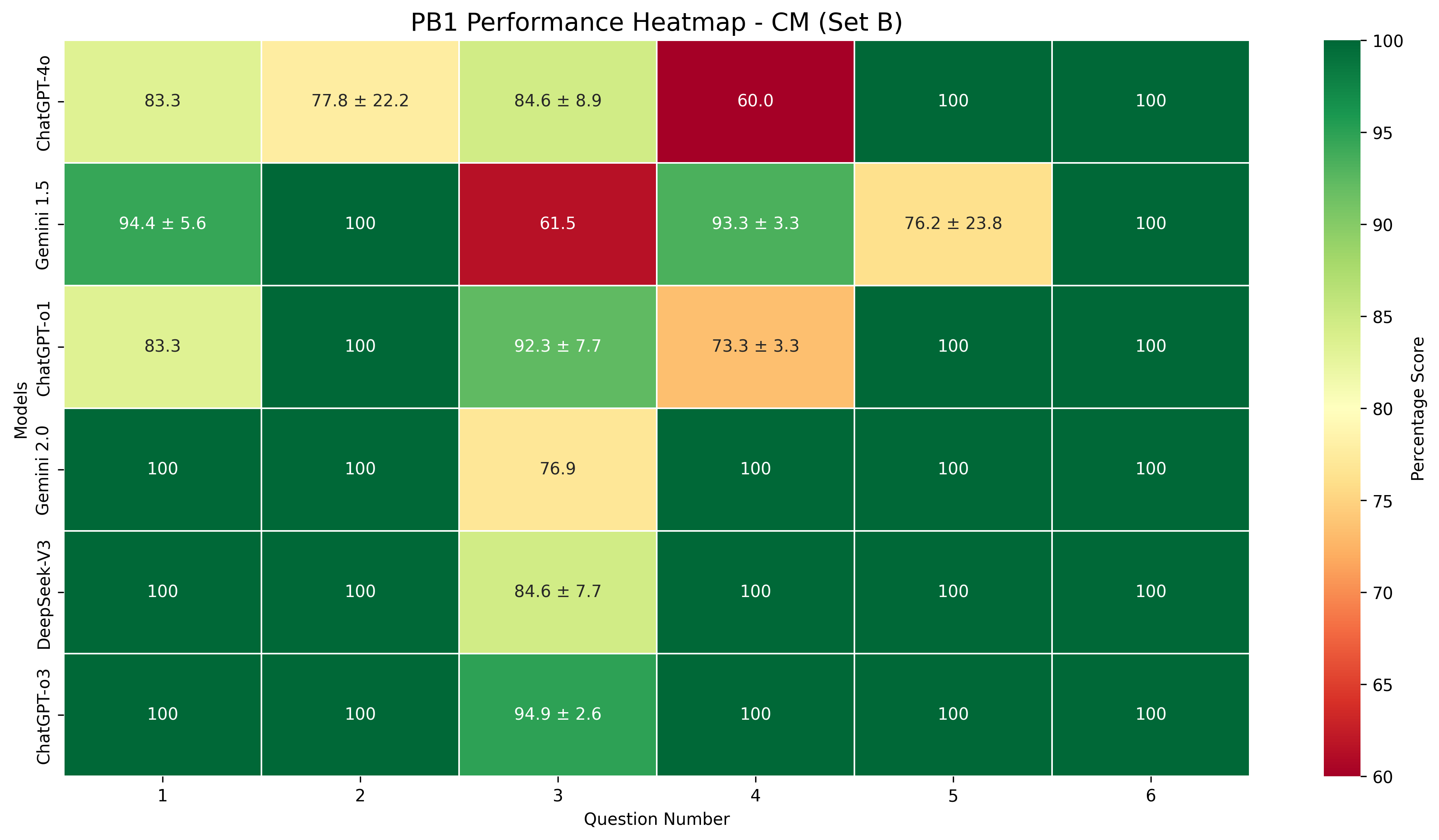}
    \caption{Heat map showing PB1 model performance in Classical Mechanics (Set B)}
    \label{fig:pb1b_cm_heat_map}
\end{figure}

\begin{figure}[htbp] 
    \centering
    \includegraphics[width=0.7\linewidth]{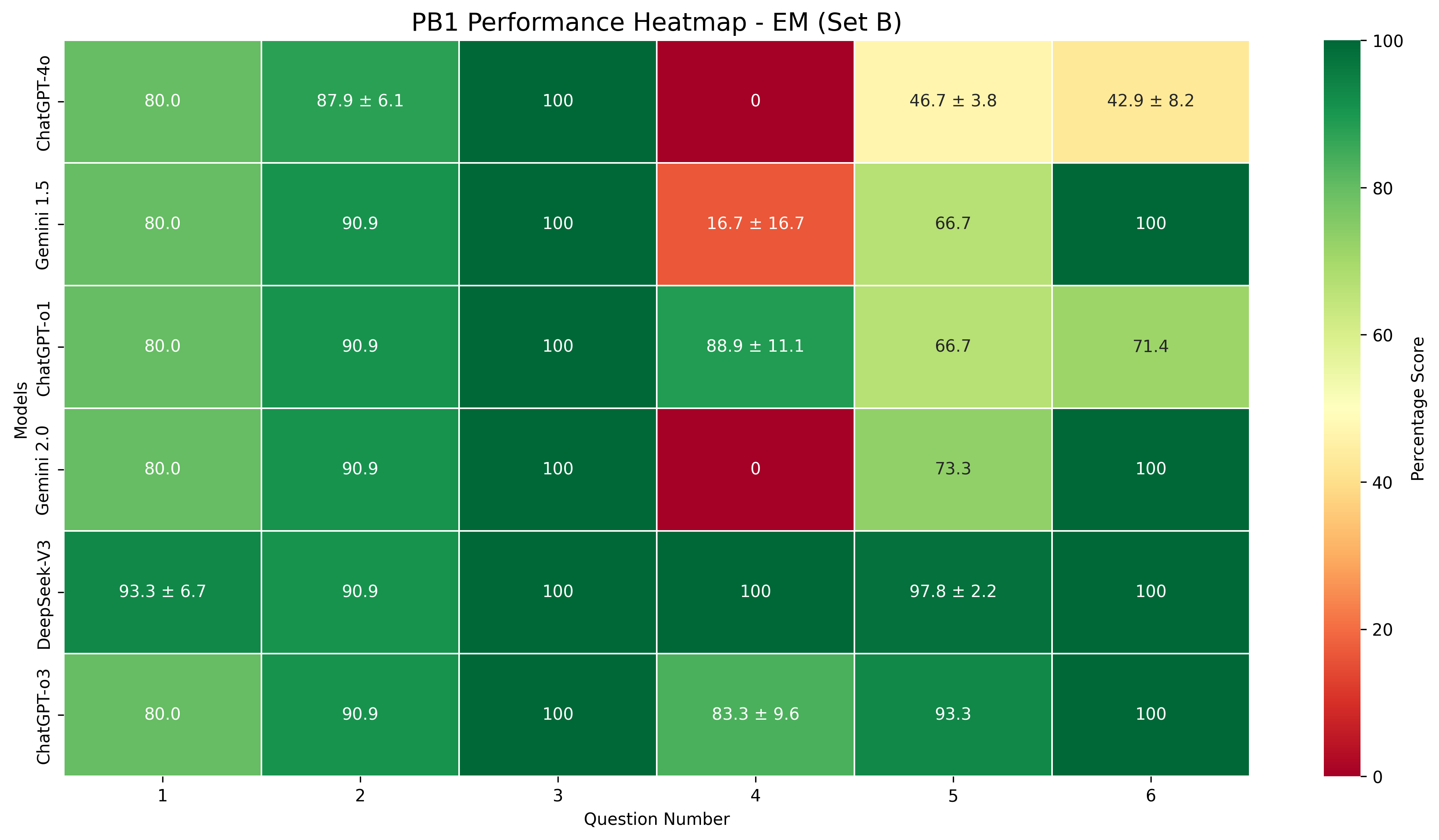}
    \caption{Heat map showing PB1 model performance in Electromagnetism (Set B)}
    \label{fig:pb1b_em_bar_chart}
\end{figure}

\begin{figure}[htbp] 
    \centering
    \includegraphics[width=0.7\linewidth]{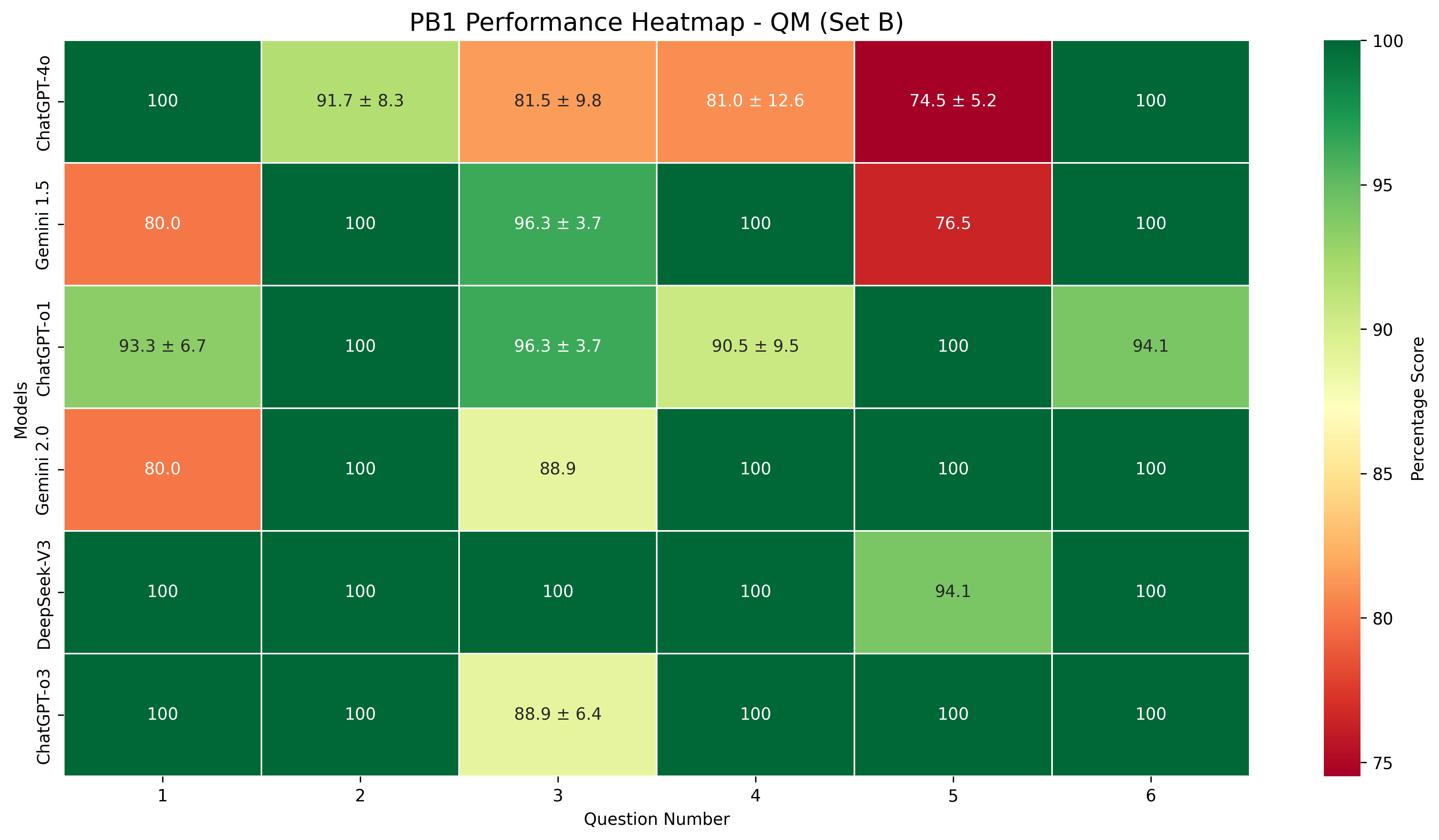}
    \caption{Heat map showing PB1 model performance in Quantum Mechanics (Set B)}
    \label{fig:pb1b_qm_heat_map}
\end{figure}

\begin{table*}[tbhp]
\centering
\begin{tabular}{lccc}
\toprule
\textbf{Model} & \textbf{Classical Mechanics} & \textbf{Quantum Mechanics} & \textbf{Electromagnetism} \\
\midrule
ChatGPT-4o & $83.02 \pm 6.13$ & $87.01 \pm 4.38$ & $58.50 \pm 15.11$ \\
Gemini 1.5 Pro & $85.53 \pm 6.32$ & $90.96 \pm 4.46$ & $75.51 \pm 12.91$ \\
ChatGPT-o1 & $91.19 \pm 4.52$ & $96.05 \pm 1.56$ & $80.27 \pm 5.15$ \\
Gemini 2.0 Flash & $94.34 \pm 3.85$ & $96.61 \pm 3.47$ & $75.51 \pm 15.44$ \\
DeepSeek-V3 & $96.23 \pm 2.56$ & $98.31 \pm 0.98$ & $96.60 \pm 1.61$ \\
ChatGPT-o3 & $98.74 \pm 0.85$ & $98.31 \pm 1.85$ & $91.84 \pm 3.40$ \\
\bottomrule
\end{tabular}
\caption{PB1 Bar Chart data Summary Table (Set B)}
\label{app:pb1b_summary}
\end{table*}

\end{document}